\documentclass[11pt,a4paper]{article}
\usepackage[left=1in, right=1in, top=1in, bottom=1in]{geometry}

\pdfoutput=1
\usepackage[ocgcolorlinks,citecolor=black,colorlinks=true,linkcolor=black,urlcolor=blue]{hyperref}
\usepackage{enumitem}
\usepackage{amssymb,amsmath,amsthm}
\usepackage{graphicx,floatflt,psfrag} 

\newcommand{\Emph}[1]{\emph{#1}}

\newtheorem{MyTheorem}{Theorem}[section] 
\newtheorem{proposition}[MyTheorem]{Proposition}
\newtheorem{lemma}[MyTheorem]{Lemma}
\newtheorem{corollary}[MyTheorem]{Corollary}
\newtheorem{conjecture}[MyTheorem]{Conjecture}
\newtheorem{question}[MyTheorem]{Question}
\newtheorem{claim}[MyTheorem]{Claim}

\newenvironment{EnumRom}
{
\begin{enumerate}
[leftmargin=1.7em, 
topsep=0.3em, 
parsep=0.2em, 
itemsep=0em, 
labelsep=0.2em, 
label={(\roman*)}]
}
{
\end{enumerate}
}

\newenvironment{EnumAlph}
{
\begin{enumerate}
[leftmargin=1.7em, 
topsep=0.3em, 
parsep=0.2em, 
itemsep=0em, 
labelsep=0.2em, 
label={(\alph*)}]
}
{
\end{enumerate}
}

\newsavebox{\smallProofsym}
\savebox{\smallProofsym}  
{%
\begin{picture}(3.5,6)                                   
\put(0,0){\framebox(6,6){}}
\put(0,2){\framebox(4,4){}}
\end{picture}
}%
\newcommand{\smalleop}{\mbox{} \hfill \usebox{\smallProofsym}}

\newenvironment{MyProof}[1]
[Proof]
{
\smallskip

\noindent
\emph{#1.}
}
{
\smalleop
\smallskip
}

\newlength{\mytextwidth}
\newcommand{\placefig}[2]
        {\includegraphics[page=#1,width=#2]{./figures-revision.pdf}}
        
\newcommand{\placefigOLD}[2]
        {\includegraphics[width=#2]{./#1.pdf}}

\newcommand{\RR}{\mathbb{R}}

\newcommand{\NN}{\mathbb{N}}
\newcommand{\ZZ}{\mathbb{Z}}
\renewcommand{\SS}{\mathbb{S}}
\newcommand{\TT}{\mathcal{T}}
\newcommand{\LL}{\mathcal{L}}
\newcommand{\PP}{\mathcal{P}}
\newcommand{\A}{\mathcal{A}}

\newcommand{\st}{\mid}

\newcommand{\conv}[1]{{\sf conv}#1}

\newcommand{\OTaff}[1]{{\sf O\hspace{-0.1em}T}^{_\mathrm{aff}}_{\!#1}}
\newcommand{\OTproj}[1]{{\sf O\hspace{-0.1em}T}^{_\mathrm{proj}}_{\!#1}}
\newcommand{\LOTaff}[1]{{\sf L\hspace{-0.1em}O\hspace{-0.1em}T}^{_\mathrm{aff}}_{\!#1}}
\newcommand{\LorOTaff}[1]{{\sf (\hspace{-0.08em}L\hspace{-0.08em})O\hspace{-0.1em}T}^{_\mathrm{aff}}_{\!#1}}

\newcommand{\g}{{\sf G}}
\newcommand{\gr}{{\sf G^r}}
\newcommand{\G}[1]{\g_{#1}}
\newcommand{\f}{{\sf F}}
\newcommand{\giso}{\overline{G}}

\newcommand{\pbrcx}[1]{\ensuremath{\left[#1\right]}}
\newcommand{\Ex}[1]{\mathbb{E}\pbrcx{#1}}
\newcommand{\Var}[1]{\textrm{Var}\pbrcx{#1}}
\newcommand{\Prob}[1]{\mathbb{P}\pbrcx{#1}}
\newcommand{\mcs}[1]{{\sf mcs}_{#1}}
\newcommand{\Mcs}[2]{{\sf mcs}_{#1}(#2)}
\newcommand{\act}{\triangleleft}

\newcommand{\sgn}{{\chi}} 
\newcommand{\id}{{\sf id}}
\newcommand{\Sym}[1]{{\sf Sym}(#1)}

\newcommand{\Lab}[1]{\overline{#1}} 
\newcommand{\TheNumber}{4- \mbox{$\frac{8}{n^2 - n +2}$}}

\newcommand{\pth}[1]{\left( #1 \right)}
\newcommand*{\eqdef}{\stackrel{_\text{\tiny{def}}}{=}}
\newcommand*{\equivalentdef}{\stackrel{_\text{\tiny{def}}}{\Longleftrightarrow}}

\newcommand{\ff}[3]{f_{#1,#3}\pth{#2}}

\newcommand{\cH}{{\Sigma}} 
\newcommand{\gC}{{C}} 

\newcommand{\ie}{{i.e.,}\ } 
\newcommand{\eg}{{e.g.,}\ } 
\newcommand{\cf}{{cf.}\ }
\newcommand{\etal}{{et\,al.}}

\newcommand{\vs}{{vs.}\ }

\newcommand{\FigRef}[1]{Figure\,\ref{#1}}

\usepackage{color}
\definecolor{blue}{rgb}{0,0,1}
\definecolor{red}{rgb}{1,0,0}
\definecolor{grey}{rgb}{0.6,0.6,0.6}
\definecolor{purple}{rgb}{0.6,0,1}

\newcommand{\grey}[1]{{\color{grey} #1}}

\title{Convex Hulls of Random Order Types\thanks{This research started at the Banff Workshop ``Helly and Tverberg Type Theorems'', October 6-11, 2019, at the Casa Matem\'atica Oaxaca (CMO), Mexico.}}

\begin{document}

\author{
  Xavier Goaoc\footnote{Supported by grant ANR-17-CE40-0017 of the French National Research Agency (ANR project ASPAG) and Institut Universitaire de France.}\\
  Universit\'e de Lorraine, CNRS, INRIA\\
  LORIA,  F-54000 Nancy, France\\
  {\tt xavier.goaoc@loria.fr}
  \and Emo Welzl\footnote{Supported by the Swiss National Science Foundation within the collaborative DACH project Arrangements and Drawings as SNSF Project 200021E-171681}\\
  Dept.\ of Computer Science\\
  ETH Z\"urich, Switzerland
  \\ 
  {\tt emo@inf.ethz.ch}
  }
\date{}
\maketitle


\begin{abstract}
  We establish the following two main results on order types of points
  in general position in the plane (realizable simple planar order
  types, realizable uniform acyclic oriented matroids of rank~$3$):
  \begin{EnumAlph}
  \item 
  \label{it:test}
  The number of extreme points in an $n$-point order type,
    chosen uniformly at random from all such order types, is on
    average $4+o(1)$. For labeled order types, this number 
    has average $\TheNumber$ and variance at most $3$.

  \item The (labeled) order types read off a set of $n$ points sampled
   independently from the uniform measure on a convex planar domain,
   smooth or polygonal, or from a Gaussian distribution are
   concentrated, \ie such sampling typically encounters only a
   vanishingly small fraction of all order types of the given
   size.
   \end{EnumAlph}
    Result~(a) generalizes to arbitrary dimension $d$ for labeled
    order types with the average number of extreme points $2d+o(1)$
    and constant variance. We also discuss to what extent our methods
    generalize to the abstract setting of uniform acyclic oriented
    matroids.  Moreover, our methods allow to show the following
    relative of the Erd\H{o}s-Szekeres theorem: for any
    fixed $k$, as $n \to \infty$, a proportion $1 - O(1/n)$ of the
    $n$-point simple order types contain a triangle enclosing a convex
    $k$-chain over an edge.
    
    For the unlabeled case in (a), we prove that for any antipodal, finite subset of the $2$-dimensional
    sphere, the group of orientation preserving bijections is cyclic,
    dihedral or one of $A_4$, $S_4$ or $A_5$ (and each case is
    possible). These are the finite subgroups of $SO(3)$ and our proof
    follows the lines of their characterization by Felix Klein.
    
   \paragraph{keywords}
   order type; oriented matroid; Sylvester's Four-Point Problem;
   random polytope; sampling random order types; projective plane; excluded pattern; Hadwiger's
   transversal theorem; hairy ball theorem; finite subgroups of $SO(3)$.
   
   \vfill
   
   \paragraph{Acknowledgements}
   The authors thank Boris Aronov for helpful discussions, Gernot Stroth for help on the group theoretic aspects of the paper, and Pierre Calka for help on probabilistic geometry. Moreover, the referees made many suggestions helping us to improve the presentation.
  \vfill
\end{abstract}


\section{Introduction}

Geometric algorithms are often designed \emph{over the reals}, taking
advantage of properties of continuity, closure under arithemic
operations, and geometric figures of $\RR^d$, but implemented \emph{in
  discrete floating point arithmetic}. As documented by, \eg Kettner
\etal\,\cite{kettner2008classroom}, even mild numerical approximations
suffice to provoke spectacular failures in basic geometric algorithms
over simple, non-degenerate inputs. An established approach to address
this issue, carried out for example in the CGAL
library~\cite{cgal:eb-19a}, is to design geometric algorithms that
branch according to predicates of bounded complexity that depend
solely and directly on the numbers in the input of the algorithm
(rather than on numbers resulting from intermediate calculations of
the algorithm); this encapsulates the handling of numerical issues in
the correct evaluation of signs of functions, and since these
functions are typically polynomials, their sign can be efficiently
certified by computer algebra methods such as interval arithmetic and
root isolation (\eg Descartes' rule of sign or Sturm sequences). As a
result, such geometric algorithms effectively operate on a
combinatorial abstraction of the geometric input, as their courses are
determined not by the numerical values given in input, but by the
output of the predicate functions.

\bigskip

One of the simplest geometric predicates is the planar orientation
predicate. The \Emph{orientation} $\sgn(p,q,r)$ of an ordered triple $(p,q,r)$
of points in $\RR^2$ is defined as $1$ (resp. $-1$, $0$) if $r$ is to
the left of (resp. to the right of, on) the line through $p$ and $q$, oriented from
$p$ to $q$. Note that $\sgn(p,q,r)$ equals the sign of the determinant
\begin{equation}
\label{eq:signdet}
{\left|\begin{array}{ccc}
x_p & y_p & 1 \\
x_q & y_q & 1 \\
x_r & y_r & 1
\end{array}
\right|
=
}
\left|\begin{array}{ccc} x_p - x_r & y_p-y_r \\ x_q - x_r &
y_q-y_r\end{array}\right|~,
\end{equation} 
so it evaluates like
a polynomial in the coordinates
of $p$, $q$ and $r$. An algorithm that relies solely on orientation
predicates, for instance Knuth's planar convex hull
algorithms~\cite[$\mathsection 10$ and~$\mathsection 11$]{knuth1992axioms}, will behave identically on two input
point \emph{sequences} $(p_1,p_2,\ldots, p_n)$ and $(q_1,q_2, \ldots, q_n)$ such
that
\begin{equation}\label{eq:samelot}
\forall 1 \le i,j,k \le n, \quad \sgn(p_i,p_j,p_k) = \sgn(q_i,q_j,q_k).
\end{equation}
It is therefore natural to consider such point sequences to be equivalent; this is done by declaring
that they \emph{have the same labeled order type}. This is an
equivalence relation, and a \emph{labeled order type} is an
equivalence class for that relation. An even coarser grouping is
obtained when one identifies point \emph{sets} $P$ and $Q$ for which
there exists a bijection $f\!:P \to Q$ that preserves orientations; an
equivalence class for this coarser relation is called an \Emph{order
  type}. The order type of a point set determines many of its
properties.\footnote{To give a few examples: the face lattice of its
  convex hull, the graphs that can be straight-line embedded onto it,
  including the triangulations it supports, the maximum depth of a
  point from the set with respect to Tukey or simplicial depth, and
  the range space it defines over halfspaces.}
  
\bigskip

Order types, labeled or not, were introduced by Goodman and
Pollack~\cite{goodman1983multidimensional} to study higher-dimensional
analogues of sorting, just like \emph{uniform oriented matroids} were
devised, independently, by Bland in his PhD
thesis~\cite{blandcomplementary} to study the simplex algorithm, by
Folkman and Lawrence~\cite{folkman1978oriented} to study face lattices
of polytopes, and by Las Vergnas~\cite{lvlj1975matroides} to study
questions in graphs and combinatorics, and later rediscovered by
Knuth~\cite{knuth1992axioms} to study convex hull algorithms. These
two structures are actually closely related. The orientation
predicate, and therefore the notion of (labeled) order type can be
defined in any \emph{topological affine
  plane}~\cite{salzmann1967topological}, that is, in any geometry
defined by a system of simple, connected, unbounded curves (called
\emph{pseudolines}) satisfying the usual incidence axioms (any two
points are on exactly one pseudoline, and any two pseudolines
intersect in at most one point), and some continuity
conditions~\cite[$\mathsection 1$]{salzmann1967topological}. An order
type is called \emph{abstract} if it can be constructed in a
topological affine plane, and \emph{realizable} if it can be
constructed in the usual, euclidean, affine plane. The
Faulkman-Lawrence representation theorem~\cite{folkman1978oriented}
asserts that abstract order types coincide with the 
relabeling classes of acyclic uniform oriented matroids of
rank~$3$.\footnote{More generally, abstract and realizable order types
  can be defined in dimension $d$ and the abstract ones coincide with
  the relabeling classes of acyclic uniform oriented matroids
  of rank $d+1$.} These two structures, abstract \vs realizable, do,
however, behave very differently from a computational point of view:
abstract order types can be characterized by a handful of axioms on up
to five points, whereas deciding if a given abstract order type is
realizable is
$\exists\RR$-complete~\cite{shor1991stretchability,Schaefer09}. The
reason for that is Mn\"ev's universality theorem~\cite{Mnev90}, which
essentially states that for any semi-algebraic set $S$, there is a
planar order type whose space of realizations is homotopy equivalent
to $S$. This universality propagates to some structures determined by
order types, for instance polytopes, even simplicial
ones~\cite{adiprasito2017universality}, or Delaunay
triangulations~\cite{lvlj1975matroides}.

\bigskip

A geometric algorithm or conjecture can sometimes be \emph{tested} by
trying it on a large number of (pseudorandomly generated) candidate
point sets. If the algorithm/conjecture actually depends on the order
type of the input point set, this is merely a way of trying it on
candidate order types.\footnote{For example, the largest point set in
  general position with no empty convex hexagon is known to have size
  between $29$ and $1716$~\cite{overmars2002finding,gerken2008empty};
  it is tempting to try and improve the lower bound by testing order
  types of size $30$ or so.}  The first result of this paper
(Theorem~\ref{t:concentration}) is that many standard models of random
point sets explore very inefficiently the space of (labeled) order
types. To our knowledge, this is the first theoretical result
on the quality of \emph{any} method for generating random (labeled)
order types.

\bigskip

\begin{figure}[htb]
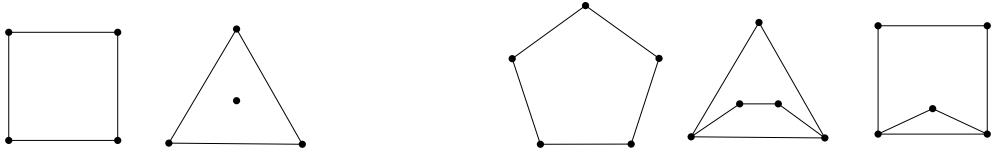

\centerline{
\placefig{1}{0.25\textwidth} 
\hspace{0.15\textwidth} 
\placefig{2}{0.4\textwidth}
}
\caption{Left: The two simple 4-point order types. Right: The three simple 5-point order types.}
\label{f:n5h5-ProjEquiv}
\end{figure}

We establish this concentration result by proving, and this is our
main result, some sharp bounds on the expected number of
\emph{extreme} points in a typical (labeled) order type; extreme
points are points that appear as vertices of the convex hull of the
point set. (Since the number of extreme points is the same for all
representatives of an order type, we speak of the number of extreme
points of the order type; we do the same for every notion independent
of the choice of representative, \eg the size.)  Here we consider only
\Emph{simple (labeled) order types}, \ie with no three points on a line; by
``typical'' we mean chosen equiprobably among all simple (labeled)
order types of a given size~$n$. As an illustration, for $n=4$, the
only two simple order types are the convex quadrilateral and the
triangle with an interior point, so the quantity we are after is
$\frac{4+3}{2} = \frac72$. For $n=5$, it is $\frac{5+4+3}{3} = 4$, see
\FigRef{f:n5h5-ProjEquiv}.

\subsection{Main results}

Let $\OTaff n$ ($\LOTaff n$) denote\footnote{We use `aff' here in order to
  discriminate from the \emph{projective}  order types, which we will
  have to consider later in the course of our investigation.} the set
of simple (simple labeled, resp.) $n$-point order types. For $n \in \NN$, let $\mu_n$ be a
probability measure on $\LorOTaff n$. We say that the family
$\{\mu_n\}_{n \in \NN}$ \emph{exhibits concentration} if there exist
subsets $A_n \subseteq \LorOTaff n$, $n \in \NN$, such that $\mu_n(A_n)
\to 1$ and $|A_n|/|\LorOTaff n| \to 0$. In plain English, families of
measures that exhibit concentration typically explore a vanishingly
small fraction of the space of simple (labeled) order types. Devillers
\etal\,\cite{devillers2018order} conjectured that the order types of
points sampled uniformly and independently from a unit square exhibit
concentration. We prove this conjecture and more:

\begin{MyTheorem}\label{t:concentration}
   Let $\mu$ be a probability measure on $\RR^2$ given by one of the
   following: (a) the uniform distribution on a smooth compact convex
   set, (b) the uniform distribution on a convex compact polygon, (c)
   a Gaussian distribution. The family of probabilities on $\LorOTaff
   n$ defined by the (labeled) order type of $n$ random points chosen
   independently from $\mu$ exhibits concentration.
\end{MyTheorem}

\noindent
Another standard model of random point sets, called the
Goodman-Pollack model, is the random $2$-dimensional projection of an
$n$-dimensional simplex; it is statistically equivalent to points
chosen independently from a Gaussian
distribution~\cite[Theorem\,1]{baryshnikov1994regular}, so the
distribution on random order types it produces in the plane also
exhibits concentration.

\bigskip

We establish Theorem~\ref{t:concentration} by comparing probability
distributions on order types through one statistic: the number of
extreme points. This statistic is already well understood for
distributions induced by random point sets, as it corresponds to the
typical number of vertices in models of random polytopes that are
standard in stochastic geometry. We establish it here for the
combinatorial model. For \emph{labeled} order types, we prove:

\begin{MyTheorem} 
  \label{t:avgl}
  For~$n\ge 3$, the number of extreme points in a random simple
  labeled order type chosen uniformly among the simple, labeled order
  types of size $n$ in the plane has average $\TheNumber$ and variance
  less than $3$.
\end{MyTheorem}

\noindent
For non-labeled order types our statement is less precise:

\begin{MyTheorem}\label{t:avg}
  For~$n\ge 3$, the number of extreme points in a random simple order
  type chosen uniformly among the simple order types of size $n$ in
  the plane has average $4+O\pth{n^{ -1/2+\varepsilon}}$ for any
  $\varepsilon >0$.
\end{MyTheorem}

\noindent
Our proof of Theorem~\ref{t:avgl} extends to arbitrary dimension
(Theorem~\ref{t:avgl-d}), but not our proof of Theorem~\ref{t:avg}. A
large part of our methods and results extend to abstract order
types. In particular, Theorem~\ref{t:avgl} holds in the abstract
setting with the same bound (Theorem~\ref{t:avgl-p}), also in
arbitrary dimension (Theorem~\ref{t:avgl-dp}). The proof of
Theorem~\ref{t:avg} does not completely carry over to the abstract
setting, but our methods yield a similar statement
(Theorem~\ref{t:avg-p}) with an upper bound of $10+o(1)$. 

Whether these methods generalize to order types with collinearities is a
  natural question; we see no easy answer, and consider this to be
  beyond the scope of this paper. Note, however, that for making the conclusion as in Theorem~\ref{t:concentration}, the result for general position is more relevant. Actually, we conjecture that simple order types constitute only a (probably vanishingly) small proportion of all order types (potentially with collinearity), quite contrary to the situation for random order types sampled geometrically as described in Theorem~\ref{t:concentration}.

\subsection{Approach, terminology and further results}

The gist of our method to establish Theorems\,\ref{t:avgl}
and~\ref{t:avg} is to divide up the simple planar order types
into classes, and average the number of extreme points inside
each class.

\subsubsection{Setting and terminology}
\label{s:SettingTerminology}

The division of order types into classes leverages a classical
  correspondence between points and lines in the plane~$\RR^2$, and
  points and great circles on the origin-centered unit sphere $\SS^2$
  in~$\RR^3$. A \Emph{great circle} is the intersection of the sphere
with a plane containing the origin $\mathbf{0}$, an \Emph{open
  hemisphere} is a connected component of the sphere in the complement
of a great circle, and a \Emph{closed hemisphere} is the closure (in
$\SS^2$) of an open one. We call a finite set of points on the sphere
an \Emph{affine set} if it is contained in an open hemisphere. The
\Emph{sign}, $\sgn(p,q,r)$, of a triple $(p,q,r)$ of points on the
sphere is the sign, $-1$, $0$, or $1$, of the determinant of the
matrix $(p,q,r) \in \RR^{3\times 3}$. A bijection $f: S \rightarrow
S'$ between finite subsets of the sphere is \Emph{orientation
  preserving} if $\sgn(f(p),f(q),f(r)) = \sgn(p,q,r)$ for every triple
of points in $S$. Two affine sets have the \Emph{same affine
order type} if there exists an orientation preserving bijection
between them. An \Emph{affine} \Emph{order type} is the equivalence
class of all affine sets that have the same affine order type.

\bigskip

\newcommand{\tgplane}{t}

The plane $\RR^2$ together with its orientation function can be mapped
to any open hemisphere~$\Gamma$ together with~$\sgn$, therefore
relating order types (in $\RR^2$) to affine order types (in
$\SS^2$). Indeed, let $\tgplane$ denote the plane tangent to $\SS^2$
in the center of $\Gamma$. Every affine transform from $\RR^2$ to
$\tgplane$ is of the form
\[ \pth{\begin{array}{c} x\\y \end{array}} \mapsto
A\pth{\begin{array}{c} x\\y\\1 \end{array}}\]
where $A \in \RR^{3 \times 3}$ is non-singular. Let us fix such a
transform with $\det A >0$, and compose it with a central projection
of $\tgplane$ onto $\Gamma$ from $\mathbf{0}$ (which amounts to normalizing
the vector from $\tgplane$). It is apparent from Equation~(\ref{eq:signdet})
that the orientation $\chi(p,q,r)$ of three points in $\RR^2$
coincides with the sign $\chi$ of their images in $\Gamma$. In
  particular, every such map sends every line to a semi-great
  circle, and a segment to a great-circle arc. Conversely, any open hemisphere can be mapped to $\RR^2$ so that the sign $\chi$ corresponds to the orientation function, semi-great circles are mapped to lines, and great-circle arcs are mapped to segments.

\bigskip

\begin{figure}[htb]
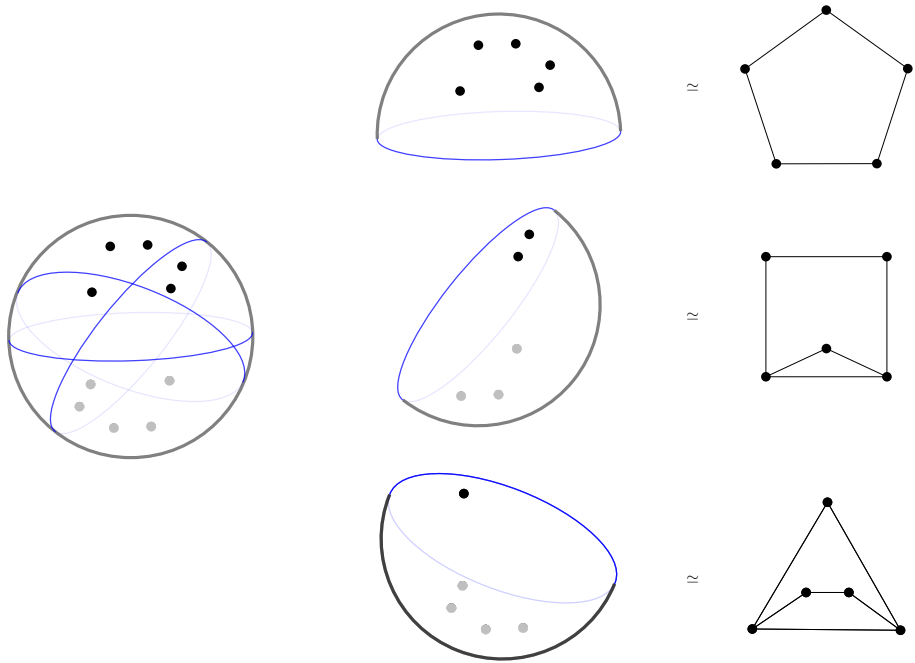

\centerline{ \placefig{4}{0.75\textwidth} } \caption{A projective set
  of size $10$ (left) containing the three simple affine order types
  of size~$5$.}
\label{f:n5-projective}
\end{figure}

We divide up the affine order types into classes as follows. Two
points $p$ and $q$ on the sphere are called \Emph{antipodal} if
$q=-p$. A finite subset $P$ of the sphere is a \Emph{projective set}
if $p \in P \Leftrightarrow -p \in P$.
Starting from an affine
  $n$-point set $A$, we obtain the class of (the affine order type of)
  $A$ as the order types of all the affine $n$-point sets that are
  contained in its \Emph{projective completion} $A \cup
  -A$.\footnote{
  The reader familiar with projective geometry may
      check that two affine sets have the same projective completion
      if and only if there is a projective map that sends (a
      realization of) one to (a realization of) the other. In other
      words, each of our classes is the orbit of an order type under
      the action of projective maps.} We illustrate this idea in
  \FigRef{f:n5-projective} and formalize it properly in
  Section~\ref{s:hemisets}.

\bigskip

This division into classes hints at yet another notion of order
  types, this time for projective point sets. Formally, two
projective sets have \Emph{the same projective order type} if there
exists an orientation preserving bijection between them. A
\Emph{projective order type} is the equivalence class of all
projective sets that have the same projective order type.
We will
  represent the class of the affine order types of an affine set $A$
  by the projective order type of $A \cup -A$. The definitions of
\emph{labeled} affine and projective order types are similar: the
ordering determines the bijection that is required to preserve
orientations. It will sometimes be convenient to write a point
sequence as $A_{[\lambda]}$, where $A$ is the point set and $\lambda:A
\to \{1,2,\ldots,n\}$, $n=|A|$, the bijection specifying the ordering.

\bigskip

We take all our points on the origin-centered unit sphere $\SS^2$ in
$\RR^3$, except for occasional mentions of the origin $\mathbf{0}$,
and restrict our attention to affine and projective sets in general
position. An affine set is in \Emph{general position} if no three
points are coplanar with ${\bf 0}$; a projective set $P$ is in
\Emph{general position} if whenever three points in $P$ are coplanar
with $\mathbf{0}$, two of them are antipodal.

\bigskip

Let $S$ be a finite subset of the sphere. A \Emph{permutation} of $S$
is a bijection $S \to S$ and a \Emph{symmetry} of $S$ is an
orientation preserving permutation of $S$. The symmetries of $S$ form
a group, which we call the \Emph{symmetry group} of $S$. This group
determines the relations between labeled and non-labeled order
types: two orderings $S_{[\lambda]}$ and $S_{[\mu]}$ of a point set
$S$ determine the same labeled order type if and only if $\mu^{-1}
\circ \lambda$ is a symmetry of $S$.

\subsubsection{Further results}

\begin{figure}[htb]
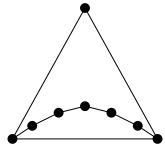

\centerline{
\placefig{41}{0.13\textwidth}
}
\caption{Eight points with three extreme points and the five inner
points forming a convex chain together with one edge of the convex hull.}
\label{f:ConvexChainOverEdge}
\end{figure}

Given two order types $\omega$ and $\tau$, we say that $\omega$
\Emph{contains} $\tau$ if any point set that realizes $\omega$
contains a subset that realizes $\tau$. (Of course this needs only be
checked for a single realization of $\omega$.) By the Erd\"os-Szekeres
theorem~\cite{erdos1935combinatorial}, almost all order types contain
the order type of $k$ points in convex position (since, for $n$ large
enough, \emph{all} order types have $k$ points in convex position). The
relation between affine and projective order types reveals the
following relative:

\begin{MyTheorem}\label{t:avoid}
  For any integer $k \ge 3$, the proportion of order types of size $n$
  that contain $k$ points with $3$ extreme points and the $k-3$ inner
  points forming a convex chain together with one
  edge of the convex hull (see \FigRef{f:ConvexChainOverEdge}) is
  $1-O(1/n)$.
\end{MyTheorem}

 A crucial ingredient in our proof
of Theorem~\ref{t:avg} is a classification of the symmetry groups of
the affine and projective sets. Here it is for affine sets. (The
definitions of layers, sometimes called onion layers, and lonely point are given in
Section~\ref{s:ConvexityOnSphere}.)

\begin{MyTheorem}\label{t:classAff}
  The symmetry group of any affine set $A$ in general position is
  isomorphic to the cyclic group $\ZZ_k$ for some $k \in \NN$ that
  divides the size of every layer of $A$ other than its lonely point
  (if $A$ has one). In particular, $k$ divides $|A|$ (if $A$ has no
  lonely point) or $|A|-1$ (if $A$ has a lonely point); the latter can
  happen for $k$ odd only.
\end{MyTheorem}

\begin{figure}[htb]
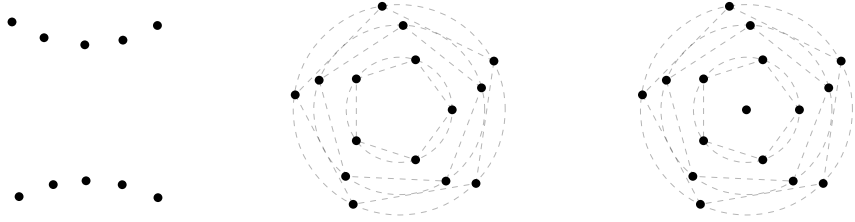

\centerline{
\placefig{5}{0.7\textwidth}
}
\caption{Left: For any even $n \ge 6$, there exists an affine set of $n$
  points with symmetry group $\ZZ_2$: take two sufficiently flat
  convex chains of $n/2$ points each, facing each other {(so-called double chain, \cite{GNT00})}. Center and
  Right: For any $3 \le k \le n$ where $k$ divides $n$ or for any odd
  $k$ where $k$ divides $n-1$, there exists an affine set of $n$
  points with symmetry group $\ZZ_k$: just pile up regular $k$-gons
  inscribed in concentric circles.}
\label{f:example-Zk}
\end{figure}

\noindent
For all values of $k$ and $n$ satisfying $k \mid n$, or $k$ odd and $k \mid n-1$, with the exception of $(k,n) = (2,4)$, there
exists an affine order type of size $n$ with $\ZZ_k$ as symmetry group
(see \FigRef{f:example-Zk}). 

\bigskip

We also prove that the symmetry groups of projective sets are finite
subgroups of $SO(3)$.

\begin{MyTheorem}\label{t:classProj}
  The symmetry group of any projective set of $2n$ points in general
  position is a finite subgroup of $SO(3)$. In particular, it is one
  of the following groups: $\ZZ_1$ (trivial group), $\ZZ_m$ (cyclic
  group), $D_{m}$ (dihedral), with $m \mid n$, or $m \mid n-1$, $S_4$
  (octahedral = cubical), $A_4$ (tetrahedral), and $A_5$
  (icosahedral).
\end{MyTheorem}

\noindent
We give examples of projective point sets with symmetry groups of each of the types identified in Theorem~\ref{t:classProj} (see Section~\ref{s:Gallery}).

\subsection{Related work}

We now briefly discuss previous works related to our results.

\subsubsection{Counting, enumerating and sampling order types}

The space of order types is generally not well understood. To begin
with, its size is not known, not even asymptotically. The most precise
bounds are: there are $n^{4n}\phi(n)$ labeled order types, where
$2^{-c n} \le \phi(n) \le 2^{c'n}$ for some positive constants
$c,c'$~\cite{goodman1986upper,alon1986number}. Factoring out the
labeling requires to account for symmetries; we show that in the
plane, every unlabeled order type corresponds to at least {$(n-1)!$}
{(and clearly at most $n!$)} different labeled ones
(Corollary~\ref{c:labeling}). There is no known efficient algorithm
for enumerating order types; in practice, they have been tabulated up
to size~$11$~\cite{AAK02, AK07}, for which they are already counted in
billions.\footnote{Recently, \emph{abstract} order types have been
  counted up to size $13$ by Rote and Scheucher,
  \path!https://oeis.org/A006247!.}

\bigskip

Random sampling of order types is also quite unsatisfactory. First,
the standard methods in discrete random generation such as Boltzmann
samplers are unlikely to work here, as they require structural results
(such as recursive decompositions) that usually make counting a
routine task. It is of course easy to produce a random order type by
merely reading off the order type of $n$ random points; standard
models include points chosen independently from the uniform
distribution in a square or a disk, from a Gaussian distribution, as
well as points obtained as a random $2$-dimensional projection of a
$n$-dimensional simplex~\cite{bokowski1992distribution}. No random
generation method is known to be both efficient (say, taking
polynomial time per sample) and with controlled bias, and our
Theorem~\ref{t:concentration} is the first negative result in this
direction. This sad state of affairs can perhaps be explained by two
fundamental issues: when working with order types symbolically (say as
orientation maps to $\{-1,0,1\}$), one has to work around the
NP-hardness {(actually, $\exists\RR$-completeness)} of membership (\ie
realizability) testing~{\cite{shor1991stretchability, Mnev90,
    Schaefer09}}. When working with explicit point sets, one has to
account for the exponential growth of the worst-case number of
coordinate bits required to realize an order type of size
$n$~\cite{goodman1990intrinsic}. It is an open question whether
\emph{most} order types can be realized using small (polynomial-size)
coordinates (see Caraballo \etal\,\cite{caraballo2018number} for
recent progress).

\subsubsection{Random polytopes and Sylvester's problem}

Counting extreme points relates to the study of face vectors of random
polytopes, a classical line of research in stochastic geometry
initiated by Sylvester in 1865, who asked for ``the probability that
$4$ points in the plane are in convex position''. A standard model of
a random polytope~$K_n$ is the convex hull of $n$ random points chosen
uniformly and independently in some fixed convex body $K$. In this
setting, the number of extreme points, \ie  of vertices
of $K_n$, is well understood. Its average is asymptotically
pro\-portional to $(1 + o(1))n^{\frac{d-1}{d+1}}$ if $K$ is smooth and
to $(1 + o(1))\log^{d-1} n$ if $K$ is a poly\-tope~\cite{rs-1,rs-2}
(see~\cite[$\mathsection 2.2.2$]{Reitzner-survey}), and up to
multiplicative constants these are the two
extremes~\cite[Theorems~1--3]{BaranyLarman}. There are also estimates
on the variance, concentration inequalities, central limit theorems,
and large deviation inequalities. We refer the interested reader to
the survey of Reitzner~\cite{Reitzner-survey}.

\bigskip

This model of a random polytope naturally generalizes to
arbitrary probability measures, or even to the convex hull
of random 
dependent point sets such as determinantal point
processes. Much less is known in this direction, aside from the
occasional extensively-studied model such as Gaussian polytopes
(see~\cite[$\mathsection 2.3$]{Reitzner-survey}). In a sense, what we
investigate is the average number of extreme points in a random
polytope for a \emph{combinatorially defined} probability distribution
on point sets.

\bigskip

The study of random polytopes also relates to the $\epsilon$-net
theory for halfspaces through the use of floating
bodies~\cite{BaranyLarman} (see also~\cite{Har11}
and~\cite[$\mathsection 3.2$]{barany2019random}). It also relates to
graph drawing:  Blaschke proved that
the probability that $4$ points chosen uniformly in a convex domain
are in convex position is minimized when the domain is a triangle; for
arbitrary planar probability measures, this merely asks for the limit
as $n \to \infty$ of the rectilinear crossing number of the complete
graph.

\subsubsection{Symmetry groups of oriented matroids}

The symmetry groups of 
  oriented matroids of rank $2$ and $3$ were previously
classified by Miyata~\cite{Miy13}. Although phrased for realizable
order types, our proof of Theorem~\ref{t:classAff} extends to abstract
ones and offers 
an alternative to Miyata's
proof 
in the case of acyclic matroids~\cite[$\mathsection
  6$]{Miy13}. As we spell out in Section~\ref{s:generalizations}, some
of our other proofs also extend to the abstract setting.

\subsubsection{Order types of random point sets}

Several recent works have studied order types of random point
sets~\cite{cardinal2020chirotopes,devillers2018order,fabila2017order,han2019extremal,van2019smoothed},
but they do not address the \emph{equiprobable} distribution on
$n$-point order types. The recent work of Chiu \etal\,\cite{CFS19}
comes closer, as they have looked at the average size of the $j$th
level in a random planar arrangement of $n$ lines, chosen by fixing a
projective line arrangement of size $n$ and equiprobably choosing a
random cell to contain the south-pole. This is similar to what we do,
but let us stress that they do not take symmetries into account, so
the actual distribution on planar arrangements they consider is not
equiprobable (not even among those contained in the projective
arrangement).

\subsubsection{Order types with forbidden patterns}

Order types with forbidden patterns were previously investigated in
several directions. The Erd\H{o}s-Szekeres theorem was
strengthened for order types with certain forbidden
patterns~\cite{nevsetvril1998ramsey,karolyi2006erdHos,karolyi2012erdHos}. Han \etal\,\cite{han2019extremal} studied the patterns
contained in random samples. Eppstein~\cite{eppstein2018forbidden} offers a beautiful small book about forbidden configurations, which concentrates on patterns in degenerate position (see Theorems 8.13, 8.16, 9.7, 11.22, and 12.3, and Lemma 15.14 in \cite{eppstein2018forbidden}). We are not aware of previous results on the number of order types with a forbidden pattern in general position such as Theorem~\ref{t:avoid}.

\subsection{Open problems}
\label{s:OpenProblems}

In our opinion, the most prominent open problem is the design of a
method for generating pseudorandom order types that is both efficient
(say, taking polynomial time per sample) and with controlled
bias. Our methods reveal that this problem should perhaps be
  approached by sampling \Emph{projective order types} first, an idea
  that we discuss in Section~\ref{s:outlook}. Here, let us say that
  one approach we believe \emph{does not} work is the
following (with the terminology of Theorem~\ref{t:concentration}):

\begin{conjecture}\label{c:weak}
 Let $\mu$ be a probability measure on $\RR^2$ for which every line is negligible and such that the expected
  number of extreme points among $n$ random points chosen
  independently from $\mu$ goes to infinity as $n \to \infty$. The
  family of probabilities on $\LorOTaff n$ defined by the (labeled)
  order type of $n$ random points chosen independently from $\mu$
  exhibits concentration.
\end{conjecture}

We actually believe that a stronger conjecture holds.

\begin{conjecture}\label{c:strong}
 Let $\mu$ be any probability measure on $\RR^2$ for which every
   line is negligible. The family of probabilities on $\LorOTaff n$
 defined by the (labeled) order type of $n$ random points chosen
 independently from $\mu$ exhibits concentration.
\end{conjecture}

We have only weak indicators for Conjecture~\ref{c:strong}. As it is easily seen, a distribution that exhibits perfect uniform distribution on all $\LorOTaff n$, $n \in \NN$, for random points chosen independently from $\mu$ is not possible, since random order types do not satisfy a ``reducibility'' condition which is true for any i.i.d.\ sampling, namely that removing random points from a random configuration gives a random configuration: For example, if we sample a random $5$-point order type, and then remove one of the five points at random, then we get the convex position $4$-point order type with probability $\frac{1}{3}\left(1 + \frac{1}{5} +  \frac{3}{5} \right) = \frac{3}{5}$ (check in \FigRef{f:n5h5-ProjEquiv}), and not $\frac{1}{2}$, as we get it for a random $4$-point order type (see \FigRef{f:n5h5-ProjEquiv}). This irreducibility also implies, for instance, that for any distribution $\mu$ on $\RR^2$ there are two order types of size $6$ whose probabilities differ by a factor of more than $1.8$, see Goaoc \etal\,\cite[Prop.\,2]{goaoc2018limits}. Clearly, none of this implies concentration as $n$ grows.

\medskip
One approach to bypass the $\exists\RR$-completeness of testing
realizability of order types is to work in a class of abstract order
types that is not too large (having in mind that the number of
abstract order types, $e^{\Omega(n^2)}$, grows much faster than the
number of realizable ones, $e^{O(n \log n)}$). A natural way to filter out
abstract order types is to forbid them from containing patterns
violating certain ``affine theorems''.

\begin{question}
Is it true that for any \emph{fixed} (abstract) order type $\tau$, the number
  of (abstract) order types of size $n$ that \emph{do not} contain $\tau$ is
  vanishingly small as $n \to \infty$?
\end{question}

\noindent
The answer is positive for $\tau$ the order type of points in convex
position (the Erd\"os-Szekeres theorem~\cite{erdos1935combinatorial}),
a triangle with one interior point (Carath\'eodory's theorem) and a
triangle with a convex chain over an edge
(\FigRef{f:ConvexChainOverEdge}, Theorem~\ref{t:avoid}). The question may seem quite bold given the
limited number of observations, but it is also motivated by an
analogous phenomenon for permutations: the Marcus-Tardos
theorem~\cite{marcus2004excluded} asserts that for every fixed
permutation $\pi$, the number of size-$n$ permutations that \emph{do
  not} contain $\pi$ is at most exponential in $n$
(see~\cite{marcus2004excluded} for the definition of containment).

\bigskip

The paper by Aloupis \etal\,\cite{AILOW14} addresses the complexity of
order type isomorphism via so-called canonical labelings, improving
bounds by Goodman and Pollack \cite{goodman1983multidimensional}. They
describe an $O(n^d)$ time algorithm for computing the automorphisms of
an order type (what we will call the symmetry group of orientation
preserving permutations) for a set of $n$ points in $\RR^d$ (or an
acyclic oriented matroid of rank $d+1$ given by an orientation
oracle), \cite[Theorem\,4.1]{AILOW14}. While in~\cite{AILOW14} 
evidence is given that $O(n^d)$ is optimal for deciding whether two point sets
have the same order type, it is not excluded that the symmetry group
of a point set can be computed faster, at least for small $d$.

\subsection{Paper organization}

We recall some background material in Section~\ref{s:background}. The
paper is then organized in three parts:

\begin{itemize}
\item Sections~\ref{s:hemisets} and~\ref{s:labeled} deal with labeled
  affine order types. Section~\ref{s:hemisets} clarifies the relation
  between affine and projective order types, between their symmetry
  groups, and between the affine subsets of a projective sets and the
  cells of its dual arrangement. Section~\ref{s:labeled} proves
  Theorem~\ref{t:avgl} by relating the number of extreme points in a
  random affine order type to the number of edges in a random cell
  of an arrangement of great circles, and by analyzing such
  arrangements via double counting and the zone theorem.\bigskip
  
\item Sections~\ref{s:poles} and~\ref{s:non-labeled} deal with affine
  order types. Section~\ref{s:poles} proves that every symmetry of a
  projective set stabilizes exactly two subsets contained in a closed hemisphere --
  a combinatorial analogue of the property that any rotation in
  $\RR^3$ fixes two points of the sphere. This allows us, in
  Section~\ref{s:non-labeled}, to extract some information on
  projective symmetry groups by adapting the analysis of Klein leading
  to the classification of finite subgroups of $SO(3)$. We then
  analyze affine symmetries, proving Theorem~\ref{t:classAff}, and
  establish Theorem~\ref{t:avg}.\bigskip
  
\item The last five sections are independent complements to
  Theorems~\ref{t:avgl} and~\ref{t:avg}.
  Section~\ref{s:concentration} relates concentration results on
  extreme points to concentration on the distribution of order types,
  and proves Theorem~\ref{t:concentration}. Section~\ref{s:avoid} uses
  the projective setup to extend, in some sense, the Erd\"os-Szekeres
  theorem and prove Theorem~\ref{t:avoid}. Section~\ref{s:classProj}
  completes the study of projective symmetries into the
  characterization of Theorem~\ref{t:classProj} and discusses some of
  its extensions. Section~\ref{s:generalizations} presents
  generalizations of Theorems~\ref{t:avgl} and~\ref{t:avg} to higher
  dimensions and to abstract order types (that is, acyclic uniform oriented
  matroids). Section~\ref{s:outlook} discusses how projective order
  types may help sampling (labeled) order types efficiently.
\end{itemize}

\section{Background}
\label{s:background}

We recall here some notions in finite group theory and in discrete
geometry on $\SS^2$ (duality, arrangements, convexity).

\subsection{Groups}

The elements of group theory we use deal with a subgroup $G$ of the
group of permutations of a finite set $X$. The identity map, the
neutral element in $G$, is denoted by $\id$ or $\id_X$. We will study
such a group $G$ through its action on $X$ or some set of subsets of
$X$. The \Emph{orbit} $G(x)$ of $x \in X$  is the image of $x$ under
$G$, \ie $G(x) \eqdef \{g(x) \st g \in G\}$. Any two elements
have disjoint or equal orbits, so the orbits partition $X$. The
\Emph{stabilizer} of an element $x \in X$ is the set of permutations
in $G$ having $x$ as a fixed point, \ie $G_x \eqdef \{g \in G
\st g(x) = x\}$. The \Emph{orbit-stabilizer theorem asserts that for any group $G$ acting on a set $X$,} $|G| =
|G(x)|\cdot|G_x|$ for every $x \in X$. We write $\simeq$ for group
 isomorphism. 

\subsection{Duality and arrangements on $\SS^2$}
\label{s:DualityArr}

On the sphere, the \Emph{dual} of a point $p$ is the great circle $p^*$
contained in the plane through~${\bf 0}$ and orthogonal to the line
$\mathbf{0}p$. For any finite subset $S$ of the sphere, we write $S^*$
for the arrangement of the family of great circles $\{p^* \st p \in
S\}$.

Let $P$ be a projective set of $2n$ points in general position. Since antipodal points
have the same dual great circle, $P^*$ is an arrangement of $n$ great
circles. Observe that $P$ is in general position if and only if no
three great circles in $P^*$ have a point in common. Any two great
circles intersect in two points, so $P^*$ has $2{n \choose 2}$
vertices. {Every vertex is incident to four edges; the total number of edges is therefore $4{n \choose 2}$.} By Euler's formula,
$P^*$ has $2{n \choose 2}+2$ faces of dimension $2$, which we call \Emph{cells}.

Let us recall that many combinatorial quantities on arrangements of
great circles on $\SS^2$ are essentially twice their analogues for
arrangements of lines in $\RR^2$. Indeed, starting with an arrangement
$P^*$ of $n$ great circles in general position, we can add another
great circle $C_\infty$, chosen so that $P^* \cup \{C_\infty\}$ is
also in general position, and consider the two open hemispheres
bounded by $C_\infty$. Each open hemisphere can be mapped to $\RR^2$
by a central projection onto a plane parallel to $C_\infty$, so that
the half-circles of $P^*$ are turned into lines, and the two line
arrangements are combinatorially equivalent by antipodality. In this
way, we can for instance obtain the following version of the zone
theorem from the bound given in \cite{BEPY90} for the zone of a line
in an arrangement of lines:\footnote{In~\cite{BEPY90} it is shown that
  the cells in the zone of a line $h_0$ in an arrangement of $n+1$
  lines in the plane has edge-complexity at most $\lfloor
  19n/2\rfloor-1$. For translating this bound to the zone of a great
  circle in an arrangement of $n$ great circles on $\SS^2$, (i) we
  replace $n$ by $n-1$, (ii) we double for the two sides of
  $C_\infty$, and (iii) we subtract 8 for the edges that get merged
  along $C_\infty$ (note that the infinite edges on $h_0$ get merged
  and contribute 1 on each of their sides). Note that the unpublished
  manuscript
  \path!http://www2.math.technion.ac.il/~room/ps_files/zonespl.pdf! by
  Rom Pinchasi improves the bound in \cite{BEPY90} by 2 to $\lfloor
  19n/2\rfloor-3$.}

\begin{MyTheorem}[Zone Theorem]
\label{t:Zone}
Let $P^*$ be an arrangement of $n$ great circles on $\SS^2$ and let $p^* \in P^*$. Let $Z(p^*)$ denote the \Emph{zone} of $p^*$, \ie the set of cells of the arrangement incident to $p^*$. For a cell $c$, let $|c|$ denote the number of edges incident to $c$. Then $\sum_{c \in Z(p^*)} |c| \le 19(n-1) - {{10}}$.
\end{MyTheorem}

\subsection{Convexity on the sphere}
\label{s:ConvexityOnSphere}

A point $p\in A$ is \Emph{extreme} in an affine set $A$ if there
exists a great circle $C$ that strictly separates $p$ from $A
\setminus \{p\}$; that is, $p$ and $A \setminus \{p\}$ lie in two
different connected components of $\SS^2 \setminus C$. An ordered pair
$(p,q) \in A^2$, $p \neq q$, is a \Emph{positive extreme edge} of $A$ if for all $r
\in A\setminus \{p,q\}$ we have $\sgn(p,q,r) = +1$. Assuming general
position and $|A| \ge 2$, a point $p \in A$ is extreme in $A$ if and only if there
exists $q \in A$ such that $(p,q)$ is a positive extreme edge; in that
case, the point $q$ is unique. 

A \Emph{CCW order} of the extreme points of $A$ is an order
$(p_0,p_1,\ldots, p_{h-1})$ of its extreme points such that for all
$i=0,1,\ldots, h-1$, $(p_i,p_{i+1})$ is a positive extreme edge
(indices $\bmod\, h$). The \Emph{convex hull} of $A$ is
\[
\conv(A) \eqdef \bigcap_{\text{closed hemisphere\,} \cH \supseteq A} \cH
\]
which equals, for $A$ in general position and $|A| \ge 3$,
\[ 
\{r \in \SS^2 \st \forall \hbox{ positive extreme edges } (p,q),
\sgn(p,q,r) \ge 0\}.
\]

An affine set $A$ is \Emph{in convex position} if every point is
extreme in $A$. The (onion) \Emph{layer sequence} of $A$ is a sequence
$(A_0,A_1,\ldots,A_\ell)$ of subsets of $A$, partitioning $A$, where
$A_0$ is the set of extreme points in $A$, and $(A_1,A_2, \ldots,
A_\ell)$ is the layer sequence of $A \setminus A_0$ (if $A=\emptyset$,
then the layer sequence is empty). The $A_i$'s are called the
\Emph{layers} of $A$. If the innermost layer $A_\ell$ consists of a
sole point, then that point is called \Emph{lonely}.
There is one or no lonely point.

\section{Hemisets: relating affine and projective order types}
\label{s:hemisets}

Any affine set $A$ naturally defines a projective set $A \cup -A$,
which we call its \Emph{projective completion}. Going in the
  other direction, we define a \Emph{hemiset} of a projective set $P$
  as the intersection of $P$ with a \Emph{closed} hemisphere, and call
  a hemiset of $P$ an \Emph{affine hemiset} of $P$ if it is contained
  in an open hemisphere (or, equivalently for general position, a hemiset that contains no antipodal pair). With these definitions, we have: 

\begin{lemma}\label{l:hemiset}
  A projective set $P$ is the projective completion of an affine
  set $A$ if and only if $A$ is an affine hemiset of $P$.
\end{lemma}
\begin{MyProof}
  Let $P$ be a projective set and let $A$ be an affine set. If $P = A
  \cup -A$, then any open hemisphere $\Sigma$ that contains $A$ has no
  point of $P$ on its boundary, and the closure of $\Sigma$ intersects
  $P$ in $A$. Conversely, if $A = \Sigma \cap P$ for some closed hemisphere
  $\Sigma$, then every point $p \in P \setminus A$ must be in
  interior of $-\Sigma$, so that $-p \in P \cap \Sigma = A$ and $P = A
  \cup -A$. 
\end{MyProof}

\bigskip

\noindent
We note that an affine set is in general position if and only if
its projective completion is. Although we are primarily interested
in affine hemisets, it will be instrumental to consider also hemisets
that are not affine. Note that for an open hemisphere to cut out an
affine set that completes to $P$, it must be bounded by a great circle
that avoids $P$. For instance, the set of vertices of the cross
polytope $P \eqdef\{(\pm1,0,0), (0,\pm1,0), (0,0,\pm1)\}$ intersects
some open hemispheres in a single point.

\paragraph{Notation.}~
Now seems a good time to introduce or recall our notation. For $n \ge 3$ we write
$\LOTaff n$ for the set of simple labeled affine order types of
size~$n$, $\OTaff n$ for the set of simple affine order types of
size~$n$, and $\OTproj n$ for the set of simple projective order types
of size $2n$. For an affine point set $A$ with affine order type
$\omega$, we write $\LOTaff A = \LOTaff \omega$ for the set of the
labeled affine order types of the orderings of $A$. For a projective point set $P$ with projective
order type $\pi$, we write $\LorOTaff P = \LorOTaff \pi$ for the set
of affine (labeled) order types of the affine hemisets of $P$.

\subsection{Symmetries acting on hemisets}

To understand how affine order types relate to projective order types,
an important idea is that the symmetries of a projective point set $P$
act on the (affine) hemisets of $P$. This action also carries the
following structure. We define the \Emph{layer sequence} of a hemiset
$B$ of a projective set $P$ as the sequence $(B_{-1}, B_0, B_1,
\ldots, B_\ell)$ of subsets of $B$, where $B_{-1} \eqdef B \cap -B$,
and $(B_0, B_1, \ldots, B_\ell)$ is the layer sequence of the affine
set $B \setminus B_{-1}$. In particular, $B_{-1} = \emptyset$ if and
only if $B$ is an affine hemiset.\footnote{When the context makes it
  clear that we are dealing with affine sets, we may drop the term
  $B_{-1}$ for affine sets to fall back on the definition of layer
  sequence given in Section~\ref{s:ConvexityOnSphere}.} If the
innermost layer $B_\ell$ consists of a sole point, then that point is
called \Emph{lonely}.

\begin{proposition}\label{p:act}
  Let $g:P \to P'$ be an orientation preserving bijection between two
  projective sets in general position, $|P|=|P'| \ge 6$.
  \begin{EnumRom}
  \item \label{i:iact} $g$ maps hemisets of $P$ to hemisets of $P'$
    and affine hemisets of $P$ to affine hemisets of $P'$.
  \item \label{i:iiact} If a hemiset $B$ of $P$ has layer sequence
    $(B_{-1}, B_0, B_1, \ldots, B_\ell)$, then its image $g(B)$ has
    layer sequence $\pth{g(B_{-1}), g(B_0), g(B_1), \ldots,
      g(B_\ell)}$.
  \end{EnumRom}
\end{proposition}

The rest of this section is devoted to the proof of
Proposition~\ref{p:act}. We start with a basic lemma.\footnote{The
  lemma basically states that if the points on $\SS^2$ are considered
  as vectors in $\RR^3$, then orientation preserving bijections map
  sets of convexly dependent vectors to sets of convexly dependent
  vectors.}

\begin{lemma}
\label{l:EmbracingPreservation}
  Let $g: S \to S'$ be an orientation preserving bijection between two
  subsets $S$ and $S'$ of the sphere, with $S$ not contained in a great circle. 
\begin{EnumRom}
\item
\label{i:iEmbracingPreservation} 
If $\{p,-p\}$ is a pair of antipodal points in $S$, then $g(-p)=-g(p)$.
 \item
\label{i:iiEmbracingPreservation} 
If $X$ is a set of points in $S$ whose convex hull (in $\RR^3$) contains $\mathbf{0}$ in its interior, then the convex hull of $g(X)$ contains $\mathbf{0}$ in its interior.
\end{EnumRom}
\end{lemma}
\begin{MyProof}
  \ref{i:iEmbracingPreservation}
  We have $g(-p)\neq g(p)$ since $g$ is bijective. If $g(-p)$ and
  $g(p)$ are not antipodal, then they span a unique great circle
  $\gC$. For $r \in S'$ we have $0 = \sgn(p,-p,g^{-1}(r)) =
  \sgn\pth{g(p),g(-p),r}$, \ie all points in $S'$ lie on $\gC$, and
  therefore all points in $S$ lie on a great circle, contrary to our
  assumption.
\smallskip

\noindent
\ref{i:iiEmbracingPreservation} The convex hull of $X$ contains $\mathbf{0}$ in its interior if and only if there exists a pair of \emph{non-antipodal} points in $X$ and for any two non-antipodal points $p$ and $q$ in $X$, the plane spanned by $p$, $q$, and $\mathbf{0}$ has points $r'$ and $r''$ in $X$ on opposite sides, \ie $0 \neq \sgn(p,q,r') = - \sgn(p,q,r'')$. Clearly, also with \ref{i:iEmbracingPreservation}, this property is preserved by an orientation preserving bijection.
\end{MyProof}

\bigskip

This readily gives a more local characterization of (affine) hemisets:

\begin{corollary}\label{c:characterizations2}
Let $P$ be a projective set in general position with $|P|\ge 6$. A subset $B \subseteq
  P$ is a hemiset of $P$ if and only if (a) $B$ contains at least one
  point of every antipodal pair in $P$, and (b) 
  the convex hull of $B$
  does not contain 
  $\mathbf{0}$ in its
  interior. Moreover, a hemiset $B$ of $P$ is affine if and only if
  (c) $|B|=|P|/2$.
\end{corollary}
\begin{MyProof}
  Conditions~(a) and~(b) are clearly necessary so let us argue they
  are sufficient. 
  Condition~(b) shows that $B$
  is contained in a closed halfspace with $\mathbf{0}$ on its
  boundary, \ie there is a closed hemisphere $\cH \supseteq
  B$. Suppose $\cH \cap P \neq B$, \ie there is a point $p \in \cH
  \cap P$ not in $B$. Since $-p \in B$ by (a), $p$ and $-p$ must lie
  on the boundary of $\cH$ and, therefore, by the general position
  assumption, there are at most two such points $p$. An appropriate
  perturbation of $\cH$ yields a closed hemisphere $\cH'$ with $\cH'
  \cap P = B$ and thus $B$ is indeed a hemiset. From~(a) it follows that a
  hemiset $B$ of $P$ is affine if and only if $|B|=|P|/2$.
\end{MyProof}

\bigskip

The fact that symmetries of a projective point set $P$ act on its
hemisets and on its affine hemisets is now apparent.

\begin{proof}[Proof of Proposition~\ref{p:act}]
 Statement~\ref{i:iact} follows from the observation that Conditions
  (a), (b) and~(c) from Corollary~\ref{c:characterizations2}
  are preserved under orientation preserving bijections.

  Let us now consider a hemiset $B$ of $P$ with layer sequence
    $(B_{-1}, B_0, B_1, \ldots, B_\ell)$. Let $B'\eqdef g(B)$ and let
    $(B_{-1}', B_0', B_1', \ldots, B_{\ell'}')$ denote the layer
    sequence of the hemiset $B'$. By
    Lemma~\ref{l:EmbracingPreservation}\ref{i:iEmbracingPreservation}, for any $p \in B \cap -B$
    we have $g(\{-p,p\}) = \{-g(p),g(p)\}$ so $g(B \cap -B) \subseteq
    B'\cap -B'$. In particular, $|B \cap -B| \le |B'\cap -B'|$. By a
    similar argument we have $g^{-1}(B' \cap -B') \subseteq B\cap -B$,
    therefore $|B \cap -B| = |B' \cap -B'|$ and $B_{-1}' =
    g(B_{-1})$. Now, $g$ maps the affine set $B\setminus B_{-1}$ to
    the affine set $B' \setminus B_{-1}'$. Since $g$ is order
    preserving, it must map every positive extreme edge to a positive
    extreme edge, and therefore $g(B_0) = B_0'$ (here, again, we use
    $g^{-1}$ for one of the inclusions). By induction, for every $i
    \ge -1$, $g$ maps $B \setminus \cup_{j=-1}^{j=i} B_j$ to $B'
    \setminus \cup_{j=-1}^{j=i} B_j'$, and therefore maps $B_{i+1}$,
    the extreme points of the former, to $B_{i+1}'$, the extreme
    points of the latter. Statement~\ref{i:iiact} follows.
\end{proof}

\subsection{Orbit and stabilizer of a hemiset}

Given a projective set $P$ with symmetry group $\g$ and a subset $S$
of $P$, we write $\g_{S}$ for the stabilizer of $S$ in the action of
$\g$ on subsets of $P$. We also write $\g(S)$ for the orbit of $S$
in that action. (Note that in the following lemma, we do allow $S$ to contain antipodal pairs.) 

\begin{lemma}\label{l:actaff}
  Let $P\eqdef S \cup -S$ for a finite $S \subseteq \SS^2$ not contained in a great circle and let $\g$ denote the symmetry group of $P$.
  \begin{EnumAlph}
  \item 
  \label{i:iactaff}
The symmetry group of $S$ is isomorphic to $\g_{S}$.
  \item 
\label{i:iiactaff}
Given $S' \subseteq P$, there is an order preserving bijection from $S$ to $S'$ if and only if $S' \in \g(S)$.
  \end{EnumAlph}
\end{lemma}
\begin{MyProof}
  Let $\f$ denote the symmetry group of $S$. Note that since $S$ is not contained in a great circle, by Lemma~\ref{l:EmbracingPreservation}\ref{i:iEmbracingPreservation} any $f \in \f$ preserves antipodality for any antipodal pair occurring in $S$. Since $P = S \cup -S$, we can extend any $f \in \f$ to a permutation
  $\hat{f}$ of $P$ by setting $\hat{f}(p) \eqdef f(p)$ for $p \in S$
  and $\hat{f}(p) \eqdef -f(-p)$ for $p \notin S$. Let $\hat{\f}
  \eqdef \{\hat{f} \st f \in \f\}$. We have that $\hat{\f}$ is
  isomorphic to $\f$ since $\widehat{f_1 \circ f_2} = \widehat{f_1} \circ
  \widehat{f_2}$ for any two symmetries $f_1, f_2$ of $S$. Moreover, any element $g \in \hat{\f}$ fixes~$S$
  and, conversely, any symmetry $g:P \to P$ that fixes $S$ writes $g =
  \widehat{g|_{S}}$ (by Lemma~\ref{l:EmbracingPreservation}\ref{i:iEmbracingPreservation}).  Then, $\hat{\f} = \g_S$ and statement~\ref{i:iactaff} follows.

  For statement~\ref{i:iiactaff}, note that for any orientation
  preserving bijection $f:S \to S'$,  the extension~$\hat{f}$ of $f$
  to~$P$ also preserves orientations, and is therefore in $\g$. It
  follows that $S' \in \g(S)$. The reverse inclusion is immediate since every symmetry of $\g$ preserves orientations.
\end{MyProof}

\bigskip
With Lemma~\ref{l:actaff}, specialized to affine hemisets of a
projective set $P$, the orbit-stabilizer theorem readily implies:

\begin{corollary}\label{c:repetition}
  Let $P$ be a projective set of $2n$ points, $n \ge 3$, in general
  position and $A$ an affine hemiset of $P$. Let $\f$ and $\g$ denote
  the symmetry groups of $A$ and $P$, respectively. There are
  $|\g|/|\f|$ affine hemisets of $P$ with the same affine order type as
  $A$.
\end{corollary}

\subsection{How many points determine an order preserving bijection?}
\label{subs:HowMany}

We conclude this section with a basic fact about order preserving
bijections and symmetries  (see, \eg \cite{AILOW14} for similar observations).

\begin{claim} 
\label{c:TwoPtsDetSymmetry}For $P$ a projective set in general position and for $S \subseteq P$ not contained in a great circle, let $f\!: S \rightarrow S$ be a symmetry of $S$ with $f(p) = p$ and $f(q) = q$ for some $p, q \in S$, $q \notin \{p,-p\}$. Then $f = \id_S$.
\end{claim} 
\begin{MyProof}  For $r \in S \setminus \{p,q\}$ we want to show $f(r) = r$. Suppose first that $\sgn(p,q,r) = 0$, \ie $r \in \{-p,-q\}$; $r = -p$, say. Then, by Lemma~\ref{l:EmbracingPreservation}\ref{i:iEmbracingPreservation}, $f(r) = f(-p) = - f(p) = -p = r$.

Suppose next that $\sgn(p,q,r) \neq 0$; $\sgn(p,q,r) = 1$, say. Let $k$ be the smallest positive integer with $f^k(r) = r$. We need to show $k=1$. Obviously, for $R \eqdef \{p,q,r,f(r),\ldots f^{k-1}(r)\}$, $f|_R$ is an orientation preserving bijection on $R$. $R$ is an affine set with $(p,q)$ a positive extreme edge of $R$, thus $p$ is extreme in $R$ and there is a unique positive edge $(r',p)$ for some $r' \in R$. $(q,p)$ cannot possibly be a positive extreme edge of $R$ since $\sgn(q,p,r) = -1$. Hence, $(r',p) = (f^i(r),p)$ for some $i$. $f|_R$ must map this edge to a positive extreme edge of $R$, which, since $f(p) = p$, shows $f^{i+1}(r)= f^i(r)$, forcing $k=1$.
\end{MyProof}

\begin{lemma}\label{l:unique-if-not-lonely}
  Let $P$ and $P'$ be projective sets in general positions, with $|P|
  = |P'| \ge 6$.  Let $B$ and $B'$ be hemisets of $P$ and $P'$,
  resp., and let $p \in B$ and $p' \in B'$. Unless $p$ is lonely in $B$, there is at most one order preserving bijection $B \rightarrow B'$ that maps $p$ to $p'$.
\end{lemma}

\begin{MyProof}
  Let $f_1$ and $f_2$ be order preserving bijections $B \rightarrow B'$ with
  $f_1(p) =f_2(p) = p'$.  Then $f \eqdef f_1 \circ f_2^{-1}$ is a symmetry of $B$ with $f(p)=p$. We have $f_1 = f_2$ if and only if $f = \id_B$. Assuming that $p$ is not lonely, we want to show $f = \id_B$. Note right away that hemisets of projective sets $P$ in general position with $|P|\ge 6$ cannot be contained in a great circle.
  
For $i=-1,0,1,\ldots$, we let $B_i$ denote the $i$th layer of $B$. By Proposition~\ref{p:act}, $f$ preserves layers (\ie $f(B_i) = B_i$ for all $i$).

Let us first deal with the case where $p \in B_i$ with $i \neq -1$. Since $p$ is not lonely, there is a unique point $q$ such that $(p,q)$ is a positive extreme edge of~$B_i$. Clearly, its image $(f(p),f(q))$ is a positive extreme edge of $B_i$. Since $f(p)=p$, we have $f(q)=q$. Since $B$ cannot be contained in a great circle, Claim~\ref{c:TwoPtsDetSymmetry} shows $f= \id_B$.

Next we assume $p \in B_{-1}$. Then $-p \in B_{-1} \subseteq B$ and $f(-p)=-f(p)=-p$ (Lemma~\ref{l:EmbracingPreservation}\ref{i:iEmbracingPreservation}). Set $B'' \eqdef B \setminus \{-p\}$. Since $-p \not\in B''$, $p$ is not in the layer $-1$ of $B''$. Actually, $p$ has to be in layer $0$. If $p$ is not lonely in $B''$, the argument in the previous paragraph shows that $f|_{B''}=\id_{B''}$ which entails $f = \id_B$. If $p$ is lonely in $B''$, then $B''=\{p\}$ or $B''=\{q,-q,p\}$ for some point $q$. But then $B=\{p,-p\}$ or $B=\{q,-q,p,-p\}$ which is not possible for hemisets as postulated in the assertion.
\end{MyProof}

Lemma~\ref{l:unique-if-not-lonely} implies that for any hemiset
  $B$ of a projective set of at least $6$ points in general position,
  only $\id_B$ fixes a non-lonely point. Moreover, if $B$ is
  non-affine then it has at most $4$ symmetries; see
  Lemma~\ref{l:non-affine-hemiset} for more.

\section{Analysis of labeled affine order types}
\label{s:labeled}

Perhaps surprisingly, Corollary~\ref{c:repetition} is all we need to
prove Theorem~\ref{t:avgl}. Once this is done, the reader
  interested in proving Theorem~\ref{t:concentration} for labeled
  order types only can skip Sections~\ref{s:poles} and~\ref{s:non-labeled}
  and proceed to Section~\ref{s:concentration}.

\subsection{The two roles of affine symmetries}

The number of symmetries of an affine order type determines both its
number of labelings, and how often it occurs among the affine hemisets of a
projective completion of one of its realizations. These two roles
happen to balance each other out nicely:

\begin{proposition}\label{p:uniform}
  Let $P$ be a projective set of $2n$ points, $n \ge 3$, in general
  position. Let $R$ be a random affine hemiset chosen uniformly among
  all affine hemisets of $P$. Let $\lambda$ be a random permutation $R
  \to \{1,2,\ldots,n\}$ chosen uniformly among all such permutations. The labeled
  affine order type of $R_{[\lambda]}$ is uniformly distributed in $\LOTaff{P}$.
\end{proposition}
\begin{MyProof}
  Let $N$ denote the number of affine hemisets of $P$. Let
  $\omega_1,\omega_2, \ldots, \omega_k$, $k \le N$, denote the order types of the
  affine hemisets of $P$, without repetition (that is, the $\omega_i$
  are pairwise distinct). Let $\g$ denote the symmetry group of $P$
  and let $\f_i$, $1 \le i \le k$, denote the symmetry group of~$\omega_i$. Let $\rho$ denote the affine order type of $R$. By
  Corollary~\ref{c:repetition}, we have
  \[ \Prob{\rho = \omega_i} = \frac{|\g|/|\f_i|}{N}.\]
  Next, the number of distinct labelings of the order type of an
  affine set $A$ is $n!/|\f_A|$, since two labelings $A_{[\lambda]}$ and
  $A_{[\mu]}$ of $A$ have the same labeled order type if and only if
  $\mu^{-1} \circ \lambda$ is a symmetry of $A$. Let $\Lab{\rho}$
  denote the labeled affine order type of $R_{[\lambda]}$. For any
  $\Lab{\sigma} \in \LOTaff {\omega_i}$, we have
  \[ \Prob{\Lab{\rho} = \Lab{\sigma} \mid \rho = \omega_i} = \frac{|\f_i|}{n!}.\]
  Altogether, for any $\Lab{\sigma} \in \displaystyle\bigcup_{i=1}^k
  \LOTaff {\omega_i} = \LOTaff{P}$, we have
  \[ \Prob{\Lab{\rho} = \Lab{\sigma}} = \frac{|\g|}{N n!}\]
  and the distribution is uniform as we claimed. This also shows that $|\LOTaff{P}| = \frac{N n!}{|\g|}$ which will come handy later in the paper.
\end{MyProof}

\subsection{Hemisets and duality}
\label{s:duality}

The following dualization will make counting easy.

\begin{lemma}\label{l:duality-basic}
  There is a bijection $\phi$ between the affine hemisets of a
  projective point set~$P$ and the cells of the dual arrangement
  $P^*$, such that a point $p$ is extreme in an affine hemiset~$A$ if
  and only if the great circle $p^*$ supports an edge of $\phi(A)$.
\end{lemma}
\begin{MyProof}
  For any point $p$ we write $p^+$ for the hemisphere centered in $p$,
  that is, the closed hemisphere containing $p$ and bounded by
  $p^*$. For any closed hemisphere $\cH$ we write $\cH^+$ for its center,
  that is, the point $q$ with $\cH= q^+$. Now, a point $p$ is in a
  closed hemisphere $\cH$ if and only if the scalar product $\langle p,
  \cH^+ \rangle$ is nonnegative. Thus, $p$ lies in $\cH$ if and only if
  $\cH^+$ lies in $p^+$. It follows that two hemispheres $\cH_0$ and $\cH_1$
  intersect $P$ in the same hemiset if and only if $\cH_0^+$ and $\cH_1^+$
  lie in the same cell of $P^*$. Moreover, as $\cH^+$ moves in the
  cell the hemisphere $\cH$ also moves while enclosing the same set
  of points; the boundary of $\cH$ touches a point $p$ if and only if
  $\cH^+$ touches $p^*$.
\end{MyProof}

\bigskip

For example, we now see that a projective set of $2n$ points, $n \ge
3$, in general position has $2\binom{n}2+2$ distinct affine hemisets
(see Section~\ref{s:DualityArr}). Also, it should be clear from the
final computations of the proof of Proposition~\ref{p:uniform} that if
that projective point set has symmetry group $\g$, then it supports
$\pth{2\binom{n}2+2} \frac{n!}{|\g|}$ distinct labeled affine order
types.

\subsection{Counting extreme points: expectation and variance}

We can now prove Theorem~\ref{t:avgl} on the expectation and variance
of the number of extreme points in a random labeled affine order type.

\begin{lemma}\label{l:avgl-cond}
  Let $P$ be a projective set of $2n$ points, $n \ge 3$, in general
  position. If $X_P$ denotes the number of extreme points in a labeled
  affine order type chosen uniformly among those supported by $P$,
  then
  \[ \Ex{X_P} = \frac{4n(n-1)}{n(n-1)+2} =\TheNumber \qquad \hbox{and} \qquad \Ex{{X_P}^2} \le \frac{\displaystyle 19n(n-1) - {{10}}n}{\displaystyle n(n-1) + 2} < 19. \]
\end{lemma}
\begin{MyProof}
  By Proposition~\ref{p:uniform} and Lemma~\ref{l:duality-basic}, $X_P$ has the same distribution as
  the number of edges in a cell chosen uniformly at random in
  $P^*$. The arrangement $P^*$ has $2\binom{n}2+2$ cells and
  $4\binom{n}{2}$ edges. Since every edge bounds exactly two cells, it
  follows that
  \[ \Ex{X_P} = \frac{8\binom{n}{2}}{2\binom{n}2+2} = {\frac{4n(n-1)}{n(n-1)+2}} = \TheNumber.\]
  Moreover, the random variable ${X_P}^2$ has the same distribution as
  the square of the number of edges in a random cell chosen
  uniformly in $P^*$. Let $F_2(P^*)$ denote the set of cells of
  $P^*$ and for $c \in F_2(P^*)$ let $|c|$ denote its number of
  edges. We thus have
  \[ \pth{2\binom{n}2+2}\Ex{{X_P}^2} = \sum_{c \in F_2(P^*)} |c|^2.\]
  In the right-hand term, every edge $e$ of $P^*$ is counted
  $|c_1|+|c_2|$ times, where $c_1$ and $c_2$ are its two adjacent
  cells. {For any point $p \in P$, the contribution of the edges
    supported by $p^*$ to that sum equals $\sum_{c \in Z(p^*)} |c| \le
    19(n-1) - {{10}}$ (following notation and bound in
    Theorem~\ref{t:Zone}). Altogether,
  \[ \pth{2\binom{n}2+2}\Ex{{X_P}^2} \le n(19(n-1)-{{10}})\]
  and $\Ex{{X_P}^2} \le \frac{\displaystyle 19n(n-1) - {{10}}n}{\displaystyle n(n-1) + 2} < 19$.}
\end{MyProof}

\bigskip
Here comes the announced proof.

\begin{MyProof}[Proof of Theorem~\ref{t:avgl}]
  Let $\Lab{\rho}$ be a simple labeled order type chosen uniformly at
  random in $\LOTaff{n}$. Let $X_n$ denote the number of extreme
  points in $\rho$, {where $\rho$ denotes} the unlabeling of~$\Lab{\rho}$
  and let $\pi$ be the projective completion of $\rho$. By Lemma~\ref{l:avgl-cond}, we have
\[ \forall \pi' \in \OTproj n, \quad \Ex{X_n \mid \pi = \pi'} = \frac{4n(n-1)}{n(n-1)+2} \quad \hbox{and} \quad \Ex{{X_n}^2 \mid \pi = \pi'} \le \frac{19n(n-1) - {{10}}n}{n(n-1) + 2}.\]
  The formula of total probability therefore yields
  \[  \Ex{X_n} = \frac{4n(n-1)}{n(n-1)+2} \quad \hbox{and} \quad \Ex{{X_n}^2} \le \frac{19n(n-1) - {{10}}n}{n(n-1) + 2}.\]
  From there, $\Var{X_n} = \Ex{{X_n}^2} - \Ex{X_n}^2 <3$. (A bound of $3 + o(1)$ is readily seen from $\Ex{X_n} = 4 + o(1)$ and $\Ex{{X_n}^2} < 19$.)
\end{MyProof}

\bigskip

As a consequence, we obtain for instance the following estimates.

\begin{corollary}
  For $h\ge 6$, the proportion of simple labeled affine $n$-point order types with at least $h$ convex hull vertices is at most ${3}/(h-4)^2$.
\end{corollary}
\begin{MyProof}
  By the Bienaym{\'e}-Chebyshev inequality, for any real $t>0$ and any
  random variable~$X$ with finite expected value and non-zero
  variance, we have
  \[ \Prob{|X-\Ex{X}| \ge t\sqrt{\Var{X}}} \le \frac1{t^2}.\]
  Together with Theorem~\ref{t:avgl}, this implies the statement.
\end{MyProof}

\bigskip

Here is a more direct\footnote{The machinery we set up for our
      proof of Theorem~\ref{t:avgl} is needed in the analysis of the unlabeled setting, which was our initial goal.} way to
  prove Theorem~\ref{t:avgl} which we learned from Arnau Padrol. We
  can define a labeled projective $2n$-point set $\bar P$ as a
  projective set where the antipodal pairs are labeled from $1$ to $n$
  (antipodal points receive the same label). Any affine hemiset of
  $\bar P$ determines a labeled affine order type. It turns out that
  for $n \ge 4$ these labeled affine order types are pairwise
  distinct: there is no multiplicity!~\footnote{Indeed, consider
      two labeled affine order types $\bar A$ and $\bar A'$ of $\bar
      P$. The map $\phi$ that sends every point of $\bar A$ to the
      point in $\bar A'$ with the same label can be described as
      follows: for $p \in \bar A$, we have $\phi(p) = p$ if $p \in
      \bar A'$ and $\phi(p) = -p$ otherwise. Since $\bar A \neq \bar
      A'$, at least one point is antipodal to its image. Now, for $n
      \ge 4$, in any vector in $\{\pm 1\}^n$ with at least one $-1$
      entry, there exist three entries for which the number of $-1$ is
      odd. This fails for $n=3$.} Thus, the number of extreme points
  in a random labeled affine order type supported by $\bar P$ has the
  same distribution as the number of edges in a random $2$-cell chosen
  uniformly from $\bar P^*$.

\section{Poles of projective symmetries}
\label{s:poles}

To analyze non-labeled affine order types, we again relate, for a
projective point set $P$, the number of extreme points in a random
order type of $\OTaff P$ to the average number of edges in a random
cell of $P^*$. The issue is, however, that we no longer have
Proposition~\ref{p:uniform}: to count every affine order type of
$\OTaff P$ only once, and not as many times as there are hemisets of
$P$ realizing it, will require some control over the structure of
the symmetries of affine and projective sets.

\bigskip

We draw inspiration from Klein's classical characterization of the
finite subgroups of $SO(3)$. An easily accessible exposition of
Klein's proof can be found in \cite{Sen90}, whose line we follow
here. This proof analyzes how a finite subgroup of $SO(3)$ acts on the
(finite) set of points fixed by at least one of its nontrivial
members. The notion of \emph{pole hemisets} that we now define plays
the role of these fixed points.

\bigskip

Let $P$ be a projective point set and $\g$ its symmetry group. Given a
nontrivial symmetry $g \in \g$, a \Emph{pole} of $g$ is a hemiset $B$
such that $g(B)=B$. A \Emph{pole} of $P$ is a pole of some nontrivial
symmetry of $P$.  We say that two hemisets
$B_0$ and $B_1$ of $P$ are \emph{antipodal} if $B_0=-B_1$. The
following will be instrumental to mimick Klein's proof and
to classify the structure of symmetry groups of projective sets.

\begin{proposition}\label{p:poles}
  Let $P$ be a projective set of $2n$ points in general position, with
  $n \ge 3$. Every symmetry $g \neq \id$ of $P$ has
  exactly two poles and they are antipodal.
\end{proposition}

\noindent
The rest of this section is devoted to the proof of
Proposition~\ref{p:poles}. A first, at this point unmotivated step is
to clarify some properties of order preserving and order reversing
bijections of affine sets.

\subsection{Preparation: reflections of affine sets}
\label{s:Reflections}

A bijection $f: S \rightarrow S'$ between sets on the sphere is \Emph{orientation reversing} if $\sgn(f(p),f(q),f(r)) = -\sgn(p,q,r)$ for every triple in $(p,q,r) \in S^3$. A permutation $f$ of a set $S$ on the sphere \Emph{goes across} a great circle $\gC$ on the sphere, if, for all $p \in S$, either $f(p)=p$ and $p \in \gC$, or $p$ and $f(p)$ are strictly separated by $\gC$. The first ingredient of the proof of Proposition~\ref{p:poles} is:

\begin{proposition}\label{p:goesacross}
  Every orientation reversing permutation $f$ of an affine set $A$ in
  general position goes across some great circle $\gC$.
\end{proposition}

It will be convenient to transport the affine set $A$ under consideration to the plane $\RR^2$ as discussed in
Section~\ref{s:SettingTerminology} and show the equivalent claim that every orientation reversing bijection $f$ goes across some line $\ell$, \ie  for all points $p$ we have either $f(p)=p$ and $p$ lies on $\ell$, or $\ell$ strictly separates $f(p)$ from $p$. 

\begin{lemma}\label{l:involute}
  If $f$ is an orientation reversing permutation of a finite set $A \subseteq \RR^2$ in general position, then $f^2 = \id$.
\end{lemma}
\begin{MyProof}
   Note that $(p,q)$ is a positive extreme edge of $A$ if and
  only if $(f(q),f(p))$ is a positive extreme edge of $f(A)$. Hence,
  $f$ maps each layer of $A$ to itself and it suffices to prove the
  statement for $A$ in convex position. So let $(p_0,p_1,\ldots,
  p_{n-1})$ be a CCW extreme points order of~$A$ and let $t$ be such that $f(p_0) =
  p_t$. Since $f$ reverses orientation, for all $0 \le i \le n-1$ we
  must have $f(p_i) = p_{t-i}$ (indices $\bmod\, n$). It follows that $f^2(p_i) = p_i$.
\end{MyProof}

Let $f$ be an orientation reversing permutation of $A$. Since $f^2= \id_A$, $\{\id_A, f\}$ is a group and its action partitions $A$ into orbits of size $1$ or~$2$, which we call
\Emph{$f$-orbits}. For $p \in A$ we write $[p] \eqdef \{p,f(p)\}$ and
$\bar{p} \eqdef \conv([p])$, which is a segment or a single point; in the latter case we call $[p]$ a \Emph{point-orbit}. Let $\TT = \TT(A,f) \eqdef \{\bar{p} \st p \in A\}$. Our task is to prove that there exists a line that intersects every element in $\TT$. Note that if such a line transversal exists, then the general position ensures that one exists that is disjoint from the endpoints of  segments in $\TT$. 

In order to prove Proposition~\ref{p:goesacross} for a set $A$ in general position and an orientation reversing permutation $f$ of $A$, we discriminate three cases depending on the number of point-orbits of $f$.

\paragraph{Two point-orbits.} Suppose there are two point-orbits $[p]$ and $[q]$, \ie $f(p) = p$ and $f(q) = q$. Then the line $\ell_{pq}$ through $p$ and $q$ hits all segments $\bar r$ in $\TT$ since
\[
\sgn(p,q,r) = -\sgn(f(p),f(q),f(r)) = - \sgn(p,q,f(r))
\]
and thus $r$ and $f(r)$ have to lie on opposite sides of $\ell_{pq}$ (on $\ell_{pq}$ is outruled by general position).

\paragraph{One point-orbit.} Suppose $[p]$ is the only point-orbit and let $[q]$ and $[r]$ be two distinct $f$-orbits different from $[p]$. For the line  $\ell_{pq}$ through $p$ and $q$ observe that the product $\sgn(p,q,r) \cdot \sgn(p,q,f(r))$ is $-1$ if and only if the line $\ell_{pq}$ hits $\bar r$.
We have that $\ell_{pq}$ hits $\bar r$ if and only if $\ell_{pf(q)}$ hits $\bar r$ since
\begin{eqnarray*}
\sgn(p,f(q),r) \cdot \sgn(p, f(q),f(r)) 
&=& -\sgn(f(p),f^2(q),f(r)) \cdot -\sgn(f(p), f^2(q),f^2(r))\\
&=& \sgn(p,q,f(r)) \cdot \sgn(p, q,r) ~,
\end{eqnarray*}
and we have that $\ell_{pq}$ hits $\bar r$ if and only if $\ell_{pr}$ does \Emph{not} hit $\bar q$ since
\begin{eqnarray*}
\sgn(p,r,q) \cdot \sgn(p, r,f(q)) 
=& \sgn(p,r,q) \cdot -\sgn(f(p), f(r),f^2(q)) &\\
=& \sgn(p,r,q) \cdot -\sgn(p, f(r),q)  
&= -\sgn(p,q,r) \cdot \sgn(p, q,f(r)) ~.
\end{eqnarray*}

Hence, either $\ell_{pq}$ hits $\bar r$ or $\ell_{pr}$ hits $\bar q$ (but not both). W.l.o.g.\ let $\ell_{pq}$ hit $\bar r$ and thus $\ell_{pf(q)}$ hits~$\bar r$. Then all lines through $p$ passing through $\bar q$ must hit $\bar r$. This holds, since if we rotate the line through $p$ and $q$ to the line through $p$ and $f(q)$ so that $\bar q$ is always hit, we can never encounter an endpoint of $\bar r$, otherwise $\ell_{pr}$ or $\ell_{pf(r)}$ hits $\bar q$, which we excluded for $\ell_{pq}$ hitting $\bar r$.

Consequently, the set of lines $L_{p \bar q}$ through $p$ and $\bar q$ is a subset of the set $L_{p \bar r}$ of lines through $p$ and $\bar r$. It follows that the sets $L_{p \bar s}$, $\bar s \in \TT \setminus \{\bar p\}$, are totally ordered by inclusion and the minimal set in this order exhibits a line hitting all elements in $\TT$. This concludes the argument for Proposition~\ref{p:goesacross} in the one point-orbit case.

\paragraph{No point-orbit.} Suppose there is no point-orbit of $f$. We will employ Hadwiger's \emph{transversal theorem}~\cite{Had57}: \emph{a finite family of pairwise disjoint, convex, subsets of the plane has a line transversal if and only if they can be ordered such that every three members can be intersected by a directed line in the given order.} 

We start with a few observations about the relative position of segments in $\TT$. 

\begin{claim}\label{c:RelPos}
\begin{EnumRom} Let $\bar p, \bar q$, and $\bar r$ be three distinct segments in $\TT$.
\item \label{i:Disjoint}The line supporting $\bar p$ is disjoint from  $\bar q$.
\item \label{i:CHEdges} Exactly two of the segments $\bar p, \bar q$, and $\bar r$ are edges of $\conv([p] \cup [q] \cup [r])$.
\item \label{i:3EdgeTrans} The segments $\bar p, \bar q$, and $\bar r$ have a line transversal. 
\end{EnumRom}
\end{claim}
\begin{MyProof}
\ref{i:Disjoint} We have
$
\sgn(p,f(p),q) = - \sgn(f(p),\overbrace{f^2(p)}^{=p},f(q)) = \sgn(p,f(p),f(q))$,
and therefore $q$ and $f(q)$ are on the same side of the line through $p$ and $f(p)$.
\medskip

\noindent 
\ref{i:CHEdges} For each of the three $f$-orbits of $A'\eqdef [p] \cup [q] \cup [r]$, either both of its points are extreme in~$A'$ or none is. Hence, the orbits define a matching on the extreme points of $A'$. Since, by~\ref{i:Disjoint}, no two of the segments $\bar p, \bar q$, and $\bar r$ cross, it follows that at least two segments are edges of $\conv(A')$.

Now suppose all three segments are edges of $\conv(A')$. Since, moreover, the segments are disjoint, all oriented triangles $(x,y,z)$ with $x \in \bar p$, $y \in \bar q$, and $z \in \bar r$ have the same orientation, \ie they have the same sign $\sgn(x,y,z)$. This contradicts $\sgn(p,q,r) = - \sgn(f(p),f(q),f(r))$. 
\medskip

\noindent
\ref{i:3EdgeTrans}
W.l.o.g.\ let $\bar p$ and $\bar q$ be edges of $\conv(A')$. If $\conv(A')$ is a quadrilateral, then $\bar r$ is in the interior of $\conv(A')$. For every given pair of opposite edges of a convex quadrilateral, every interior point has a line passing through it and the given pair of edges. This establishes the claim.

The only case left is that of $\conv(A')$ being a convex hexagon. Since $\bar r$ is not an edge of  $\conv(A')$, it is a diagonal separating $\bar p$ and $\bar q$. The claim is obvious in this case.
\end{MyProof}

\bigskip

Let $\vec \TT$ be the set $\TT$ where every segment is directed in some way; we denote the segment directed from $p$ to $f(p)$ by $\vec p$. We say that $\vec q \in \vec \TT$ is left of $\vec p \in \vec \TT$ if $[q]$ lies to the \Emph{left of} $\vec p$, \ie $\sgn(p,f(p),q) = \sgn(p,f(p),f(q)) = +1$. If $\sgn(p,f(p),q) = \sgn(p,f(p),f(q)) = -1$ we say that $\vec q$ is \Emph{right of} $\vec p$.  Claim~\ref{c:RelPos}\ref{i:Disjoint} ensures that $\vec q$ is either left or right of $\vec p$. However, we cannot assume that $\vec q$ left of $\vec p$ implies $\vec p$ right of $\vec q$.

\begin{figure}
  \centerline{
    \placefigOLD{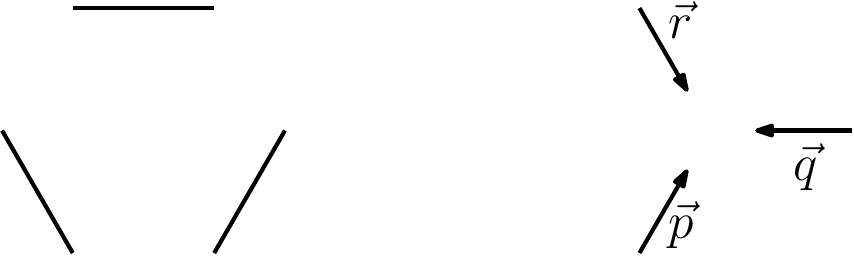}{0.4\textwidth}
    }
  \caption{Left: Three segments that cannot be directed in a consistent manner. Right: Three consistently directed segments with $\vec p$ left of $\vec q$ left of $\vec r$ left of $\vec p$. Note that while these segments satisfy Claim~\ref{c:RelPos}\ref{i:Disjoint}, they are in contradiction with Claim~\ref{c:RelPos}\ref{i:CHEdges}.}
  \label{f:SegmentAnomalies}
\end{figure}

We will proceed as follows. First, we show that we can indeed choose directed versions $\vec \TT$ of the segments in $\TT$ such that $\vec q$ is left of $\vec p$ if and only if $\vec p$ is right of $\vec q$, for all $\vec q, \vec p \in \vec \TT$. We call these \Emph{consistent directions}. Then we show that the relation ``left of'' is transitive. This induces a total order on $\vec \TT$, which will be the basis for the use of Hadwiger's transversal theorem. Let us point out that even sets of segments satisfying Claim~\ref{c:RelPos}\ref{i:Disjoint} do not necessarily allow a consistent way of choosing directions, and moreover, even consistently directed segments do not necessarily imply transitivity as described above (see \FigRef{f:SegmentAnomalies} for examples).

\newcommand{\p}{p_0}
We now choose a set $\vec \TT$ of directions for the segments in $\TT$: Orient one of the segments arbitrarily, say orient ${\bar \p} \in \TT$ as ${\vec \p}$. Then orient each other segment as $\vec q$ so that the direction is consistent with ${\vec \p}$. Note here that $\vec q$ is consistent with ${\vec \p}$ if and only if $\sgn(\p,f(\p),q) \cdot \sgn(q,f(q),\p) = -1$. 

\begin{claim} Every pair $\vec q, \vec r \in \vec \TT\setminus \{\vec \p\}$, $\vec q \neq \vec r$,  is consistently directed.
\end{claim}
\begin{MyProof}
Suppose $\vec q$ and $\vec r$ are not consistent with each other. On the one hand, this means\begin{eqnarray}
\sgn(\p,f(\p),q) \cdot \sgn(q,f(q),\p) &=& -1~,\nonumber\\
\sgn(\p,f(\p),r) \cdot \sgn(r,f(r),\p) &=& -1~,\mbox{~ and} \label{eq:Prods1}\\
\sgn(q,f(q),r) \cdot \sgn(r,f(r),q) &=& ~1~. \nonumber
\end{eqnarray}
On the other hand, by Claim~\ref{c:RelPos}\ref{i:CHEdges}, we know that two of $\bar \p, \bar q, \bar r$ are edges of $\conv([\p] \cup [q] \cup [r])$ and the third one is not. Again with Claim~\ref{c:RelPos}\ref{i:Disjoint} in mind, this fact can be expressed as
\begin{eqnarray}
\label{eq:Prods2}
\mbox{among} \left\{
\begin{array}{c}
\sgn(\p,f(\p),q) \cdot \sgn(\p,f(\p),r)\\
\sgn(q,f(q),r) \cdot \sgn(q,f(q),\p)\\
\sgn(r,f(r),\p) \cdot \sgn(r,f(r),q)
\end{array}
\right\} \mbox{two are $+1$, and one is $-1$.}
\end{eqnarray}
The six $\sgn$-terms in (\ref{eq:Prods2}) are the same as the terms used in (\ref{eq:Prods1}). According to (\ref{eq:Prods1}), their overall product is $+1$, according to (\ref{eq:Prods2}) it is $-1$, which gives the desired contradiction.
\end{MyProof}

\begin{claim}\label{c:transitiv}
Let $\vec p, \vec q, \vec r \in \vec \TT$  be such that $\vec p$ is left of $\vec q$ and $\vec q$ is left of $\vec r$. Then $\vec p$ is left of $\vec r$, $\bar q$ is not an edge of $\conv([p]\cup[q]\cup[r])$, and every transversal meets $\bar q$ in between $\bar p$ and $\bar r$.
\end{claim}
\begin{MyProof}
Since $\vec q$ is left of $\vec r$, we have $\vec r$ is right of $\vec q$, by consistency of directions. Since $\vec p$ is left of $\vec q$ and $\vec r$ is right of $\vec q$, the segment $\bar q$ is not an edge of $\conv([p] \cup [q] \cup [r])$. By Claim~\ref{c:RelPos}\ref{i:CHEdges}, $\bar r$ is an edge of this convex hull, and hence $\vec q$ left of $\vec r$ implies that also $\vec p$ left of $\vec r$. Since $\bar p$ and $\bar r$ are disjoint edges of $\conv([p] \cup [q] \cup [r])$, every transversal meets $\bar q$ in between $\bar p$ and $\bar r$.
\end{MyProof}

We can now conclude the proof of Proposition~\ref{p:goesacross} for the case of no point-orbit.
Define a relation $\preceq$ on $\vec \TT$ by
\[
\vec p \preceq \vec q ~~\equivalentdef~~ \mbox{$\vec p$ left of $\vec q$ ~or~ $\vec p=\vec q$}~.
\]
This is a total order: It is obviously reflexive; transitivity is shown in Claim~\ref{c:transitiv}; what we called consistency implies antisymmetry ($\vec p \preceq \vec q$ and $\vec q \preceq \vec p$ implies $\vec p = \vec q$) and connectedness ($\vec p \preceq \vec q$ or $\vec q \preceq \vec p$).

Whenever $\vec p \preceq \vec q \preceq \vec r$ for three distinct elements in $\vec \TT$, there is a directed line meeting the segements $\bar p$, $\bar q$, and $\bar r$ in this order (Claims~\ref{c:RelPos}\ref{i:3EdgeTrans} and \ref{c:transitiv}). Hadwiger's transversal theorem entails a transversal of all segments in $\TT$.

\subsection{Uniqueness of poles}

With reflections of affine sets under control with
Proposition~\ref{p:goesacross}, we now turn to the proof of
Proposition~\ref{p:poles}. We start with the uniqueness, which
  easily follows from the following remarkable property of
  hemisets.\footnote{In the affine case, the proposition basically
    states that no nontrivial symmetry can respect a nontrivial
    partition of the point set by a line.}

\bigskip

\begin{proposition}\label{p:no-nontrivial-split}
  Let $P$ be a projective set in general position, $|P| \ge 6$. Let $B$ be a hemiset of  $P$. Let $g \neq \id_B$ be a
  symmetry of $B$, and let $\cH$ be a closed hemisphere with $g(B \cap
  \cH) = B \cap \cH$.  Then $B$ is contained in $\cH$ or in
    $-\cH$.
\end{proposition}

It is perhaps worthwhile to mention that while many of the basic lemmas (\eg Proposition~\ref{p:act} and Lemma~\ref{l:unique-if-not-lonely}) have appropriate generalizations to higher dimensions (along the lines of our proofs or also \cite{AILOW14}), this proposition fails for higher dimensions: a set in $\RR^3$, or -- in our terminology -- an affine set $A$ on $\SS^3$ can have a nontrivial symmetry (rotation) which stabilizes a nontrivial intersection of $A$ with a hemisphere.

\begin{MyProof}
  Let us first consider the case where $B$ is an affine set. If $\emptyset \neq B
  \cap \cH \neq B$ then there must be a unique positive extreme edge
  $(p_0,p_1)$ of $B$ with $p_0 \not\in \cH$ and $p_1 \in \cH$. Since
  $g$ is a symmetry, $(g(p_0),g(p_1))$ is a positive extreme edge of
  $B$. By the assumption $g(B \cap \cH)=B \cap \cH$, we have $g(p_0)
  \not\in \cH$ and $g(p_1) \in \cH$. It follows that $(g(p_0),g(p_1))
  = (p_0,p_1)$ and thus, with Claim~\ref{c:TwoPtsDetSymmetry}, we
  conclude $g= \id_{A}$.

  \bigskip

  Now, consider a hemiset $B$ of $P$ and assume that $E \eqdef B \cap
  -B \neq \emptyset$. Let $B' \eqdef B \setminus E$ (an affine
  set). By Proposition~\ref{p:act}\ref{i:iiact}, $g$
  maps $E$ to $E$ and $B'$ to $B'$. Moreover, by Lemma\,\ref{l:unique-if-not-lonely}, $g \neq \id_B$ implies $g|_E \neq \id_E$.

 Let us first argue (from $g|_E \neq \id_E$) that $E \subseteq \cH$
 (which immediately shows that $E$ lies in the boundary of $\cH$ and
 thus also implies that $E \subseteq -\cH$). For every $p \in E$,
 $\cH$ must contain $p$ or $-p$, say $p$. If $-p$ is in the orbit of
 $p$ under $g$, then $-p \in \cH$ as well because $g(B \cap \cH)=B
 \cap \cH$. The alternative is that $E = \{p,-p,q,-q\}$ (remember that
 $P$ is in general position) and, up to exchanging $q$ and $-q$, that
 $p \stackrel{g}{\mapsto} q \stackrel{g}{\mapsto} p$. But then, taking
 any $r \in B'$,
  \[ \sgn(p,q,r) = \sgn(g(p),g(q),g(r)) = \sgn(q,p,g(r)) = - \sgn(p,q,g(r)),\]
  which is impossible since $g(B') = B'$ and all points of $B'$ are on
  the same side of the great circle through $p$ and $q$. 

  Given that we know that $E \subseteq \cH$ and $E \subseteq -\cH$, we then have two cases. If $|B'| \ge 2$, then $g|_{B'}$ is
  nontrivial by Lemma~\ref{l:unique-if-not-lonely}. We already know
  that the proposition holds in the affine case, so it applies to
  $B'$, which must be contained in $\cH$ or in $-\cH$. If $|B'|=1$,
  then $\cH$ or $-\cH$ will always contain the given 1-element set
  $B'$. Altogether, $B = B' \cup E$ is also contained in $\cH$ or
  in $-\cH$.
\end{MyProof}

\begin{corollary}
  If $B_0$ and $B_1$ are poles of $g$, then $B_1 = \pm B_0$.
\end{corollary}
\begin{MyProof}
  This follows from Proposition~\ref{p:no-nontrivial-split} with $B \eqdef B_0$ and $\cH$ a closed hemisphere with $P \cap \cH = B_1$.
\end{MyProof}

\subsection{Existence of poles}

Now, let us argue that $g$ has some pole. Since $|P| \ge 6$ and $P$ is
in general position, $g$ preserves antipodality
(Lemma~\ref{l:EmbracingPreservation}\ref{i:iEmbracingPreservation}) and acts on the hemisets of $P$
(Proposition~\ref{p:act}\ref{i:iact}); in particular, for any hemiset $B$ of $P$,
$g(-B) = -g(B)$. As spelled out in Lemma~\ref{l:duality-basic}, for
any projective set $P$, the faces of the great circle arrangement
$P^*$ are in correspondence with the hemisets of $P$. In this
correspondence, a hemiset with $k$ antipodal pairs corresponds to a
face of dimension $2-k$. We therefore have:

\begin{claim}\label{c:permarr}
  Any symmetry $g$ of a projective set $P$ induces a dimension
  preserving permutation $\bar{g}$ of the faces of the arrangement of
  $P^*$, where also incidences are preserved: if face $F$ is incident
  to face $F'$, then face $\bar g(F)$ is incident to face $\bar
  g(F')$.
\end{claim}

\noindent
This combinatorial map extents into a continuous map.

\begin{claim}
  There exists a continuous injective map $\gamma\!: \SS^2 \to \SS^2$
  such that for any $x \in \SS^2$ and any face $F$ of $P^*$, $x$ is in
  $F$ if and only if $\gamma(x)$ is in $\bar{g}(F)$.
\end{claim}
\begin{MyProof}
  We start by setting $\gamma(v) \eqdef \bar g(v)$ for every vertex
  $v$ of $P^*$. Next, for every edge $e$ of $P^*$, note that
  $\gamma$ maps the vertices of $e$ to the vertices of $\bar g(e)$; we
  extend it to a continuous (actually, ``linear'') map from $e$ to
  $\bar g(e)$. Last, for every cell $c$ of $P^*$, $\gamma$ already
  defines a continuous injective map from the boundary of $c$ to the
  boundary of $\bar g(c)$ and can be extended into a continuous injective map $c
  \to \bar g(c)$. Observe that $\gamma$ agrees with $\bar g$ as stated.
\end{MyProof}

\bigskip

Now enters the so-called hairy ball theorem\footnote{It is often
  formulated in terms of vector fields on $\SS^2$, with the assertion
  at hand a simple corollary.}: \emph{If $d$ is even and $\psi: \SS^d
  \rightarrow \SS^d$ is a continuous function, then there exists at
  least one $x_0 \in \SS^d$ such that either $\psi(x_0)=x_0$ or
  $\psi(x_0) = -x_0$.} Hence, there exists $x_0 \in \SS^2$ such that
$\gamma(x_0) \in \{x_0, -x_0\}$. Let $B$ denote the hemiset 
corresponding, via Lemma~\ref{l:duality-basic}, to the face containing
$x_0$ ($B$ is the intersection of $P$ with the closed hemisphere centered in $x_0$). Since $\gamma$ agrees with $\bar g$, $\gamma(x_0)$ lies in the
face corresponding to the hemiset $g(B)$, that is, $g(B)$ is the intersection of $P$ with the hemisphere centered in  $\gamma(x_0)$.

\bigskip

When $\gamma(x_0) = x_0$ these faces coincide and $g(B)=B$ is a pole
of $g$. Then also $g(-B)=-B$ and we have our two poles.

\bigskip
  
Let us prove that poles exist also when $\gamma(x_0) = -x_0$. In that
case, $g(B) = -B$. Let $g_R\!: P \rightarrow P$ be the auxiliary
function $g_R(p) \eqdef - g(p)$. Observe that $g_R$ is orientation
\emph{reversing}, that $g_R(B) = B$, and that $g_R \neq \id_P$ (since $P$ contains three points $p$, $q$, $r$ with $\sgn(p,q,r) \neq 0$). Our intention is to build our poles
for $g$ from a great circle that $g_R$ goes across. If $B$ is affine,
we apply Proposition~\ref{p:goesacross} to find a great circle $C$
such that the restriction ${g_R}|_{B}$ goes across~$C$. The
antipodality of $g_R$ ensures that ${g_R}|_{-B}$ also goes across
$C$. The closed hemispheres bounded by $C$ determine two poles of $g$.

\bigskip

When $B$ is not affine, a similar argument works once the points in $B
\cap -B$ have been properly handled.  Let $E = B \cap - B$ be the set
of antipodal pairs of $B$, all of which are on $x_0^*$. By general
position of $P$, $|E| \le 4$. We cannot have $E = \{p,q,-p,-q\}$ with
$g$ acting by $p \mapsto q \mapsto -p \mapsto -q \mapsto p$. Indeed, this would imply
that for any $r \in B \setminus E$,
\[\sgn(p,q,r) = \sgn(g(p),g(q),g(r)) = \sgn(q,-p,g(r)) = \sgn(p,q,g(r)) ~, \]
which is impossible because the great circle through $p$ and $q$
separates $B$ from $g(B)=-B$. Next, if $E = \{p,q,-p,-q\}$ with $g(p)
= q$ and $g(q)=p$, then we can perturb $x_0$ into a nearby position
$x_1$ whose corresponding hemiset $B'$ is either $B \cup \{p,-q\}$ or
$B \cup \{-p,q\}$. We may have $\gamma(x_1) \neq \pm x_1$, but we do
not care as we still have $g(B')=-B'$. Since $B'$ is now affine, we
can find our poles as we did above, using a circle that ${g_R}|_{B'}$
goes across. Any pair $\{p,-p\}$ in $E$ with $g(p)=-p$ can be pushed
into $B'$ by a similar perturbation argument. We can therefore assume
that we are left with some great circle $x_1^*$ determining two
hemisets $B'$ and $-B'$ such that $g(B')=-B'$ and such that $E' = B'
\cap -B'$ consists of one or two pairs $\{p,-p\}$ with $g(p)=p$. We
compute $B_0$ and $B_1$ by applying, as above,
Proposition~\ref{p:goesacross} to the affine set $B' \setminus E'$ to
find two hemisets of $P\setminus E'$ fixed by $g$, say $B_0$ and
$-B_0$. The hemisets $B_0$ and $-B_0$ are affine so they can be
defined by a great circle $C$ that contains no point of $E$. For every
pair $\{p,-p\} \subseteq E'$, we add $p$ to the set, $B_0$ or $-B_0$,
on the same side as $p$ of $C$ and add $-p$ to the other. The
resulting sets $B_0$ and $-B_0$ are poles of $g$. This concludes the
proof of Proposition~\ref{p:poles}.

\section{Analysis of affine order types}
\label{s:non-labeled}

With the notion of pole hemisets and Proposition~\ref{p:poles} at our
fingertips, we can now analyze the average number of extreme points of
affine order types.

\subsection{Orbit types}
\label{s:OrbitTypes}

We start by gaining some insight on the projective symmetry groups
through their action on poles (carrying over Felix Klein's analysis of
finite subgroups of $SO(3)$, as presented in \cite{Sen90}).  Let $\g$
be a group. We say that $\g$ has \Emph{orbit type}\footnote{As
  defined, a group $\g$ could have more than one orbit type. As we
  will see later, in Proposition~\ref{p:mcs}, it turns out that every
  projective symmetry group has a unique orbit type.}
$[\mu_1,\mu_2,\ldots,\mu_k]$, $\mu_1 \le \mu_2 \le \ldots \le \mu_k$,
if there exists a projective point set $P$ with symmetry group $\g$
such that the action of $\g$ on the poles of $P$ defines $k$ orbits of
sizes $\mu_i$, $i=1,2,\ldots,k$.

\begin{proposition}\label{p:OrbitType}
  Let $\g$ be the symmetry group of a projective set of at least $6$ points
  in general position. If $\g$ is nontrivial, then its possible orbit types are 
  $[1,1]$, $[4,4,6]$, $[6,8,12]$,
  $[12,20,30]$ or $[2,N/2,N/2]$, where $N \eqdef |\g|$.
\end{proposition}
\begin{MyProof}
Let $P$ be a projective set of $2n$ points in general position with
symmetry group $\g$. We let $N \eqdef |\g|$ and assume $N \ge 2$. We
are going to count in two ways the pairs $(g,B)$, with $g \in
\g\setminus\{\id\}$ and $B$ a pole of $g$. For the first count,
note that every $g \in \g\setminus\{\id\}$ has exactly two poles by
Proposition~\ref{p:poles}. Hence, the number of pairs is $2(|\g|-1)
= 2N-2.$

\bigskip

The second count is less direct. Let $\PP$ denote the set of poles of
$P$. Recall that for every $B \in \PP$, $\g(B)$ denotes its orbit and
$\g_B$ its stabilizer under $\g$. Note that by the definition of a
pole, $|\g_B| \ge 2$. We number the orbits of $\PP$ from $1$ to $K$
and let $\mu_i$ be the size of the $i$th orbit. By the
orbit-stabilizer theorem, for every $B \in \PP$, $|\g| = |\g_B|\cdot
|\g(B)|$. It follows that every hemiset in the $i$th orbit has a
stabilizer of the same size; we let $\gamma_i$ denote that size (so
$\mu_i \gamma_i = N$). Now, a hemiset $B \in \PP$ occurs in a pair
$(g,B)$ exactly for the nontrivial permutations in the stabilizer
$\g_B$, that is, $|\g_B|-1$ times. The number of pairs is therefore
$\sum_{i=1}^K \mu_i (\gamma_i-1) = K N - \sum_{i=1}^K \mu_i$.

\bigskip

Equating the two counts, dividing by $N$, and rearranging terms gives
$\sum_{i=1}^K \frac{1}{\gamma_i} = K - 2 + \frac{2}{N}~.$ This
immediately restricts the range of possible values of $K$. Since each
$\gamma_i$ is at least $2$ (by definition of a pole), $K$ must be less than $4$. Since $N \ge 2$, $K > 1$\grey{.} and the parameters thus satisfy

\begin{eqnarray}
 \hbox{either } K=2 \qquad \hbox{and} & \qquad \qquad
\frac{1}{\gamma_1} + \frac{1}{\gamma_2} = \frac{2}{N}
~\Leftrightarrow~ \mu_1 + \mu_2 = 2,\label{eq:2Orbits}\\ 
 \hbox{or } K=3 \qquad \hbox{and} &\qquad \qquad \frac{1}{\gamma_1} + \frac{1}{\gamma_2} + \frac{1}{\gamma_3} = 1 + \frac{2}{N} ~\Leftrightarrow~ \mu_1 + \mu_2 + \mu_3 = N + 2.
\label{eq:3Orbits}
\end{eqnarray}

\noindent
For $K=2$ clearly, the only positive integer solution of $\mu_1 +
\mu_2 = 2$ is $\mu_1= \mu_2 =1$, and the orbit type that $P$ allows
for $\g$ is $[1,1]$.

\bigskip

For $K=3$, let us recall that all $\mu_i$ are divisors of $N$. Since
all $\gamma_i$ are at least 2, all $\mu_i$ are at most $N/2$. Let us
assume they are ordered $\mu_1 \le \mu_2 \le \mu_3 \le N/2$. We have
$\mu_3 > N/3$ (otherwise $\mu_1 + \mu_2 + \mu_3 \le N$,
  contradicting (\ref{eq:3Orbits})), so $\mu_3 = N/2$ is determined
and we are left with $\mu_1 + \mu_2 = N/2 + 2$. We have $\mu_2 > N/4$
(otherwise $\mu_1 + \mu_2 \le N/2$), so $\mu_2 \in \{N/2,N/3\}$. If
$\mu_2 = N/2$, then the orbit type is $[2,N/2,N/2]$. If $\mu_2 = N/3$
then we must have $\mu_1 = N/6 + 2$. Since $\mu_1$ divides $N$, the
only feasible choices are
\begin{eqnarray*}
\mu_1 = N/3  &\Rightarrow& N=12 \mbox{~~and~~} [\mu_1,\mu_2,\mu_3]=[4,4,6]\\
\mu_1 = N/4  &\Rightarrow& N=24 \mbox{~~and~~} [\mu_1,\mu_2,\mu_3]=[6,8,12]\\
\mu_1 = N/5  &\Rightarrow& N=60 \mbox{~~and~~} [\mu_1,\mu_2,\mu_3]=[12,20,30]
\end{eqnarray*}
This completes the proof.
\end{MyProof}

\subsection{More on affine symmetries}
\label{s:symaff}

Next, we clarify the symmetries of affine sets. 

\begin{MyProof}[Proof of Theorem~\ref{t:classAff}]
  Let $A$ be an affine set with layer sequence
  $(A_0,A_1,\ldots,A_\ell)$ and symmetry group $\f$. Note that by Proposition~\ref{p:act}, any
  $f\in \f$ preserves the layer sequence, that is $f(A_i) = A_i$. Moreover, for any non-lonely point $p \in
  A$, the stabilizer $\f_p$ is reduced to $\{\id\}$ by
  Lemma~\ref{l:unique-if-not-lonely}, so $|\f(p)| = |\f|$ by the
  orbit-stabilizer theorem. Now, consider a layer $A_i$ not reduced to
  a single point. Any map $f \in \f$ maps a positive extreme edge of
  $A_i$ to another one. The orbits under $\f$ partition $A_i$ into
  classes of equal sizes. Since $|\f| = |\f(p)|$ for any $p\in A_i$,
  $|\f|$ divides $|A_i|$.

  \bigskip

  It is left to show that $\f$ is cyclic. Fix $p \in A_0$. The set
  $\f(p)$ is in convex position, in fact a subset of $A_0$, so let
  $(p_0=p, p_1, p_2, \ldots, p_{k-1})$ be some CCW extreme points order of
  $\f(p)$. Let $f \in \f$ be the permutation with $f(p) = p_1$. We
  then have $f(p_i) = p_{i+1}$ for $i=0,1,\ldots,k-1$ (indices
  $\bmod\, k$) since $f$ preserves positive extreme edges. From
  $f^i(p_0) = p_i$, it follows that $\{f^0,f^1,\ldots,f^{k-1}\}$ are
  all distinct. Since $|\f| = |\f(p)|=k$, $\f$ is generated by $f$.

  \bigskip

  Finally, assume that $A$ has a lonely point $q$ (hence $f(q) = q$) and that $\f$ has
  even order. There is an element $f \in \f$ of order 2, \ie $f^2 = \id$ (choose $f = f_0^{k/2}$ for a generator $f_0$ of
  $\f$). For any other point $p$, we have $f(p) \neq p$ by
  Lemma~\ref{l:unique-if-not-lonely}, so
  \[ \sgn\pth{q,p,f(p)} = \sgn(\underbrace{f(q)}_{=q}, f(p), p) = -\sgn\pth{q, p, f(p)}\] 
  implies that $\sgn(q,p,f(p)) = 0$, contradicting the assumption that
  $A$ is in general position.
\end{MyProof}

\bigskip

\begin{corollary}\label{c:labeling}
  Let $A$ be an affine set of $n$ points with symmetry group $\f$. The
  orderings of $A$ realize exactly $\frac{n!}{|\f|} \ge (n-1)!$
  pairwise distinct labeled affine order types.
\end{corollary}
\begin{MyProof}
  Let $\f$ denote the symmetry group of $A$. Recall that two labelings
  $A_{[\lambda]}$ and $A_{[\mu]}$ of $A$ determine the same labeled
  order type if and only if $ \mu^{-1} \circ \lambda$ is a symmetry of
  $A$. The labelings of $A$ therefore determine $n!/|\f|$ labeled
  affine order types. Theorem~\ref{t:classAff} implies $|\f| \le n$,
  so this number is always at least $(n-1)!$.
\end{MyProof}

\bigskip

We also refine the upper bound on the number of affine order
types with many symmetries. 

\begin{proposition}
\label{p:OTsym}
  There is a constant $c_0$, such that for all $1 \le k \le n$, there are at most $\pth{c_0 \frac{n}{\sqrt{k}}}^{4n}$ simple, affine order types of size $n$ with $k$ symmetries.
\end{proposition}
\begin{MyProof}
Let $\OTaff{n,k}$ denote the set of simple, affine order types of size $n$ with $k$ symmetries. By Theorem~\ref{t:classAff}, either $k$ divides $n$ and none of the order types in $\OTaff{n,k}$ has a lonely point, or $k$ divides $n-1$ and all do.

Let $A$ be an affine point set with order type in $\OTaff{n,k}$. Again by Theorem~\ref{t:classAff}, the symmetry group $\f$ of $A$ is cyclic. We let $f$ be the generator of $\f$ such that for every non-lonely point $p \in A$, the points $p$, $f(p)$, $f^2(p)$, \ldots, $f^{k-1}(p)$ in its orbit appear in this order (counterclockwise) in the layer of $A$ that contains $p$.

  We call a labeling $(p_0,p_1, \ldots, p_{n-1})$ of $A$ a \Emph{standard labeling} if $p_i = f(p_{i-1})$ for all $0 \le i \le n-1$ with  $i \bmod k \neq 0$. Note that this simply means that for each $a$, $0 \le a \le \lfloor \frac{n}{k}\rfloor-1$, we have
  \[
  (p_{ak}, p_{ak+1}, p_{ak+2}, \ldots, p_{ak+k-1}) = (p_{ak}, f(p_{ak}), f^2(p_{ak}), \ldots f^{k-1}(p_{ak}))
  \]
  and if $n$ is not a multiple of $k$,  then $p_{n-1}$ is the unique lonely point in $A$. The points $p_{ak}$, $a=0,1,\ldots,\lfloor n/k \rfloor -1$ are called \Emph{anchors} in the given standard labeling. Note that if a lonely point exists, it is not an anchor point. For every non-lonely point $p$, there is an $i_p$, $0 \le i_p \le k-1$, such that $f^{i_p}(p)$ is an anchor.
  
It follows that the orientations $\sgn(p^*,q,r)$ for $(p^*,q,r) \in A^3$, $|\{p^*,q,r\}|=3$, $p^*$ an anchor, determine all orientations of all triples $(p,q,r) \in A^3$, $|\{p,q,r\}|=3$, since 
\[
\sgn(p,q,r) = 
\left\{ 
\begin{array}{ll}
\sgn(f^{i_p}(p), f^{i_p}(q), f^{i_p}(r)) & \mbox{if $p$ is not a lonely point, and} \\
- \sgn(f^{i_q}(q), p, f^{i_q}(r)) & \mbox{if $p$ is a lonely point (hence $f^{i_q}(p) = p$).}
\end{array}
\right.
\]
 
 We represent the space of
  all $n$-point affine sets by $\RR^{2n}$, equipped with the
  coordinate system $(x_0,y_0,x_1,y_1,$ \ldots,
  $x_{n-1},y_{n-1})$, where $p_i=(x_i,y_i)$. Let $\PP_{n,k}$ be
  the family of polynomials
  \[ \begin{aligned}
    \PP_{n,k} \eqdef \left\{ \left|\begin{matrix}x_{ak} & x_{i} & x_{j}\\ y_{ak} & y_{i} & y_{j}\\ 1 & 1 & 1 
  \end{matrix}\right|
    \st  0 \le a \le \lfloor \frac{n}k\rfloor-1, 0 \le i,j \le n-1, |\{ak,i,j\}| = 3\right\}.
  \end{aligned}\]
We let $m \eqdef |\PP_{n,k}| = \lfloor \frac{n}{k} \rfloor (n-1)(n-2) < \frac{n^3}{k}$ and order the
    polynomials in $\PP_{n,k}$ as $P_1, P_2, \ldots, P_m$. The number of standard labelings of order types in  $\OTaff{n,k}$ (and thus $|\OTaff{n,k}|$) is at most the number of sign vectors
    \[\{\pth{\textrm{sign} \pth{P_1(x)},\textrm{sign} \pth{P_2(x)}, \ldots, \textrm{sign} \pth{P_m(x)}}  \in \{-1,+1\}^m \st x \in \RR^{2n}\}\]
   of the polynomials in $\PP_{n,k}$. 
     By Warren's
    theorem~\cite[Theorem\,3]{warren1968lower}, $m\le m'$ real polynomials in
    $v \le m'$ variables, each of degree at most $2$, determine at most
    $\pth{\frac{8em'}{v}}^{v}$ sign vectors. Here $v=2n$ and we can choose $m' =\frac{n^3}{k}$, so
  \[ |\OTaff{n,k}| \le \pth{\frac{4e\,n^2}{k}}^{2n}
  \]
    and the condition $v \le m'$  holds for all $n \ge 2$  and $k\le n$. The claimed bound on $|\OTaff{n,k}|$ follows with $c_0 = 2\sqrt{e}$. 
\end{MyProof}

\noindent {\bf Remark.} We mention that the proof above carries many redundancies which can be exploited. Improvements we see, however, would not be relevant when we apply it in the proof of Theorem~\ref{t:avg} in Section~\ref{sub:CountExtrPts} below.  Let us still briefly sketch an improvement by a factor of $\frac{1}{\lfloor n/k\rfloor!k^{\lfloor n/k\rfloor-1}}$. For that, note that the upper bound we derived is actually for the number of standard labelings of order types with $k$ symmetries. How many \emph{distinct} such \emph{standard-}labeled order types does a given element in $\OTaff{n,k}$ have? There are $\lfloor n/k\rfloor!$ ways to order the non-lonely orbits, and there are $k^{\lfloor n/k\rfloor}$ choices for the anchors (the first elements of the respective orbits). Note, however, that for a standard labeling $(p_0,p_1, \ldots, p_{n-1})$, all labelings $(f^i(p_0),f^i(p_1), \ldots, f^i(p_{n-1}))$, $i=0,1,\ldots,k-1$, yield the same labeled order type, all standard labelings. Hence we get exactly $\lfloor n/k\rfloor!k^{\lfloor n/k\rfloor}/k$ standard labelings for each order type in $\OTaff{n,k}$. 

Let us take a step back here and review, why did we have to divide by $k$ here? If we have a \emph{point set}, then, for a standard labeling $(p_0,p_1, \ldots, p_{n-1})$, all labelings $(f^i(p_0),f^i(p_1), \ldots, f^i(p_{n-1}))$, $i=0,1,\ldots,k-1$, are distinct orderings (in standard form) of this given point set. However, as a labeled order type, they are clearly all the same, since $f$ is a symmetry ($\sgn(p_a,p_b,p_c) = \sgn(f^i(p_a),f^i(p_b),f^i(p_c))$.

With $N! > \sqrt{2\pi N}\pth{\frac{N}{e}}^{N}$ we have 
\[\lfloor n/k\rfloor!k^{\lfloor n/k\rfloor-1} > \sqrt{2\pi \lfloor n/k\rfloor}\pth{\frac{\lfloor n/k\rfloor}{e}}^{\lfloor n/k\rfloor}k^{\lfloor n/k\rfloor-1} \ge (c_1 n)^{n/k + O(1)} ~.
\]
for\footnote{``$+O(1)$'' in the exponent stands for a negative constant.} a constant $c_1>0$ sufficiently small. The resulting bound of $|\OTaff{n,k}| \le n^{-n/k+O(1)} \left(c_2 \frac{n}{\sqrt{k}}\right)^{4n}$ is now in line with the known bound of $(cn)^{3n+o(1)}$ for $|\OTaff{n,1}|$. For the other extreme case of $k=\Theta(n)$, the bound is $(c' \sqrt{n})^{\,4n+o(1)} = (c'' n)^{2n+o(1)}$ and we do not know how close that is to the truth.    

\subsection{Counting extreme points in one projective class}

For any affine order type $\omega$ we write $h(\omega)$ for its number
of extreme points. For any projective set $P$, we define $h(P) \eqdef
\frac1{|\OTaff{P}|}\sum_{\omega \in \OTaff{P}} h(\omega)$.

\begin{proposition}\label{p:avg-proj}
  If $P$ is a projective set of $2n$ points in general position with
  $N$ symmetries, then
  $ 4 - \varepsilon_n \le h(P) \le  4 + 3 N/n$
  with $0 \le \varepsilon_n = O\pth{\frac{1}{n^2}}$. Moreover, if
  $N=1$ then $h(P) = \TheNumber$.
\end{proposition}
\begin{MyProof}
  Let $\g$ denote the symmetry group of $P$ (so $N = |\g|$). Let us
  put $M \eqdef |\OTaff{P}|$, $\OTaff{P} = \{\omega_1,\omega_2,
  \ldots, \omega_M\}$ and $H \eqdef h(P) = \frac1M \sum_{i=1}^M
  h(\omega_i)$. Let $\mu_i \eqdef |\g(\omega_i)|$; by
  Corollary~\ref{c:repetition}, $\mu_i$ is the number of affine
  hemisets of $P$ with order type $\omega_i$.

  \bigskip

  By~Lemma~\ref{l:duality-basic}, affine hemisets of $P$ are in
  bijection with cells of $P^*$, of which there are $2{n \choose
    2}+2$. Also, a point $p$ is extreme in an affine hemiset of $P$ if
  and only if $p^*$ supports an edge of the corresponding cell;
  there are $4{n \choose 2}$ edges, and each edge is adjacent to two
  cells. Altogether we obtain
\begin{equation}\label{eq:duality2}
  \sum_{i=1}^M \mu_i = 2{n \choose 2} +2 \qquad \mbox{and} \qquad
  \sum_{i=1}^M \mu_i h(\omega_i) = 8 {n \choose 2}~.
\end{equation}

\bigskip

Let $K'$ be the number of order types in $\OTaff{P}$ with nontrivial
symmetry group. We claim that $K' \le 3$. Indeed, by
Lemma~\ref{l:actaff} the order types of $\OTaff{P}$ correspond to the
orbits of affine hemisets of $P$ under $\g$. Moreover, for every
affine hemiset $A$ in the orbit of $\omega_i$, the stabilizer $\g_A$
is isomorphic to the symmetry group of $\omega_i$. Hence, when this
group is nontrivial, the orbit consists of poles of $P$; there are at
most three such orbits by Proposition~\ref{p:OrbitType}. Let us
stress that $K'$ counts only affine pole orbits, whereas
Proposition~\ref{p:OrbitType} also accounts for non-affine pole
orbits.

\bigskip

When $K'=0$, which holds, in particular, for $\g$ the trivial group,
we have $M = 2{n \choose 2} +2$ and $\mu_i = 1$ for all
$i=1,2,\ldots,M$, and we obtain
\begin{equation}\label{eq:K=0}
  H = \frac{1}{M}{\sum_{i=1}^M h(\omega_i)} = \frac{8 {n \choose
      2}}{2{n \choose 2} + 2} = \TheNumber~,
\end{equation}
as for labeled order types. This gives us the last statement.

\bigskip

So assume that $1 \le K' \le 3$ and that we have ordered $\OTaff{P}$
so that the $K'$ order types with nontrivial symmetry group are
$\omega_1, \ldots, \omega_{K'}$. We therefore have $\mu_i < N$ for $i
\le K'$ and, by Corollary~\ref{c:repetition}, $\mu_i = N$ for $i >
K'$. Equation~(\ref{eq:duality2})-right can be rewritten as
\[ \begin{aligned}
  8 {n \choose 2} = & \sum_{i=1}^M \mu_i h(\omega_i) = N
  \sum_{i=1}^M h(\omega_i) - \sum_{i\le K'} (N-\mu_i) h(\omega_i)\\
  & \Rightarrow \qquad M H  = \frac{1}{N} \pth{ 8 {n \choose 2} + \sum_{i\le K'} (N-\mu_i) h(\omega_i)}.
\end{aligned}\]
For the same reason, Equation~(\ref{eq:duality2})-left can be rewritten as
\[ \begin{aligned}
2{n \choose 2} +2 = & \sum_{i=1}^M \mu_i  = N M - \sum_{i\le K'} (N-\mu_i)
\nonumber \\
& \Rightarrow \qquad M = \frac{1}{N} \left( 2{n \choose 2} + 2 + \sum_{i\le K'} (N-\mu_i) \right)
\end{aligned}\]
Together, this gives $H = 4 + \Delta$ where $\Delta \eqdef \frac{\displaystyle-8 + \sum_{i\le K'} (N-\mu_i) (h(\omega_i)-4)}{\displaystyle2{n \choose 2} + 2 + \sum_{i\le K'} (N-\mu_i)}$.

\bigskip

On the one hand, 
\[ \Delta \le \frac{\displaystyle\sum_{i\le K'} N (h(\omega_i)-1)}{\displaystyle2{n \choose 2}} \le \frac{K' N(n-1)}{n(n-1)} \le 3\frac{N}n,\]
which proves the upper bound. For the lower bound, recall that the
order of the symmetry group of $\omega_i$ equals $N/\mu_i$ and must
divide $h(\omega_i)$. Now, if the numerator of $\Delta$ is less than
$-8$, there must exist some $i$, $1 \le i \le K'$, with $h(\omega_i) =
3$. By Proposition~\ref{p:OrbitType}, this can happen only for
$(N,\mu_i) \in \{(3,1), (6,2), (12,4), (24,8), (60,20)\}$. Hence
\[ \Delta \ge \frac{\displaystyle-8-3\cdot40}{\displaystyle2{n \choose 2}} = -\frac{128}{n(n-1)},\]
which proves the lower bound.
\end{MyProof}

\subsection{Counting extreme points in affine order types}
\label{sub:CountExtrPts}

We now build on Proposition~\ref{p:avg-proj} to prove
Theorem~\ref{t:avg}. The main issue is the factor $N/n$: projective
order types with $\Omega(n)$ symmetries may contribute substantially more than $4$ to
the average. We keep them in check using Proposition~\ref{p:OTsym} and
the following consequence of Proposition~\ref{p:OrbitType}.

\begin{corollary}\label{c:symprojaff}
  Any projective order type with $N > 60$ symmetries contains an
  affine hemiset with at least $N/2$ symmetries.
\end{corollary}

\begin{MyProof}[Proof of Theorem~\ref{t:avg}]
  The lower bound of Proposition~\ref{p:avg-proj} immediately
  implies that the average number of extreme points is at least
  $4-O(n^{-2})$. We therefore focus on the upper bound.

  \bigskip

  If two affine sets $A_1$, $A_2$ have the same affine order type, then their
  projective completions $A_1 \cup -A_1$ and $A_2 \cup -A_2$ have the
  same projective order type. Thus, the family $\{\OTaff{\pi} \st \pi
  \in \OTproj{n}\}$ partitions $\OTaff{n}$. It follows that $|\OTaff
  n| = \sum_{\pi \in \OTproj n}|\OTaff \pi|$, and
  \begin{equation}\label{eq:break}
    \sum_{\omega \in \OTaff n} h(\omega) =  \sum_{\pi \in \OTproj n} \sum_{\omega \in \OTaff \pi} h(\omega).
  \end{equation}

  \bigskip

  For $n \in \NN$ and $N_0 \in \RR$, let $\OTproj {n, \ge N_0}$ (resp.\
  $\OTproj {n, < N_0}$) denote the number of projective order types
  $\pi$ with $|\pi|=2n$ and $|\G{\pi}| \ge N_0$ (resp.\ $|\G{\pi}| <
  N_0$). For any $N_0$, $1 \le N_0 \le n$, we can inject the bounds of
  Proposition~\ref{p:avg-proj} in Equation~\eqref{eq:break} and obtain (we use $N \le \min\{2n, 60\}$ and therefore $N/n = O(1)$):
  \[\begin{aligned}
  \sum_{\omega \in \OTaff n} h(\omega) & =  \sum_{\pi \in \OTproj {n,
      < N_0}} \sum_{\omega \in \OTaff \pi} h(\omega) + \sum_{\pi \in
    \OTproj {n,  \ge N_0}} \sum_{\omega \in \OTaff \pi} h(\omega)\\
  & \le \ 4\,|\OTaff n| + \sum_{\pi \in \OTproj {n, < N_0}} 3\,\frac{N_0}n
  |\OTaff \pi| + \sum_{\pi \in \OTproj {n,  \ge N_0}} O(1) \ |\OTaff
  \pi|\\ & \le \ \pth{4+3N_0/n}|\OTaff n|  + O(n^2) |\OTproj {n,  \ge N_0}|.
  \end{aligned}\]
  We cut off at $N_0 = 2n^{2c}$, with $0<c<\frac12$ to be specified
  shortly. By Corollary~\ref{c:symprojaff}, the number of projective
  order types with at least $N_0$ symmetries is at most the number of
  affine order types with at least $\frac{N_0}2$ symmetries. By
  Proposition~\ref{p:OTsym}, the latter is at most
    \[ \sum_{k=\frac{N_0}2}^n \pth{c_0\frac{n}{\sqrt{k}}}^{4n} \le n \pth{c_0\frac{n}{\sqrt{N_0/2}}}^{4n} = {c_0}^{4n}n^{4(1-c)n+1}.\]

  \bigskip
  
  Crudely factoring out symmetries -- by dividing by $n!$ -- in the
  Goodman-Pollack lower bound of $(n!)^4/2^{3n}$ on the number of
  labeled order types~\cite[$\mathsection 5$]{goodman1986upper}, we
  get $|\OTaff n| \ge (c_2 n)^{3n+O(1)}$ for some constant $c_2$. The
  bound therefore becomes
  \[ \frac1{|\OTaff n|}\sum_{\omega \in \OTaff n} h(\omega) \le
    4+3n^{2c-1} + O(n^2) \frac{{c_0}^{4n} n^{4(1-c)n+1}}{{c_2}^{3n}
      n^{3n+O(1)}} = 4+3n^{2c-1} + n^{O(1)} {c_3}^n n^{(1-4c)n} \]
  for some constant $c_3$. Taking $c\eqdef \frac14+\frac{\log
    c_4}{\log n}$, for some $c_4 > \sqrt[4]{c_3}$, we get
  \[ \frac1{|\OTaff n|} \sum_{\omega \in \OTaff n} h(\omega) \le 4 + 3 n^{-\frac12+\frac{2\log c_4}{\log n}}+ n^{O(1)} c_3^n \,\underbrace{n^{-\frac{4 \log c_4}{\log n}n}}_{c_4^{-4n}}\le 4+O\pth{n^{-\frac12+O(\frac{1}{\log n})}}\]
  as announced.
\end{MyProof}

\section{Concentration of (labeled) order types of random point sets}
\label{s:concentration}

Let us now turn our attention to the efficiency of random sampling
methods for order types based on sampling point sets. We start by a
sufficient condition for a family of distributions on $\LorOTaff n$ to
exhibit concentration.

\begin{proposition}\label{p:cond-for-conc}
  Let $\mu_n$ be a probability distribution on $\LorOTaff n$ and let
  $Z_n$ denote the number of extreme points in a (labeled) order type
  chosen from $\mu_n$. If $\Ex{Z_n} \to_{n \to \infty} \infty$ and
  $\Var{Z_n} = o\pth{\Ex{Z_n}^2}$, then $\{\mu_n\}_{n \ge 3}$ exhibits
  concentration.
\end{proposition}
\begin{MyProof}
  We let~$\LL_n$ denote the set of (labeled) planar, simple order
  types of size $n$ with at least {$\Ex{Z_n}/2$} extreme points. On
  one hand, by Markov's inequality and Theorem~\ref{t:avg} (Theorem~\ref{t:avgl}, resp.), we have
  \[ \frac{|\LL_n|}{|\LorOTaff n|} \le \frac{4+o(1)}{{\Ex{Z_n}/2}} \to_{n \to \infty} 0,\]
  so $\LL_n$ is a vanishingly small part of $\LorOTaff
  n$. On the other hand, the Bienaym{\'e}-Chebyshev inequality ensures
  that for any real $t>0$,
  \[  \Prob{|Z_n-\Ex{Z_n}| \ge t\sqrt{\Var{Z_n}}} \le \frac1{t^2}.\]
  Let us take {$t = \frac{\Ex{Z_n}}{2\sqrt{\Var{Z_n}}}$}, so that
 
  \[  \Prob{Z_n \le {\frac{\Ex{Z_n}}{2}}} \le \Prob{|Z_n-\Ex{Z_n}| \ge {\frac{\Ex{Z_n}}{2}}} \le {\frac{4\Var{Z_n}}{\Ex{Z_n}^2}} ,\]
  which goes to $0$. This ensures that the probability that a
  (labeled) order type chosen from~$\mu_n$ lies in $\LL_n$ goes to
  $1$.
\end{MyProof}

\bigskip

Theorem~\ref{t:concentration} follows from
Proposition~\ref{p:cond-for-conc} and previous work in stochastic
geometry.

\begin{MyProof}[Proof of Theorem~\ref{t:concentration}]
  Let $\mu$ be a probability distribution on $\RR^2$ and let $Z_n$
  denote the random variable counting the extreme points in a set (or
  sequence) of $n$ random points chosen independently from $\mu$.

  When $\mu$ is the uniform probability distribution in a compact
  convex set $K$, $\Ex{Z_n}$ is $\Omega(\log
  n)$~\cite[Theorems~1--2]{BaranyLarman}. For $K$ smooth,
  Vu~\cite[Corollary~2.12]{vu2005sharp} proved that $\Var{Z_n} =
  \Theta\pth{\Ex{Z_n}}$. For $K$ a polygon, B\'ar\'any and
  Reitzner~\cite{barany2010variance} proved that $\Var{Z_n} =
  \Theta\pth{\Ex{Z_n}}$. Proposition~\ref{p:cond-for-conc} therefore
  applies.

  When $\mu$ is a Gaussian distribution on $\RR^2$, $\Ex{Z_n}$ is
  $\Omega(\sqrt{\log n})$ and $\Var{Z_n} = \Theta\pth{\Ex{Z_n}}$,
  see~\cite[$\mathsection 2.3$]{Reitzner-survey}.
\end{MyProof}

\section{Order types with excluded patterns}
\label{s:avoid}

Building on the affine-projective relation (Section~\ref{s:hemisets}),
the correspondence between affine hemisets and dual cells
(Lemma~\ref{l:duality-basic}), and the classification of affine
symmetries, we can now prove that certain order types are hard to
avoid.

\begin{MyProof}[Proof of Theorem~\ref{t:avoid}]
  Fix $k$ and let $\tau$ be the $k$-point order type with three
  extreme points, and whose $k-3$ interior points form a convex chain
  together with two of the extreme points.\footnote{For the reader familiar with this terminology, this is equivalent to saying that $\tau$
  is the order type obtained from $k$ points in convex position by
  sending a line cutting off one point to infinity.}

  Let $n$ be large enough such that any $n/2$ points in general
  position in the plane contain a convex $2k$-gon
  (see Suk~\cite{suk2017erdHos} for the most recent
  bounds). Let $P$ be a projective set of $2n$ points in
  general position. We claim that for every projective set $P$ of
  size~$2n$, there are at most two affine hemisets of $P$ (an affine
  hemiset and its antipodal set) whose order types do not contain
  $\tau$. This shows that at most two of the affine order types in
  $\OTaff{P}$ avoid $\tau$. Since $|\OTaff{P}| = \Omega(n)$ we obtain
  that the number of $n$-point affine order types that do not contain
  $\tau$ is at most $O(n^{-1})|\OTaff n|$. The fact that $|\OTaff{P}|
  = \Omega(n)$ follows from (i) that the number of affine hemisets of
  $P$ equals the number of cells of $P^*$, that is, $2{n \choose 2}+2$
  (Lemma~\ref{l:duality-basic}), (ii) an order type $\omega$ appears
  with multiplicity $|\g|/|\f|$ ($\g$ and $\f$ the symmetry groups of
  $P$ and $A$, resp., Corollary~\ref{c:repetition}), and (iii) $|\g|
  \le \max\{60,2n\}$ (Corollary~\ref{c:symprojaff}
  and~Theorem~\ref{t:classAff}).

  It remains to prove the claim. So suppose $P$ has an affine hemiset
  $A$ with no subset of order type $\tau$. Let $\cH$ be a closed
  hemisphere such that $A' \eqdef P \cap \cH$ is an affine hemiset of
  $P$ distinct from $A$ and $-A$. We want to show that $A'$ has a
  subset of order type $\tau$. Let $\gC$ be the great circle bounding
  $\cH$. Since $A \neq A' \neq -A$, there are points of $A$ on both
  sides of $\gC$. We fix a point $p \in A$ such that the side of $\gC$
  not containing $p$ has at least as many points in $A$ as the side
  containing $p$. That is, the other side of $\gC$ has at least $n/2$
  points of $A$, so it must contain a subset $D$ of $2k$ points in
  convex position. W.l.o.g.\ let us assume that $p \not\in A'$, so
  that $D \cup \{-p\} \subseteq A'$ (otherwise,
    switch from $A'$ to $-A'$ and observe that $\tau$ appears in $A'$
    if and only if it appears in $-A'$).  Let
  $q_1$ and $q_2$ denote the neighbors of $p$ on the convex hull of
  the affine set $D \cup \{p\} \subseteq A$. Note that $q_1$ and $q_2$
  are also the neighbors of $-p$ on the convex hull of $D \cup \{-p\}
  \subseteq \grey{-}A'$ . Since $A$ has no subset of order type
  $\tau$, the interior of the triangle $pq_1q_2$ must contain less
  than $k-3$ points of $D$. Then, $p \cup D$ has at least $k+5$
  extreme points, and $\{-p\} \cup D \subseteq \grey{-}A'$ contains a
  subset of order type~$\tau$.
\end{MyProof}

\section{Classification of projective symmetries and their pole orbits}
\label{s:classProj}

This section analyzes further the symmetry groups of projective sets,
and their orbit structure. While this is not essential for the
targeted results of this paper, we consider this of independent
interest. It should be made clear that the orbit type per se, as we
considered it so far, does not say much about the underlying
group. Still, together with the special properties of the groups we
have at hand, we can derive properties of the cyclic subgroups that
can occur in the symmetry groups. Building on this, we will derive the
classification.

\bigskip

Given two groups $H,G$, let us write $H \le G$ to mean that $H$ is a
subgroup of $G$. For a group $G$ and an element $g \in G$, we write $\langle g \rangle$ for the subgroup of $G$ generated by $g$. Note that if $G$ is finite then $\langle g \rangle$ is cyclic.

\subsection{From pole stabilizers to maximal cyclic subgroups}

A \Emph{maximal cyclic subgroup} of a group $G$ is a cyclic subgroup
of $G$ that is not properly contained in another cyclic subgroup of
$G$. We next relate the maximal cyclic subgroups of the symmetry
group of a projective point set to the stabilizers of its
hemisets. Before that, we should get some hold on the symmetry groups
of \emph{non-affine} hemisets. 

\begin{lemma}\label{l:non-affine-hemiset}
  Let $B$ be a non-affine hemiset of a projective set $P$ of at least $6$ points in general
  position, with $\g$ the symmetry group of $P$. The symmetry group of $B$ (and thus the stabilizer $\g_B$ of $B$) is either trivial, or cyclic of order $2$ or $4$.\end{lemma}
\begin{MyProof} Proposition~\ref{p:act}\ref{i:iiact} and Lemma~\ref{l:unique-if-not-lonely} show that the symmetry group $\f$ of a non-affine hemiset $B$ has order at most $|B \cap -B| \in \{2,4\}$. That is, we are done if $|B \cap -B|=2$, since the only group of order $2$ is cyclic. So let us assume that $B \cap -B = \{p,-p,q,-q\}$. Consider first a symmetry $g$ with $g(p) = q$. Every point of $B$ is on the same side of the great circle through $p$ and $q$, so we cannot have $g(q) = p$: indeed, for any $r \in B \setminus  \{p,-p,q,-q\}$ we would have $\sgn(p,q,r) = \sgn(g(p),g(q),g(r)) = \sgn(q,p,g(r)) = - \sgn(p,q,g(r))$, a contradiction. This implies $g(q) = -p$, and thus the symmetry $g$ is determined by Lemma~\ref{l:EmbracingPreservation}\ref{i:iEmbracingPreservation} as $p\mapsto q \mapsto -p \mapsto -q \mapsto p$. This mapping generates a cyclic group of order $4$. Similarly, if $p \mapsto -q$. Otherwise, if $p$ maps neither to $q$ nor to $-q$, the symmetry group is either trivial or of order $2$, thus cyclic. By Lemma~\ref{l:actaff}\ref{i:iactaff}, $\g_B$ is isomorphic to $\f$.
\end{MyProof}

\bigskip

We now have the following correspondence.

\begin{proposition}\label{p:mcs}
  Let $P$ be a projective set, $|P|\ge 6$,  in general
  position, with symmetry group~$\g$. 
  \begin{EnumRom}
  \item For every hemiset $B$ of $P$, the stabilizer $\g_B$ is
    trivial or a maximal cyclic subgroup of $\g$.
  \item For every maximal cyclic subgroup $C \le \g$, if nontrivial, there are
    exactly two hemisets $B_0$, $B_1$ of~$P$ such that $C = \g_{B_0} =
    \g_{B_1}$; moreover, $B_0 = -B_1$.
  \end{EnumRom}
\end{proposition}
\begin{MyProof}
  Let $B$ be a hemiset of $P$  with $\g_B \neq \{\id\}$. First,
  note that $\g_B$ is cyclic (if $B$ is an affine hemiset, by
  Lemma~\ref{l:actaff}(i) and Theorem~\ref{t:classAff}, and if $B$ is
  a not affine, by Lemma~\ref{l:non-affine-hemiset}).  We now argue
  that $\g_B$, when nontrivial, is a maximal cyclic subgroup of
  $\g$. Suppose that $\g_B \leq C \leq \g$, for $C = \langle g_0
  \rangle$ a cyclic group. By Proposition~\ref{p:poles}, $g_0$ has two
  poles, which we denote by $B'$ and $-B'$. Any $g \in G_B \setminus
  \{\id\}$ is in $C$ and therefore writes $g = g_0^i$ for some integer
  $i$. This implies that $B'$ is a pole of $g$, since $g(B') =
  g_0^i(B') = B'$, and by Proposition~\ref{p:poles} we must have
  $B'=B$ or $B'=-B$. In either case $g_0 \in G_B$ and thus $G_B =
  C$. This proves statement~(i).

  Now, let $C = \langle g_0 \rangle$ be a maximal cyclic subgroup of
  $\g$. Let $\pm B$ be the poles of $g_0$, as per
  Proposition~\ref{p:poles}. For every $g \in C$, we have $g(B) = B$,
  so $C \le \g_B$. Since $\g_B$ is cyclic, it follows that $C =
  \g_B$. The same argument gives $C = \g_{-B}$. Finally, for every
  hemiset $B'$ of $P$ distinct from $\pm B$, we must have $g_0(B')
  \neq B'$ by Proposition~\ref{p:poles}, and $C \neq \g_{B'}$. This
  proves statement~(ii). 
\end{MyProof}

A first structural consequence is that projective symmetry groups are
what is called \emph{completely decomposable}~\cite{Suz50}, that is,
they have the following property:

\begin{corollary}\label{c:decomp}
  For any two maximal cyclic subgroups $C$, $C'$ of a projective
  symmetry group $\g$ we have $C \cap C' = \{\id\}$.
\end{corollary}
\begin{MyProof}
  Any nontrivial element in $\g$ has exactly two poles by
  Proposition~\ref{p:poles} and therefore belongs to exactly one
  maximally cyclic subgroup of $\g$ by Proposition~\ref{p:mcs}~(ii).
\end{MyProof}

\bigskip

Another consequence is that the action of a projective symmetry group
on the poles of a projective point set completely reveals its number
of maximal cyclic subgroups. Given a group $G$, let $\Mcs i G$ denote
the number of maximal cyclic subgroups of cardinality $i$ of $G$.

\begin{corollary}\label{c:mcs}
  Let $P$ be a projective set, $|P|\ge 6$, in general
  position, with symmetry group~$\g$. For any $i \ge 1$, the action of
  $\g$ on the poles of $P$ has exactly $\frac{2i}{|\g|} \Mcs i \g$
  orbits of size~$|\g|/i$.
\end{corollary}
\begin{MyProof}
  Let $\PP_i$ be the set of poles of $P$ with stabilizer of
  cardinality $i$. By Proposition~\ref{p:mcs}, $|\PP_i| = 2\,\Mcs i
  \g$. The action of $\g$ on the poles of $P$ partitions $\PP_i$ into
  orbits, since two poles in the same orbit have isomorphic
  stabilizers. Each orbit in $\PP_i$ has size $|\g|/i$ by the
  orbit-stabilizer theorem, so there must be $\frac{2i}{|\g|} \Mcs i
  \g$ orbits in $\PP_i$.
\end{MyProof}

\bigskip
\noindent
By Corollary~\ref{c:mcs}, the orbit type determines the number of
maximal cyclic subgroups of each size, and vice-versa. In particular,
a projective symmetry group has a single orbit type (a fact that is
not obvious otherwise). Proposition~\ref{p:OrbitType} therefore
yields the information summarized in Table~\ref{tb:mcs}.

\begin{table}

  \[
  \begin{array}{c|ccccc|c}
    |\g|&~~~& \mbox{orbit type} && \mbox{maximal cyclic subgroup statistics} & & \g     \\
    \hline
    N && [1,1] &\Leftrightarrow&  \mcs N = 1 & &  \ZZ_N
     \\
    4 && [2,2,2] &\Leftrightarrow&  \mcs  2 = 3 & & D_2 
     \\
    N>4 && [2,N/2,N/2] &\Leftrightarrow&  \mcs  2 = N/2,\  \mcs {N/2} = 1 &  &D_{N/2}
     \\
    12&&  [4,4,6] &\Leftrightarrow& \mcs 2 = 3, \ \mcs 3 = 4 & & A_4
    \\
    24&& [6,8,12] &\Leftrightarrow& \mcs 2=6, \ \mcs 3 = 4,\  \mcs 4 = 3 & & S_4
     \\
    60&& [12,20,30] &\Leftrightarrow& \mcs 2=15, \ \mcs 3 = 10, \ \mcs 5 = 6 & & A_5 
    
  \end{array}
  \]
  \caption{Orbit types of symmetry groups with maximal cyclic subgroup statistics. The last column anticipates the implied classification to follow below in Section~\ref{s:GroupClass}.}
  \label{tb:mcs}
\end{table}

\subsection{Group classification}
\label{s:GroupClass}

We now analyze the possible group structure of $\g$, proving
Theorem~\ref{t:classProj} on the way.

\subsubsection{Infinite cases: cyclic and dihedral}

Let us first dispose of the cases where the order may be arbitrarily
large. Let $\g$ be a projective symmetry group and let $N \eqdef
|\g|$. Recall that every element $g \in \g$ generates a cyclic
subgroup $\{\id, g, g^2, \ldots \} \le \g$ and is therefore contained
in some maximal cyclic subgroup.

\bigskip

If $\g$ has orbit type $[1,1]$, then it has a single maximal cyclic
subgroup, with $|\g|$ elements. Hence, $\g \simeq \ZZ_N$. 

\bigskip

Now assume that $\g$ has orbit type $[2,N/2,N/2]$. For $N=4$, we have
$\Mcs 2 \g = 3$ so $\g$ is a group with $4$ elements that is not
cyclic. The only possibility is the dihedral group $D_2$. For $N>4$,
we have $\Mcs 2 \g = N/2$ and $\Mcs {N/2} \g = 1$. Let $g_0$ be a
generator of the maximal cyclic subgroup of order $N/2$. Let $g_1 \in
\g \setminus \langle g_0 \rangle$. Note that Corollary~\ref{c:decomp}
implies that both $g_1$ and $g_0g_1$ are of order~$2$. Thus, the
subgroup generated by $g_0$ and $g_1$ is the dihedral group $D_{N/2} =
\langle g_0,g_1 \mid g_0^{N/2} = g_1^2 = (g_0g_1)^2 = \id
\rangle$. Since $\g$ and $D_{N/2}$ have equal cardinalities, it must
be that $\g \simeq D_{N/2}$.

\subsubsection{Finite cases: shortcuts}

For the remaining three cases, a natural approach is to compare the
information of Table~\ref{tb:mcs} to the classification of finite
groups. For instance, for orbit type $[4,4,6]$, the group has $12$
elements, none of which has order more than~$3$. From
the\footnote{Here we used
  \url{https://groupprops.subwiki.org/wiki/Groups_of_order_12}.}  five
groups of size~$12$, this readily rules out the cyclic group
$\ZZ_{12}$, the dihedral group $D_{12}$, the direct product $\ZZ_6
\times \ZZ_2$, as well as the dicyclic group ${\rm Dic}_{12}$ which
has an element of order $4$. This leaves $A_4$ as the only
possibility.

For a geometer, this does not provide much insight. We thus provide an
alternative proof that trades specific knowledge of groups of size
$12$, $24$ and $60$ for some analysis of the orbits. We let $\g$ be a
projective symmetry group, let $N \eqdef |\g|$, and let $\PP$ denote
the set of poles of some projective point set in general position with
symmetry group $\g$.

\subsubsection{Finite case: $[4,4,6] \Rightarrow A_4$}

Let $P$ be some projective point set with symmetry group $\g$ of orbit
type $[4,4,6]$ and size $12$ and let $O$ denote an orbit of size $4$
in the action of $\g$ on the pole hemisets of $P$. By
Proposition~\ref{p:poles}, every $g \in \g$ fixes exactly two poles of
$P$. The group $\g$ therefore acts faithfully\footnote{The action of a
  group $G$ on a set $X$ is \Emph{faithful} if for every $g \in
  G\setminus \{\id\}$, there is some $x \in X$ such that $g(x) \neq
  x$. Given two distinct elements $f,g \in G$, we have $f(x) \neq g(x)
  \Leftrightarrow (g^{-1} \circ f)(x) \neq x$. It follows that $G$
  acts faithfully on $X$ if and only if every element of $G$
  determines a distinct permutation of $X$, that is, $G$ is isomorphic
  to a subgroup of $\Sym X$.} on $O$, and must be a subgroup of $\Sym
O \simeq S_4$. There is only one subgroup of size $12$ in $S_4$:
$A_4$.

\subsubsection{Preparation: action on pairs of antipodal orbits}
\label{s:prep}

We will classify the remaining two cases by examining the action of
$\g$ not on pole hemisets, but on pairs of pole hemisets. We prepare
this by laying out a few basic facts.

\bigskip

Let $H$ be a nontrivial subgroup $H \le \g$, and let $B \in
\PP$. First, $B' \mapsto -B'$ defines a bijection between $H(B)$ and
$H(-B)$.

\begin{claim}\label{c:even}
  For any $B \in \PP$, $|H(B)| = |H(-B)|$.
\end{claim}

\noindent
Let us say that $g$ \emph{reverses} $B$ if $g(B)=-B$. If $g$ reverses
$B$, then $g \notin \g_B$ and $g^2 \in \g_B$. By
Proposition~\ref{p:mcs}, $g^2$ is in two distinct maximal cyclic
subgroups of $\g$, and is therefore the identity by
Corollary~\ref{c:decomp}.

\begin{claim}\label{c:idemp}
  Any symmetry that reverses some pole is of order~$2$.
\end{claim}

\noindent
Any two orbits are either equal or disjoint, in particular, either $H(-B) = H(B)$ or $H(-B) \cap H(B) = \emptyset$.

\begin{claim}\label{c:even2}
  If $-B \in H(B)$ then $-B' \in H(B)$ for all $B' \in H(B)$ and
  $|H(B)|$ is even.
\end{claim}

\noindent
We can in fact consider the action $\act$ of $H$ on the set $\PP^\pm
\eqdef \{\{-B,B\} \st B \in \PP\}$ of pairs of antipodal
hemisets. When $-B \in H(B)$, the orbit of $\{-B,B\}$ under $\act$ has
$|H(B)|/2$ elements. The orbit-stabilizer theorem therefore implies:

\begin{claim}\label{c:even3}
  If $-B \in H(B)$ then there are exactly $2|H|/|H(B)|$ symmetries
  $g\in H$ that fix or reverse $B$.
\end{claim}

\subsubsection{Finite case: $[6,8,12] \Rightarrow S_4$}

Consider the next case, when $\g$ has orbit type $[6,8,12]$ and size
$24$. Let $P$ be some projective point set with symmetry group $\g$
and let $O$ denote the orbit of size $8$ in the action of~$\g$ on the
pole hemisets of $P$. There is a single orbit of size~$8$, so by
Claim~\ref{c:even}, $O$ writes $O = \{B_1,-B_1, B_2, -B_2, \ldots,
-B_4\}$. We let $O^\pm \eqdef \{\{B_i,-B_i\} \st 1 \le i \le 4\}$
and argue that $\g$ acts faithfully on $O^\pm$.

\bigskip

Assume that $\g$ acts unfaithfully on $O^\pm$, \ie that some
$g_0 \in \g$ fixes or reverses every $B_i$. Let us make the following
observations:
\begin{EnumAlph}
\item $g_0$ must reverse all $B_i$. Indeed, Proposition~\ref{p:poles}
  ensures that $g_0$ cannot fix all $B_i$ (it can fix at most one), so it must reverse some and
  is therefore of order~$2$. Then, we cannot have $g_0 \in \g_{B_i}
  \simeq \ZZ_3$ for order reason.
\item 
\label{i:ThreeReversers} Each $B_i$ is reversed by three symmetries. Indeed, $\g_{B_i} \simeq \ZZ_3$ and each $B_i$
  is fixed or reversed by six symmetries by Claim~\ref{c:even3}. 
\item $\g$ has $9$ elements of order~$2$, $3$ of which are in maximal
  cyclic subgroups of order~$4$, as revealed by Table~\ref{tb:mcs}.
\end{EnumAlph}
We claim that there exists a symmetry in $\g \setminus \{g_0\}$ and $i
\neq j$ such that $g$ reverses $B_i$ and~$B_j$. This follows from the
pigeonhole principle if no $B_i$ is reversed by an element of a
maximal cyclic subgroup of order~$4$ (if an element reverses $B_i$, it is of order $2$ by Claim~\ref{c:idemp}; there are $6$ such elements not in a maximal cyclic subgroup of order $4$; if each of them reverses at most one $B_i$, then, together with $g_0$, we get at most $6+4=10$ reversals; but by \ref{i:ThreeReversers} above, we need $4\times 3 =12$ such reversals). If say $B_1$ is reversed by
$g^2$ with $g \in \g$, then $g$ (of order $4$) neither fixes (which would require order $3$) nor reverses $B_1$ (which would require order $2$), so
w.l.o.g.\ we have $g(B_1) = B_2$. Then, $-B_1 = g^2(B_1) = g(B_2)$, and thus
$g^2(B_2) = -g(B_1)=-B_2$; the symmetry $g^2$ thus reverses $B_1$ and also
$B_2$.

\bigskip

We can now obtain our contradiction: the symmetry $g_0 \circ g$ fixes
both $B_1$ and $B_2$, but is not the identity as $g_0^2 = \id$ and $g
\neq g_0$. Thus, $g_0$ cannot exist and $\g$ acts faithfully on
$O^\pm$. It follows that $\g \le S_4$ and, since $|\g|=|S_4|$, $\g
\simeq S_4$.

\subsubsection{Finite case: $[12,20,30] \Rightarrow A_5$}

Consider the next case, when $\g$ has orbit type $[12,20,30]$ and size
$60$. Let $P$ be some projective point set with symmetry group $\g$
and let $O$ denote the orbit of size $30$ in the action of~$\g$ on the
pole hemisets of $P$. There is a single orbit of size~$30$, so by
Claim~\ref{c:even}, we have $O = \{B_1,-B_1, B_2, -B_2, \ldots,
-B_{15}\}$. Also, each $B \in O$ has a stabilizer of size
$2$. Let~$g_i$ denote the common generator of the stabilizers of $B_i$
and $-B_i$. Proposition~\ref{p:poles} ensures that $g_i \neq g_j$
whenever $i \neq j$, and by Table~\ref{tb:mcs}, $\g$ has $15$ elements of
order $2$. They are thus all accounted for.

\bigskip

We will use the subgroups $D_2 \le \g$, so let us first clarify how
they act on $P$.

\begin{lemma}\label{l:actD2}
  Let $P$ be a projective point set with symmetry group $\g$. Let $H
  \le \g$ with $H \simeq D_2$. Let $\PP_H$ be the set of poles of the
  elements of $H$. The action of $H$ on $\PP_H$ has three orbits, each
  consisting of two antipodal hemisets.
\end{lemma}
\begin{MyProof}
  We have $|H|=4$, with all elements, except for $\id$, of order $2$. There are six poles (three antipodal pairs), grouped in three orbits of size two. Suppose, for some $B \in \PP_H$, $H(B) = \{B,B_1\}$ with $B_1 \neq -B$. Let $\id \neq g_0 \in H$ and  $\id \neq g_1 \in H$ be such that  $g_0 (B) = B$ and $g_1(B_1)=B_1$; both $g_0$ and $g_1$ are of order 2 and $g_0 \neq g_1$.  We must have $g_1(B)=B_1$, since $g_1(B) \in H(B)$ and the stabilizer $H_{B}$ is of order 2 and has no elements other than  $g_0$ and $\id$. On the one hand, this shows $g_1(g_1(B)) = g_1(B_1) = B_1$. On the other hand, $g_1^2 = \id$ and therefore   $g_1(g_1(B)) = B$; contradiction. Therefore, $H(B)$ has to be $\{B,-B\}$ as announced.
\end{MyProof}

\bigskip

Now, let $H_i$ denote the subgroup of $\g$ that fixes or reverses
$B_i$. We have $|H_i|=4$ by Claim~\ref{c:even3}. Since $H_{B_i} =
\{\id,g_i\}$, every element in $H_i \setminus \{\id,g_i\}$ reverses
$B_i$, and must be of order~$2$ by Claim~\ref{c:idemp}. Thus, $H_i
\simeq D_2$.

\begin{claim}\label{c:symmetry}
  If $g_j(B_i) = -B_i$, then $g_i(B_j) = -B_j$.
\end{claim}
\begin{MyProof}
  Assume that $g_j(B_i) = -B_i$, so that $g_j \in H_i$. By
  Lemma~\ref{l:actD2}, the action of $H_i$ on the poles of its
  elements has $\{B_j,-B_j\}$ as an orbit. Thus, $g_i(B_j)$ must be
  $B_j$ or $-B_j$, and it cannot be the former since the only poles of
  $g_i$ are $\pm B_i$.
\end{MyProof}

\bigskip
\noindent
It follows that if $g_j \in H_i$, then $g_i \in H_j$. In other words,
if $H_i = \{\id, g_i,g_j,g_k\}$, then $H_j = H_i = H_k$ and each of
the $15$ elements of $\g$ of order~$2$ belongs to exactly one
subgroup~$H_i$. The set $X \eqdef \{H_i \st 1 \le i \le 15\}$ is
therefore of size $5$.

\bigskip

Now, for any $f, g \in \g$ we write $f \act g \eqdef f \circ g \circ
f^{-1}$. Observe that for every $H \in X$ and $f \in \g$, the set $f \act H \eqdef \{f
\act g \st g \in H\}$ is also an element of $X$. Indeed, $f \act g$
has same order as $g$, and $(f \act g) \circ (f \act g') = f \act (g
\circ g')$. So $\g$ acts on $X$ by $\act$.

\begin{claim}\label{c:gigj}
 $f \act g_i = g_j$ if and only if $f(B_i) \in \{B_j,-B_j\}$.
\end{claim}
\begin{MyProof}
  On the one hand, if $f \act g_i = g_j$ then $f \circ g_i = g_j \circ f$,
  so that $f(B_i) = g_j \pth{f(B_i)}$, forcing $f(B_i) \in
  \{-B_j,B_j\}$ since $g_j$ fixes only two poles
  (Proposition~\ref{p:poles}). On the other hand, if $f(B_i) =
  \epsilon B_j$ with $\epsilon \in \{+,-\}$, then $f \act g_i
  (\epsilon B_j) = f \circ g_i \circ f^{-1}(\epsilon B_j) = f \circ
  g_i(B_i) = f(B_i) = \epsilon B_j$, revealing that $f \act g_i$ is
  the symmetry of order~$2$ that fixes $\epsilon B_j$, that is $g_j$.
\end{MyProof}

\bigskip
\noindent
For any $i,j$ there exists $f \in \g$ such that $f(B_i) = B_j$, so $f
\act H_i = H_j$. Claim~\ref{c:gigj} therefore implies that the action
$\act$ of $\g$ on $X$ is transitive.

\bigskip

Let us argue that $\g$ acts faithfully on $X$. Let $H \in X$ and let
us write $H = \{\id, g_i,g_j,g_k\}$ and introduce $O_H \eqdef
\{B_i,-B_i,B_j,-B_j,B_k,-B_k\}$. Claim~\ref{c:gigj} implies:

\begin{claim}\label{c:actOH}
  $f \act H = H$ if and only if $f(O_H) = O_H$.
\end{claim}

\noindent
Thus, the action of $\langle f \rangle$ partitions $O_H$ into classes
of size~$1$, $2$, $3$ or~$6$. These sizes must divide the order of
$f$, which is $2$, $3$ or~$5$ by Table~\ref{tb:mcs}.

\begin{claim}\label{c:order25}
  If $f$ has order~$5$ then $f \act H \neq H$. If $f$ has order~$2$
  then $f \act H = H$ if and only if $f \in H$.
\end{claim}
\begin{MyProof}
  If $f$ has order~$5$ and $f \act H =H$, then $\langle f \rangle$
  must partition $O_H$ in orbits of size~$1$, forcing $f \in H$ to be
  of order at most~$2$, a contradiction.  If $f$ has order~$2$ and $f
  \act H =H$, then the action of $\langle f \rangle$ partitions $O_H$
  in singletons and pairs. There must exist $a \in \{i,j,k\}$ such
  that $f(B_a) \in \{B_a,-B_a\}$, implying that $f \in H_a = H$. The
  reverse direction is immediate.
\end{MyProof}

\bigskip

We already have that for every element $f \in \g$ of order~$2$ or~$5$,
there exists $H \in X$ such that $f \act H \neq H$. It remains to
handle elements of order~$3$. Let $S_H$ denote the stabilizer of $H$
for $\act$. Since $\act$ is transitive, $|S_H|=60/5 = 12$ and
Claim~\ref{c:order25} implies that $S_H$ has $12-4 = 8$ elements of
order~$3$. Let $\alpha$ be the number of pairs $(H,f)$ where $H \in
X$, $f \in S_H$, and $f$ is of order~$3$; we thus have $\alpha =
5\times 8 = 40$.

Now, let $O_X \eqdef \{O_H \st H \in X\}$. Like $\g$ acts on $X$,
$\g$ must act on $O_X$. For every $f$ of order~$3$, the action of
$\langle f \rangle$ on $O_X$ creates orbits of size~$1$ or~$3$. Thus,
each $f$ of order~$3$ fixes globally either two or five elements of
$O_X$. There are $20$ elements of order~$3$ in $\g$ by
Table~\ref{tb:mcs}, so $\alpha = 40$ implies that each element of
order~$3$ fixes \emph{exactly} $2$ elements of $O_X$. It follows that
for every element $f \in \g$ of order~$3$ there also exists $H \in X$
such that $f \act H \neq H$.

\bigskip

Altogether, $\g$ acts faithfully on $X$, and is therefore a subgroup
of $S_5$. It follows that $\g \simeq A_5$, the only subgroup of $S_5$
of size~$60$.

\subsection{More on orbits}

To analyze the symmetry group of a given projective point set (as in Section~\ref{s:Gallery} below), it is
convenient to have a better grasp on the possible orbits of poles. The
next lemma clarifies the conditions under which a pole $B$ may have an
orientation reversing symmetry, that is, $-B \in \g(B)$ or,
equivalently, $\g(B) = \g(-B)$.

\begin{lemma}
\label{l:NonSymmAntipodPoles}
  Let $P$ be a projective set in general position with nontrivial
  symmetry group~$\g$, and $B$ a pole of $P$. We have $\g(B) \neq
  \g(-B)$ if and only if
  \begin{EnumRom}
  \item $\g$ has orbit type $[1,1]$, or 
  \item $\g$ has orbit type $[2,N/2,N/2]$, $N/2$ is odd, and $|\g(B)|=N/2$, or
  \item $\g$ has orbit type $[4,4,6]$ and $|\g(B)| =4$.
  \end{EnumRom}
\end{lemma}
\begin{MyProof}
  Let us go through the possible orbit types of $\g$. An
    important point is that, by Corollary~\ref{c:mcs}, $\g$ has at
    most one orbit type. Hence, the orbit type of $\g$ describes the
    orbits of the poles of $P$ under the action of $\g$. Also,
  $|\g(B)| = |\g(-B)|$ by Claim~\ref{c:even}, so $\g(B) = \g(-B)$
  holds for any pole $B$ in an orbit that have a unique
  size. This takes care of all poles for orbit types $[6,8,12]$
  and $[12,20,30]$, and of the poles in the orbit of size $2$ for
  $[2,N/2,N/2]$ with $N/2>2$, and for the the poles in the orbit of
  size $6$ for $[4,4,6]$. We are left only with the following cases to
  be clarified.

  \bigskip
  
  If $\g$ has orbit type $[1,1]$, then the action of $\g$ on the poles
  of $P$ has two orbits, both of size~$1$. It follows that $\g(B) \neq
  \g(-B)$ for every pole $B$ of $P$.

  \bigskip

  If $\g$ has orbit type $[2,2,2]$ (that is, $[2,N/2,N/2]$ with
    $N=4$), then $|\g|=4$ with all elements other than $\id$ of
  order~$2$. Hence, $\g \simeq D_2$ and Lemma~\ref{l:actD2}
    implies that every orbit is of the form $\{B,-B\}$. It follows
    that $\g(B)=\g(-B)$ for any pole of $P$.

  \bigskip

  Assume that $\g$ has orbit type $[2,N/2,N/2]$ with $N/2>2$ and
    $|\g(B)|= N/2$. If $N/2$ is odd, then $\g(B) \neq \g(-B)$ by
    Claim~\ref{c:even2}. So assume $N/2$ is even and let $g_1$ be the
  unique element of order~$2$ in the cyclic subgroup of $\g$ of order
  $N/2$ (\cf Corollary~\ref{c:mcs}).  We claim that $g_1(B) =
  -B$. In order to verify this, note that $\g_B \simeq \ZZ_2$ since
  $|\g_{B}| = |\g|/|\g(B)| = N/(N/2) = 2$, so let us write $\g_B =
  \{\id,g_0\}$. Hence, $g_0$ is of order~$2$, and $g_0 \neq g_1$
  because they belong to different maximal cyclic subgroups of $\g$
  (by Proposition~\ref{p:mcs}). Now $g_2 \eqdef g_0 \circ g_1$ has to
  be some element not in the maximal cyclic subgroup of order $N/2$ of
  $\g$, hence $g_2$ is of order 2 as well. From
  \[ g_1 \circ g_0 = {g_2}^{-1}=g_2, \quad g_0 \circ g_2 = g_1, \quad g_2 \circ g_0 =  {g_1}^{-1}=g_1, \quad g_2 \circ g_1 = g_0, \quad g_1 \circ g_2 = {g_0}^{-1}=g_0,\]
  we get that $H \eqdef \{\id,g_0,g_1,g_2\}$ is a subgroup of $\g$ of
  order~$4$, each of which element has order~$2$. It follows that $H
  \simeq D_2$ and Lemma~\ref{l:actD2} ensures that $H(B) = \{B,-B\}$
  and $g_1(B) = -B$. In this case (orbit type $[2,N/2,N/2]$ with
  $N/2>2$ even and $|\g(B)| = N/2$), we therefore have $\g(B) =
  \g(-B)$.
  
  \bigskip
  
  The last case is when $\g$ has orbit type $[4,4,6]$ and
  $|\g(B)|=4$. In preparation of the argument, let us first have a
  look at a pole $A$ with $|\g(A)|=6$. Let $H$ be the subgroup
    of~$\g$ consisting of symmetries that map $A$ to $A$ or $-A$. We
    have $|H|=4$ by Claim~\ref{c:even3}. Any symmetry that maps $A$ to
    $-A$ has order~$2$ by Claim~\ref{c:idemp}. Since
    $|\g_{A}|=12/6=2$, there is exactly one nontrivial symmetry that
    fixes $A$, and it also has order~$2$. There are exactly $3$
    elements of order~$2$ in $\g$ (\cf Table~\ref{tb:mcs}), so
    together with $\id$ they form the group $H$. 
  
{We return to orbit $\g(B)$ of size $4$ with the goal of showing that $-B \not\in \g(B)$.} If $-B \in \g(B)$, then by Claim~\ref{c:even3} there
  is a group $H'$ of $6$ symmetries in $\g$ that fix or reverse $B$. The three symmetries in $H'$ reversing $B$ are of order~$2$ by Claim~\ref{c:idemp}; again, they
  are exactly the elements of order~$2$ of $\g$. {It follows that $H \le H'$, a contradiction, since $H$ is of order $4$, $H'$ is of order $6$, and $4$ does not divide $6$.}
\end{MyProof}

\subsection{Adding reflections}

It is natural to ask what happens if we include orientation reversing
permutations (see Section~\ref{s:Reflections}) in symmetries. Given a
projective set $P$, let $\g$ be the set of orientation preserving
symmetries, and let $\gr$ be the set of orientation preserving or
reversing symmetries. Clearly, $\g \le \gr$ and $\g \neq \gr$, since
the permutation $g^{\sf inv}\,: p \mapsto -p$ is an orientation
reversing permutation (hence not in $\g$, provided $|P| \ge
6$). Moreover, if $g$ and $g'$ are orientation reversing permutations,
then $g \circ g'$ is an orientation preserving permutation.
  Any symmetry preserves antipodality by Lemma~\ref{l:EmbracingPreservation}\ref{i:iEmbracingPreservation}, so
  $g^{\sf inv}$ commutes with every $g \in \g$ and we have $\gr =
  \{\id, g^{\sf inv}\} \times \g \simeq \ZZ_2 \times \g$.

For example, if $|\gr|=24$, then this group is isomorphic to $A_4 \times \ZZ_2$, $\ZZ_{12} \times \ZZ_2$, or $D_6 \times \ZZ_2$ (not $S_4$ which is not isomorphic to any of the three groups mentioned).

\subsection{Symmetries on the Sphere}
We have characterized the symmetries of affine and projective sets in general position on the sphere $\SS^2$. What about general finite subsets $Q$ in general position of $\SS^2$? This can be easily derived as follows. Given such a set $Q$, let $P \eqdef Q \cup -Q$ be the completion of $Q$ to a projective set, which is -- as a projective set -- in general position, with $\g$ the group of symmetries of $P$.

Similar to the situation for affine sets, we can let $\g$ act on the
\Emph{semisets} of $P$, \ie the subsets of $P$ which contain exactly
one point from every antipodal pair in $P$ (the fact that this is
indeed an action follows from $g(-p) = -g(p)$, see
Lemma~\ref{l:EmbracingPreservation}\ref{i:iEmbracingPreservation}). Consider the stabilizer $\g_Q$ of $Q$. Similar
to Lemma~\ref{l:actaff}, we can derive that $\g_Q$ is isomorphic to
the group of orientation preserving symmetries of $Q$, and thus this
group is a subgroup of $\g$. This shows that $\g_Q$ is among the
groups we identified for the projective sets, as they are
  closed under taking subgroups (being the finite subgroups of
  $SO(3)$).

\subsection{Gallery}
\label{s:Gallery}

\subsubsection{Small Sets}

Table~\ref{ta:Gallery} gives a summary of all projective order types with $2n$ points, $3 \le n \le 6$, their symmetry groups and their induced affine order types. We see that for each $n\le 5$ there is exactly one projective order type. For $n=6$, we have four projective order types, the completions of convex position and the three order types with 5 extreme points. These partition the twenty 6-point affine order types (note that this is 20, since we consider symmetries without reflection; with reflection it is 16).

\begin{table}[htb]
\begin{center}
\begin{tabular}{c|c|c|c|c|c|c}
icon & $\frac{|\pi|}{2}$ & $\OTaff{\pi}$ & {\small $|\OTaff{\pi}|$} & $\g$ & $|\g|$  &  $2 \binom n 2 + 2$\\
\hline
\\[-1em]
$\placefig{7}{1.2em}$ & $3$ & ${\placefig{14}{1.7em}\,}^3_8$ & $1$ & $S_4$ & $24$ & {\footnotesize $8$} \\[0.3ex]
$\placefig{8}{1.1em}$ & $4$ & ${\placefig{15}{1.6em}\,}^4_6$, ${\placefig{16}{1.7em}\,}^3_8$ & $2$ & $S_4$ & $24$ & {\footnotesize $14=6+8$} \\[0.3ex]
$\placefig{9}{1.4em}$ & $5$ & ${\placefig{17}{1.9em}}^5_{2}$, ${\placefig{18}{1.6em}\,}^1_{10}$, ${\placefig{19}{1.8em}\,}^1_{10}$ & $3$ & $D_5$ & $10$ & {\footnotesize $22 =2 + 2 \times 10$}\\[0.3ex]
$\placefig{10}{1.45em}$ & $6$ & ${\placefig{20}{1.9em}}^6_{2}$, ${\placefig{21}{1.9em}\,}^1_{12}$, ${\placefig{22}{1.6em}\,}^1_{12}$, ${\placefig{23}{1.6em}\,}^2_{6}$ & $4$ & $D_6$ & $12$ & {\footnotesize $32 = 2 + 2 \times 12 + 6$} \\[0.3ex]
$\placefig{11}{1.4em}$ & $6$ & ${\placefig{24}{1.9em}}^5_{12}$, ${\placefig{25}{1.8em}\,}^3_{20}$ & $2$ & $A_5$ & $60$ & {\footnotesize $32= 12 + 20$} \\[0.3ex]
$\placefig{12}{1.4em}$ & $6$ &  ${\placefig{26}{1.9em}}^1_{6}$ , ${\placefig{27}{1.9em}\,}^3_{2}$,   ${\placefig{28}{1.6em}\,}^1_{6}$, ${\placefig{29}{1.6em}\,}^1_{6}$,  \ldots & $6$ & $D_3$ & $6$ & {\footnotesize $32=2+ 5 \times 6$}\\[0.3ex]
$\placefig{13}{1.4em}$ & $6$ &  ${\placefig{30}{1.9em}}^1_{4}$, ${\placefig{31}{1.6em}\,}^1_{4}$, ${\placefig{32}{1.6em}\,}^1_{4}$, \ldots & $8$ & $\ZZ_4$ & $4$ & {\footnotesize $32 = 8 \times 4$}\\[0.3ex]
\hline
\end{tabular}
\end{center}
\caption{The affine order types and symmetries of projective order types $\pi$ with $2n$ points, $n=3,4,5,6$. For an affine order type $\omega$, we write $\omega^\gamma_\mu$, with $\gamma$ the size of its symmetry group, and $\mu$ the size of its orbit among the affine hemisets. The last column indicates, how the $2 \binom n 2 + 2$ affine hemisets distribute among the affine order types induced by the projective set.}
\label{ta:Gallery}
\end{table}

\begin{figure}[htb]
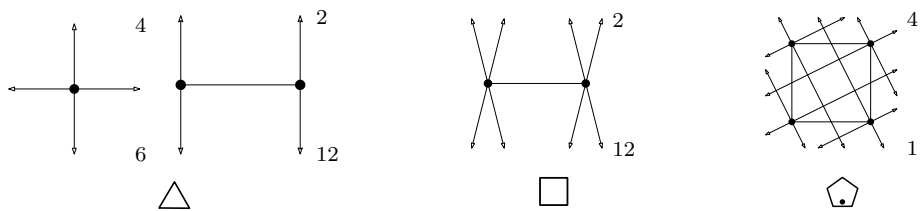

\centerline{
\begin{minipage}[c]{0.3\textwidth}
\begin{center}
${\placefig{33}{4.5em}\!}^4_6$
\hspace{0.02\textwidth}
${\placefig{34}{4.5em}\,\,}^2_{12}$
\\[1.5ex]
{\placefig{7}{1.1em}}
\end{center}
\end{minipage}
\hspace{3em}
\begin{minipage}[c]{0.15\textwidth}
\begin{center}
${\placefig{35}{4.5em}\,\,}^2_{12}$
\\[1.5ex]
{\placefig{8}{1em}}
\end{center}
\end{minipage}
\hspace{3em}
\begin{minipage}[c]{0.15\textwidth}
\begin{center}
${\placefig{36}{4.5em}\,\,}^4_1$\\[1.5ex]
{\placefig{13}{1.1em}}
\end{center}
\end{minipage}
}
\caption{Non-affine poles of projective sets. Rays indicate the connections to the antipodal pairs on the boundary of the defining closed hemispheres (points in infinity).}
\label{f:NonAffine}
\end{figure}

Let us recall that poles are hemisets, not necessarily affine hemisets. This explains, \eg that the projective set \placefig{7}{0.9em} with $\g_{\placefig{7}{0.63em}} = S_4$ exhibits in the table only 8 affine poles, all in the same orbit; the missing poles are hemisets with one or two antipodal pairs, with symmetry of size 4 or 2, resp., and thus orbits of size 6 and 12, resp., see \FigRef{f:NonAffine}(left). Similarly, the projective set \placefig{8}{0.8em} with $\g_{\placefig{8}{0.56em}} = S_4$ has $12$ non-affine poles that form a single orbit under  $\g_{\placefig{8}{0.56em}}$, see \FigRef{f:NonAffine}(center).

\begin{figure}[htb]
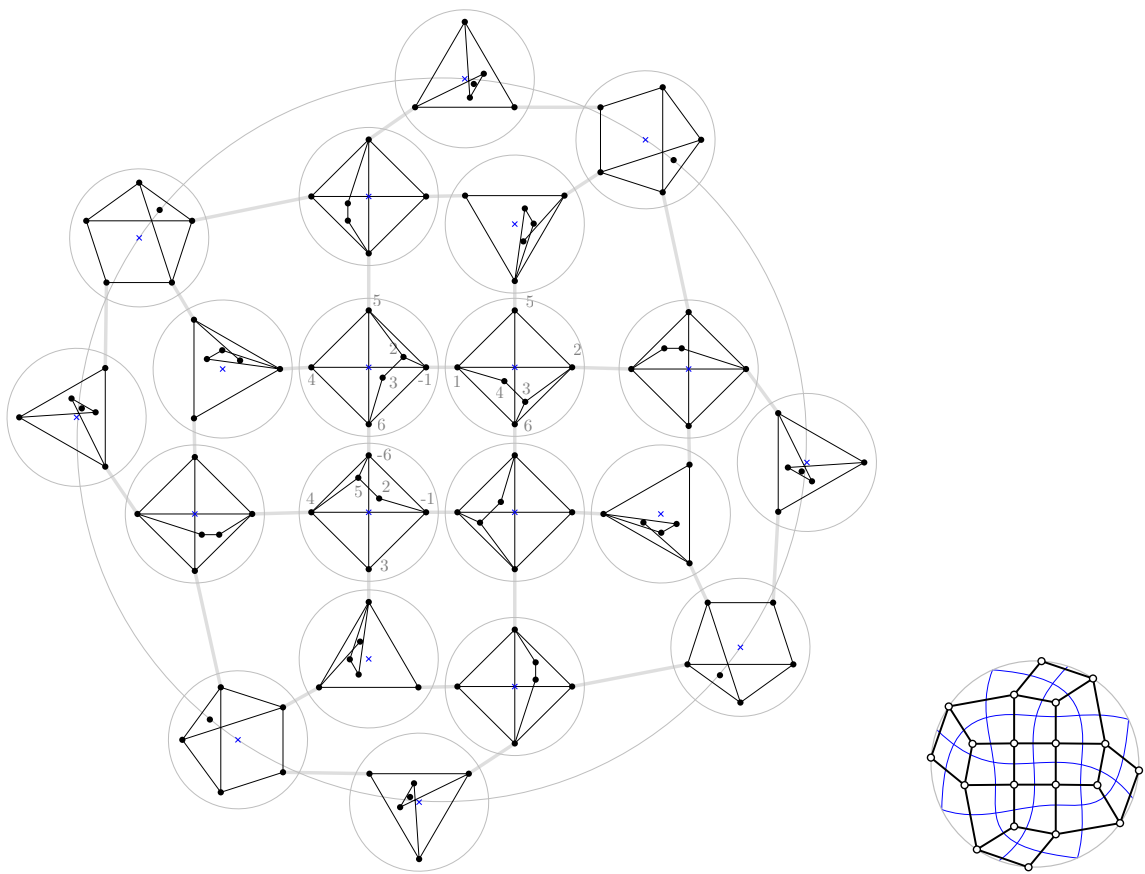

\centerline{
\placefig{42}{0.72\textwidth} 
\hspace{1em} 
\placefig{43}{0.18\textwidth}
}
\caption{A ``hemisphere'' of affine hemisets of the projective set $\mbox{\protect\placefig{13}{0.9em}}$. Each order type in $\OTaff{\mbox{\protect\placefig{13}{0.65em}}}$ occurs with multiplicity four as affine hemiset. We see five order types, with three order types missing, the reflections of the three inner order types. Pairs of affine hemisets whose dual cells share an edge, or equivalently, which can be obtained by projectively swapping a point to the other side are connected by an edge, hinged at the points swapped.}
\label{f:n6h5Brly}
\end{figure}

{The projective set \placefig{13}{0.9em} is the only one up to $n=6$ which has no affine hemiset with nontrivial symmetry (see \FigRef{f:n6h5Brly}), but there is still a non-affine hemiset (see \FigRef{f:NonAffine}(right)) with symmetry group $\ZZ_4$. }

\subsubsection{Small groups, cyclic groups}

We see that all symmetry groups have size at least 4 in
Table~\ref{ta:Gallery}, in particular, we have not yet encountered a
projective set with trivial symmetry group. So let us describe
examples with smaller symmetry groups. For that we need the following lemma:

\begin{lemma}\label{l:neighbors-dont-share}
    For any two affine hemisets $A$ and $A'
  \not\in \{A,-A\}$ of a projective set $P$ in general position with
  $|P|=2n$, we have $h(A) + h(A') \le n+4$.
\end{lemma}
\begin{MyProof}
We recall here the duality from Section~\ref{s:DualityArr} and denote by $p^*$ the great circle dual to point $p$ on $\SS^2$. Recall that $p^* = (-p)^*$, \ie $P^* \eqdef \{ p^* \st p \in P\}$ is an arrangement of $n$ great circles in general position. Every affine hemiset $A$ of $P$ corresponds to a cell in this arrangement (see Lemma~\ref{l:duality-basic}), which we denote by $A^*$. We have that cell $A^*$ is incident to $h(A)$ of the great circles in $P^*$.

For affine hemisets $A$ and $A'$ the cells $A^*$ and $A'^*$ share at most four out of the $n$ great circles in $P^*$ to which they are both incident, unless $A' \in \{A, -A\}$. This is easy to prove directly and follows from a basic fact on line arrangements which is called Gunderson's Theorem \grey{in} \cite[Theorem~V$\!$I$\!$I]{carver41}. Going back to the primal, this yields $h(A) + h(A') \le n+4$.
\end{MyProof}

\begin{figure}[htb]
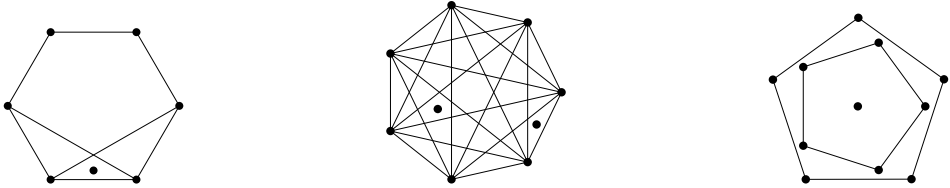

\centerline{
\placefig{37}{0.15\textwidth}
\hspace{0.15\textwidth} 
\placefig{38}{0.15\textwidth}
\hspace{0.15\textwidth}
\placefig{39}{0.15\textwidth}
}
\caption{Sets with projective completion with symmetry $\ZZ_2$ (left), $\ZZ_1$ (center), and $\ZZ_5$ (right).}
\label{f:SmallSymmetry}
\end{figure}

It follows that if $h(A) > n/2 + 2$ then no other affine hemiset
except for $-A$ has the same number of extreme points as $A$, and
therefore $\g(A) \subseteq \{A,-A\}$. If, in addition, $A$ has
symmetry group $\f$ and no orientation reversing symmetry, then $\g(A)
= \{A\}$ and $|\g| = |\g(A)|\cdot |\g_A| = |\g(A)|\cdot |\f| = |\f|$,
that is, $\g \simeq \f$. We summarize:

\begin{claim} Let $A$ be an affine subset of $\SS^2$, with $h(A) > |A|/2 +2$ and symmetry $\f$. If $A$ has no orientation reversing symmetry, then the completion of $A$ is a projective set with symmetry group isomorphic to $\ZZ_{|\f|}$. If $A$ has an orientation reversing symmetry, then the completion of $A$ is a projective set with symmetry group isomorphic to $D_{|\f|}$.
\end{claim}
This provides us immediately with many examples of projective sets with symmetry groups of size 1 or 2.\footnote{We also see why this fails for $n=6$: we have $n/2+2 = 5$ and more than 5 extreme points force convex position.} For example, suppose a 7-point set has six extreme points (note $6 > 7/2 + 2$), and the inner point placed barely inside an edge of the convex hull, see \FigRef{f:SmallSymmetry}(left). Then its symmetry group is trivial, but it exhibits an orientation reversing symmetry. Hence, the projective completion has symmetry $D_1 \simeq \ZZ_2$. If we have a 9-point set with seven extreme points (note $7 > 9/2 + 2$), then the inner two points can be easily placed so that we have no orientation reversing symmetry,  see \FigRef{f:SmallSymmetry}(center). The projective completion of such a set has trivial symmetry.

Here is a claim that provides projective sets with symmetry $\ZZ_k$, $k$ odd, see \FigRef{f:SmallSymmetry}(right).
\begin{claim} 
\label{c:OddLayers}
Let $P$ be a projective set in general position, with an affine pole $A$ with symmetry $\f \simeq \ZZ_k$, $k>1$. If $k$ is odd and $A$ has at least three layers of odd size, then $A$ has no orientation reversing symmetry and, for the symmetry group $\g$ of $P$, $\g \simeq \ZZ_k$ or $\g \simeq A_4$  (the latter can occur only for $k=3$).
\end{claim}
\begin{MyProof}  Every orientation reversing permutation of $A$ has to fix exactly one element in each odd layer, \ie it has to fix at least three elements. Obvioulsy, a permutation fixing three elements cannot be orientation reversing. The fact that $A$ has no orientation reversing symmetry implies $-A \not\in \g(A)$. By Lemma~\ref{l:NonSymmAntipodPoles}, this cannot happen if $\g$ has orbit type $[6,8,12]$ or $[12,20,30]$. Also, if $\g \simeq D_k$, then $-A \in \g(A)$, so this must be ruled out. This leaves $\ZZ_k$ or $A_4$, and the latter only for $k=3$.
\end{MyProof}

\subsubsection{Tetrahedral group}

Let $\Delta = \{p_1,p_2,p_3,p_4\}$ be the vertices of a regular
  tetrahedron inscribed in $\SS^2$, and let $\giso$ denote the set of
  rotations of $\SS^2$ that map $\Delta$ to itself; true to its name, $\giso
  \simeq A_4$ is the tetrahedral group. For any point $q \in \SS^2$ not fixed by
  any element of $\giso$, we have $|\giso(q)| = |\giso| = 12$. We fix
  a generic point $p_1'$ close to $p_1$, and close to the geodesic arc
  $p_1p_2$, but not on this arc. Let $g_1$ denote the element of
  order~$3$ in $\giso$ that fixes $p_1$ and note that the orbit of
  $p_1'$ under $\langle g_1 \rangle$ consists of three points close to
  $p_1$. Let $S_1 \eqdef \{p_1\} \cup \langle g_1 \rangle(p_1')$.

Now, let $P \eqdef \Delta \cup -\Delta \cup \giso(p_1') \cup
  -\giso(p_1')$ and let $\g$ be the symmetry group of $P$. Observe
  that $P$ is a projective set in general position with $32$
  points.

\begin{figure}[htb]
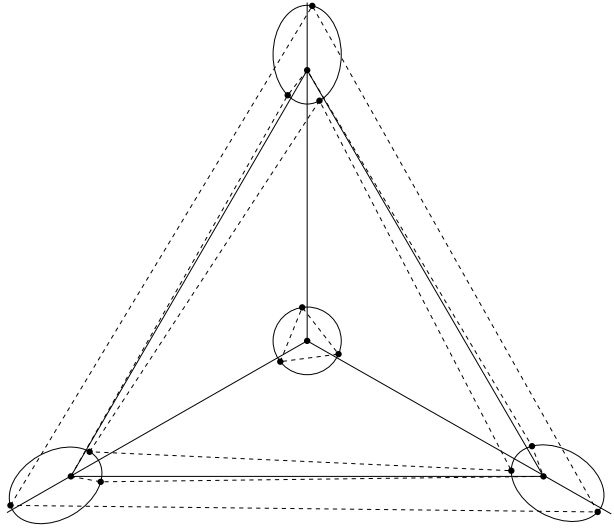

\centerline{
\placefig{40}{0.5\textwidth}
}
\caption{An 16-point set with projective completion with symmetry $A_4$. The five layers of size 1,3,3,6, and 3, resp., are indicated by dashed polygons.}
\label{f:Tetrahedral}
\end{figure}

\begin{claim}
  If $p_0'$ is chosen sufficiently close to both $p_0$ and the arc
  $p_0p_1$, then $\g \simeq A_4$.
\end{claim}
\begin{MyProof}
  We already know that $A_4 \simeq \giso \le \g$, so $\g$ cannot
    be cyclic nor dihedral. The only candidates are therefore $A_4$,
    $S_4$ and $A_5$. Observe that the orbit type of $S_4$ and $A_5$,
    and Claim~\ref{c:even}, force every hemiset $A$ to lie in the same
    orbit as $-A$. To prove the claim, it thus suffices to exhibit an
    affine hemiset of $P$ with no orientation reversing symmetry.

    Note that $P$ consists of $8$ groups of $4$ close-by points,
    each group being isometric to either $S_1$ of $-S_1$. We write
    $S_i$ for the group containing $p_i$, and $S_{-i}$ for the group
    containing $-p_i$. Let $H$ be the open hemisphere centered at
    $p_1$, and let $A \eqdef P \cap H$. The set $A$ is an affine
    hemiset of $P$ and $A = S_1 \cup S_{-2} \cup S_{-3} \cup S_{-4}$,
    see \FigRef{f:Tetrahedral}. The set $A$ has four convex layer
    of odd size and therefore, by Claim~\ref{c:OddLayers}, no
    orientation reversing symmetry. The statement follows.
\end{MyProof}

\section{Generalizations: higher dimension and abstract order types}
\label{s:generalizations}

We now examine to what extent the previous analysis generalizes to
higher dimension and to related structures.

\subsection{Arbitrary dimension}

Our methods for labeled affine order types generalize to finite
subsets of $\SS^d$, the unit sphere in~$\RR^{d+1}$.

\bigskip

Let us clarify how the notions generalize (without surprise) to higher
dimensions. We call a subset of $\SS^d$ \Emph{affine} if it is
contained in an open hemisphere; a point of an affine subset is
\Emph{extreme} if it can be cut out from the rest of the set by a
great hypersphere, that is, the intersection of $\SS^d$ with a
hyperplane through the origin $\mathbf{0}$.  A subset $P$ of $\SS^d$
is \Emph{projective} if $-p \in P$ for every $p \in P$. An affine set
is \Emph{in general position} if no $d+1$ points are coplanar with
$\mathbf 0$; a projective set is \Emph{in general position} if
whenever $d+1$ points are coplanar with $\mathbf 0$, two of them are
antipodal. The \Emph{orientation}, $\sgn(p_1,p_2, \ldots, p_{d+1})$,
of a $(d+1)$-tuple $(p_1,p_2, \ldots, p_{d+1})$ of points in $\SS^d$
is the sign, $-1$, $0$, or $1$, of the determinant of the matrix
$(p_1,p_2, \ldots, p_{d+1}) \in \RR^{(d+1)\times (d+1)}$. Two affine
(projective, resp.) sets have the same \Emph{affine}
(\Emph{projective}, resp.) \Emph{order type} if there exists an
orientation preserving bijection between them. Two affine point
sequences $(p_1,p_2,\ldots, p_{n})$ and $(q_1,q_2,\ldots, q_{n})$ are
defined to be of the same \Emph{labeled affine order type} if the map
$p_i \mapsto q_i$ preserves orientations.

\bigskip

As for $d=2$, the \Emph{projective completion} of an affine set $A$ is
the projective set $A \cup -A$. A \Emph{hemiset} of a projective set
is its intersection with a closed hemisphere, and a hemiset is
\Emph{affine} if it is contained in an open hemisphere, that is, if it
does not contain any antipodal pair. We again have that a projective
set $P$ is the projective completion of an affine set $A$ if and only
if $A$ is an affine hemiset of $P$.

In the arguments for the following theorem we only outline the
differences w.r.t.\ the 2-dimensional setting.

\begin{MyTheorem}\label{t:avgl-d}
 For~$n\ge d+1$, the number of faces of {dimension} $k-1$ in the
 convex hull of a random simple labeled order type chosen uniformly
 among the simple, labeled order types of size $n$ in $\RR^d$ has
 average $2^{k}\binom{d}{k} + o(1)$; for $k=1$, this random variable
 has variance $O(1)$. In particular, the number of extreme points
 ($0$-faces of the convex hull) has average $2d+o(1)$, with constant
 variance, and the number of facets ($(d-1)$-faces) of the convex hull
 has average $2^{d}+o(1)$.
\end{MyTheorem}
\begin{MyProof}[Proof outline]
  Let $n \ge d+1$. Let $P$ be a projective set of $2n$ points. As for
  $d=2$, the projective symmetries of $P$ act on its (affine)
  hemisets, the affine hemisets of $P$ of given order type form an
  orbit in this action, and the stabilizer of an affine hemiset is
  isomorphic to its (affine) symmetry group.

  \bigskip
  
  Let $\omega$ be the order type of an affine hemiset of $P$. Again,
  the number of (affine) symmetries of $\omega$ affects both how
  frequently $\omega$ occurs among the affine hemisets of $P$, and how
  many distinct \emph{labeled} affine order types are supported by
  $\omega$. As for $d=2$, these two effects balance each other out and
  Proposition~\ref{p:uniform} generalizes: picking uniformly a random
  affine hemiset of $P$, then picking uniformly a random ordering of
  the points of that hemiset produces a random labeled affine order
  type distributed \emph{uniformly} among all those that can be
  obtained from $P$.

  \bigskip
    
  In $\SS^d$, the dual $p^*$ of a point $p$ is the great hypersphere
  cut out by the hyperplane perpendicular to the line $\mathbf{0}p$ in
  $\mathbf{0}$. Any projective set of $2n$ points, $n \ge d+1$,
  therefore has an associated dual arrangement $P^*$ of $n$ great
  hyperspheres. Lemma~\ref{l:duality-basic} readily generalizes: there
  is a bijection $\phi$ between the affine hemisets of a projective
  point set~$P$ and the cells (\ie full-dimensional faces) of the dual
  arrangement $P^*$, such that a nonempty subset $S \subseteq A$ forms
  a face (which has to be a $(k-1)$-face, $k = |S|$) in the convex
  hull of an affine hemiset $A$ if and only if the intersection of the
  $k$ great hyperspheres $\{p^* \st p \in S\}$ supports a
  $(d-k)$-face of $\phi(A)$.

  \bigskip

  Let $\ff d n k$ denote the number of faces of codimension $k$ (\ie dimension $d-k$) in
  $P^*$. Every face of codimension $k$ of $P^*$ is contained in the
  intersection of a unique subset of $k$ of the hyperspheres, in which
  it is a cell of the induced $(d-k)$-dimensional arrangement. Hence,
  \[ \ff d n k = \binom{n}{k} \ff {d-k} {n-k} {0}.\]
  An arrangement of $n$ hyperplanes in general position in $\RR^d$ has
  $\sum_{i=0}^d {n \choose i}$
  cells~\cite[Lemma\,1.2]{edelsbook}. As explained in
  Section~\ref{s:background}, $P^*$ can be decomposed into $2$ inverted copies
  of an arrangement  of $n-1$ hyperplanes in $\RR^d$, so we
  have
  \[ \ff d n 0 = 2\sum_{i=0}^d {n-1 \choose i} \quad \text{and, more generally,} \quad \ff d n k = 2 \binom{n}{k}\sum_{i=0}^{d-k} {n-k-1 \choose i}.\] 
  The number of cells of $P^*$ that contain a given
  $j$-face is $2^{d-j}$; see~\cite[Lemma\,1.1]{edelsbook} (remark
  that by projecting along the affine span of the $j$-face, this is
  the same as counting the number of cells that contain a given vertex
  in an arrangement of hyperplanes in general position in
  $\RR^{d-j}$). The average number of faces of codimension $k$ of a
  cell of $P^*$ is therefore
  \[ \frac{2^{k} \ff d n k}{\ff d n 0} = \frac{\displaystyle 2^{k} \binom{n}{k}\sum_{i=0}^{d-k} {n-k-1 \choose i}}{\displaystyle \sum_{i=0}^d {n-1 \choose i}} = 2^{k} \frac{\displaystyle\binom{n}{k} {n-k-1 \choose d-k}}{\displaystyle\binom{n-1}{d}} + o(1),\]
  that is, $2^{k}\binom{d}{k} + o(1)$. This is also the average number of $(k-1)$-faces in the convex hull of an affine hemiset, as announced.

\bigskip

  To bound the variance, we can use the general version of the zone
  theorem~\cite{edelsbrunner1993zone}. For $p \in P$, let $Z(p^*)$ denote the zone of
  $p^*$, \ie the set of cells of $P^*$ incident to
  $p^*$. For a cell $c$, let $|c|$ denote the number
  of facets (faces of codimension $1$) that are incident to $c$. Then $\sum_{c
    \in Z(p^*)} |c| = O\pth{n^{d-1}}$ and the average squared number
  of facets in a random full-dimensional cell of $P^*$ is $O(1)$.
\end{MyProof}

\bigskip

As for \emph{unlabeled} affine order types, we do not see that any of
our results in the plane generalizes. The information we extract on
orbit types depends on the fact that every projective symmetry has
exactly two poles (Proposition~\ref{p:poles}); our proof of that fact
relies on the hairy ball theorem, which only holds in even
dimension. The analysis of reflections may be another difficulty: the
transversal theorem of Hadwiger that we used was generalized to
hyperplane transversals~\cite{goodman1988hadwiger} but with the
ordering condition rephrased (interestingly, in terms of order
types). Also, our analysis of symmetries of affine sets is specific to
the planar setting.

\subsection{Abstract order types (acyclic uniform oriented matroids)}

The order type records the orientation of every triple of points, that
is, the position of each point with respect to the line through
the other two. This can also be carried out in a more general setting
where the usual (straight) lines of the affine setting are replaced by
curves forming a pseudoline arrangement. Starting with a
\Emph{topological projective plane}~\cite{goodman1994arrangements} and
distinguishing a pseudoline as being ``at infinity'', one obtains a
\Emph{topological affine plane}, in which orientations are
well-defined: through any two points there is a unique pseudoline, and
together with the pseudoline at infinity it cuts out two connected
components (just like a line in the affine plane). The equivalence
classes of finite subsets of topological affine planes modulo
orientation preserving bijections are called \Emph{abstract order
  types}. Since the affine plane is a topological affine plane, any
order type is an abstract order type. The converse is not true, and we
refer to the survey of Goodman and
Felsner~\cite{felsner2017pseudoline} for a discussion of some of the
differences. Unlike order types, abstract order types are amenable to
combinatorial methods, and are characterized by a few simple
axioms~\cite{knuth1992axioms}; they are, in fact, equivalent to
\Emph{relabeling classes of rank~$3$ 
acyclic oriented
  matroids}, a classical combinatorial
structure~\cite{bjorner1999oriented}. More generally, order types of
point sets in $\RR^d$ enjoy a similar abstract generalization, which
turns out to be equivalent to 
relabeling classes of rank~$d+1$
acyclic oriented matroids.

\bigskip

 Our approach generalizes to abstract order types as follows. We
  work again on $\SS^2$, but now equipped with a system of
  pseudocircles, each symmetric with respect to the origin
  $\mathbf{0}$. An open pseudo-hemisphere is a connected component in
  the complement of a pseudocircle, and a closed pseudo-hemisphere is
  the closure of an open one. The abstract order types are read off
  intersections of projective sets with closed pseudo-hemispheres with
  no point on the boundary, and the notions of extreme point, extreme
  edge, convex hull, \ldots carry through. The content of
  Sections~\ref{s:hemisets} and~\ref{s:labeled} generalizes readily
  (in particular, the combinatorics of the dual arrangement and the
  bound used for the zone theorem~\cite{BEPY90} holds also for
  pseudolines), and we obtain: 

\begin{MyTheorem} 
  \label{t:avgl-p}
  For~$n\ge 3$, the number of extreme points in a random simple
  labeled abstract order type chosen uniformly among the simple,
  labeled order types of size $n$ has average $\TheNumber$ and
  variance at most {$3$}.
\end{MyTheorem}

\noindent
The extension to higher dimension for labeled order types also
generalizes to the abstract setting:

\begin{MyTheorem}\label{t:avgl-dp}
 For~$n\ge d+1$, the number of faces of dimension $k-1$ in the convex
 hull of a random simple labeled abstract order type chosen uniformly
 among the simple, labeled, $d$-dimensional abstract order types of
 size $n$ has average $2^{k}\binom{d}{k} + o(1)$; for $k=1$, this
 random variable has variance $O(1)$.
\end{MyTheorem}

In the \emph{unlabeled} setting, most of the proof of
  Theorem~\ref{t:avg} goes through, with the notable exception of the
  proof of Proposition~\ref{p:OTsym} (specific to the realizable
  setting since it reformulates orientations as signs of
  polynomials). We expect that an analogue of
  Proposition~\ref{p:OTsym} holds for abstract order types and that
  Theorem~\ref{t:avg} generalizes, but settle here for a slightly
  weaker version.

\begin{MyTheorem}\label{t:avg-p}
  For~$n\ge 3$, the number of extreme points in a random simple
  abstract order type chosen uniformly among the simple abstract order
  types of size $n$ in the plane has average~$O(1)$.
\end{MyTheorem}
\begin{MyProof}[Proof outline]
  From the beginning of Section~\ref{s:poles} to
    Corollary~\ref{c:labeling}, everything generalizes readily.  The
    only nontrivial step is the use of Hadwiger's transversal
    theorem, but Basu \etal\,\cite[Theorem\,5]{basu2004hadwiger}
    provides the required generalization. In particular, in the
    abstract setting we do have that
    \begin{EnumAlph}
    \item projective symmetries have exactly two, opposite, poles,
    \item the possible orbit types are the same in the realizable and
      abstract settings,
    \item abstract order types have the same symmetry groups as the
      realizable ones (that is, Theorem~\ref{t:classAff} holds also
      for abstract order types), and
    \item every abstract order type of size $n$ corresponds to at
      least $(n-1)!$ and at most $n!$ labeled abstract order types.
    \end{EnumAlph}
    We cannot control the number of abstract order types with many
    symmetries as in the affine setting by counting sign vectors of
    polynomials. Still, the proof of Proposition~\ref{p:avg-proj} does
    not require it, and readily goes through. In other words, the
    average number of extreme points in an abstract order type of size
    $n$, chosen uniformly conditioned on a given projective
    completion, is at most $4 + 3 N/n$ where $N$ is the number of
    projective symmetries.

  Then, all of Section~\ref{s:classProj} readily extends to the
    abstract setting. This include the correspondence between orbit
    types and maximal cyclic subgroups (Proposition~\ref{p:mcs}),
    which ensures that any abstract projective order type with $2n$
    points and $N>60$ symmetries has a cyclic subgroup of size $N$ or
    $N/2$, so that $n \ge N/2$. Altogether, for every
    sufficiently large abstract projective order type, the average
    number of extreme points in the abstract order types it contains
    is at most $10$. The statement follows.
\end{MyProof}

  \noindent
As noted in the proof outline of Theorem~\ref{t:avg-p}, the
classifications of symmetry groups (Theorems~\ref{t:classAff}
and~\ref{t:classProj}) also hold in the abstract setting.

\section{Outlook: random sampling via projective order types\label{s:outlook}}

We wrap up by continuing the discussion about sampling random order
types from Section\,\ref{s:OpenProblems}, now with the extra insights
from the results of this paper and its approach.

\bigskip

Let us clarify the algorithmic problems we consider here. We take as
input an integer $n \ge 3$ and want to output an element chosen
uniformly at random in $\LOTaff{n}$, $\OTaff{n}$ or $\OTproj{n}$,
depending on the variant of the problem. The algorithm has access to a
sequence of uniform random bits. For simplicity, we represent an
element of $\LOTaff{n}$ as the orientation map $\sgn$ from the ordered
triples from $\{1,2,\ldots, n\}$ to $\{-1,1\}$, but note that more
compact representations are possible (for instance the
$\lambda$-matrices of Goodman and
Pollack~\cite[Def.\,1.3,Cor.\,1.9]{goodman1983multidimensional} or the
encoding based on hierarchical cuttings of Cardinal
\etal\,\cite{cardinal2019subquadratic}). We represent an element of
$\OTaff{n}$ or $\OTproj{n}$ as any labeled order type it contains. To
be clear, $\omega \in \OTaff{n}$ \Emph{contains} $\Lab{\omega} \in
\LOTaff{n}$ if the latter can be obtained by ordering the vertices of
the former; $\pi \in \OTproj{n}$ \Emph{contains} $\omega \in
\OTaff{n}$ if the latter is the order type of some affine hemiset of
the former; $\pi \in \OTproj{n}$ contains $\Lab{\omega} \in
\LOTaff{n}$ if there exists $\omega \in \OTaff{n}$ that is contained
in the former and contains the latter. Let us stress that given two
orientation maps, one can decide in $O(n^2)$ time whether the labeled
affine order types they represent are contained in the same affine
order type, \ie isomorphic, see Aloupis \etal\,\cite{AILOW14}.

\subsection{Polynomial-time equivalence}

Let us first argue that any of the variants of the problem reduces to
any other variant in time polynomial in $n$.

\paragraph{From projective to labeled affine.}

Assume given an algorithm $\A$ that outputs a random projective order
type $\pi$ chosen uniformly in $\OTproj{n}$. We first describe a preliminary procedure for a uniform sampling of $\LOTaff{n}$ which allows failure, \ie the procedure may decide to output a failure symbol $\perp$ instead of a labeled affine order type: For $\pi \in \OTproj{n}$ generated by $\A$, we first determine the symmetry group $\g_\pi$ of $\pi$. With probability $\frac{1}{|\g_\pi|}$, we pick an affine hemiset
of $\pi$ uniformly at random, then an ordering of its vertices uniformly at
random, and then we output this labeled affine order type $\Lab{\rho}$ (note that $\Lab{\rho}$ is uniformly chosen in $\LOTaff{\pi}$ by Proposition~\ref{p:uniform}). With probability $1-\frac{1}{|\g_\pi|}$, we output $\perp$. The necessary operations can be performed in time polynomial in $n$, in particular, computing the symmetry group can be done in $O(n^2)$ time, along the lines of
Aloupis \etal\,\cite{AILOW14}.

Since $|\g_\pi| \le \max\{60,2n\}$, the procedure succeeds in producing an order type with probability $\Omega\pth{n^{-1}}$. Hence, if we repeat the procedure until success, $O(n)$ iterations will suffice on average. It remains to ensure that the procedure generates every $\Lab{\omega}$ with the same probability. Let $\pi_\omega$ be the completion of $\omega$ (the unlabeled affine order type underlying $\Lab{\omega}$). Then the probability of the procedure to output $\Lab{\omega}$ is given by
\[
\frac{1}{|\OTproj{n}|} \cdot \frac{1}{|\g_{\pi_\omega}|} \cdot \frac{1}{|\LOTaff{\pi_\omega}|} = \frac{1}{|\OTproj{n}|} \cdot \frac{1}{\pth{2\binom{n}2+2} n!}
\]
where we use $|\LOTaff{\pi}| = \pth{2\binom{n}2+2} \frac{n!}{|\g_\pi|}$ (see end of proof of Proposition~\ref{p:uniform}).

\paragraph{From labeled affine to affine.}

Now, assume given an algorithm $\A$ that outputs a random labeled
affine order type $\Lab{\rho}$ chosen uniformly in
$\LOTaff{n}$. Simply outputting the affine order type that contains
$\Lab{\rho}$ gives us a random generator of affine order types, but it
has some bias: indeed, an affine order type $\omega$ with symmetry
group $\f_\omega$ contains exactly $\frac{n!}{|\f_\omega|}$ distinct
labeled affine order types. Since $1 \le |\f_\omega| \le n$, we can
correct this bias using rejection, by accepting the output
$\Lab{\omega}$ of algorithm $\A$ with probability
$\frac{|\f_\omega|}{n}$. Clearly, at most $n$ iterations are needed in
expectation. Computing the symmetry group of an affine order type can be done in
  $O(n^2)$ time, as shown by Aloupis \etal\,\cite{AILOW14}.

\paragraph{From affine to projective.}

Finally, assume given an algorithm $\A$ that outputs a random affine
order type $\rho$ chosen uniformly in $\OTaff{n}$. Again, we output
the projective order type containing $\rho$ (\ie the completion of $\rho$) after correcting for bias
via rejection: a projective order type contains between $1$ and
$2\binom{n}2+2$ affine order types. The number of affine order types
contained in a given $\pi \in \OTproj{n}$ can be computed in
polynomial time by examining each affine hemiset in turn, and counting
how many distinct affine order types occur. The number of rejections
is $O(n^2)$ on average.

\paragraph{About concentration.}
The transforms listed above can turn any algorithm $\A$ simulating a
distribution on one sort of order types into an algorithm $\A'$
simulating a distribution on another sort of order types. Let us
remark, however, that when $\A$ is not uniform, our transforms may no
longer compensate exactly the imbalance due to the fact that an
$n$-point order type (affine or projective, labeled or not) may have
from $1$ to $\Theta(n)$ symmetries. We cannot exclude that (the
distribution simulated by) $\A'$ exhibits concentration, altough (the
one simulated by) $\A$ does not. However, if $\A'$ is
\Emph{sufficiently} concentrated, in the sense that a subset $A_n$ of
the order type gets hit with probability going to $1$ but represents a
fraction $o(n^2)$ of all order types, then it must be that $\A$
already exhibits concentration.

\subsection{Models from projective order types}

Starting from \Emph{any} distribution on projective order types, the
(polynomial-time) transform ``projective to labeled affine'' presented above produces a distribution on
labeled affine order types with average number of extreme points equal
to $\TheNumber$, \Emph{just like the equiprobable distribution on
  $\LOTaff n$}. In particular, the selection of an affine hemiset
equiprobably (whether or not we account for symmetries) seems
effective at breaking the ``reducibility'' barrier pointed out right
after Conjecture~\ref{c:strong}.

\bigskip

A natural distribution on projective order types is given by the
projective order type of the projective completion of $n$ points
chosen independently and uniformly on $\SS^2$. This leads to two
natural distributions on labeled affine order types:
\begin{description}
\item[Geometric projection:] pick a hemisphere uniformly at random
  among all hemispheres,  read off the order type of the affine
  hemiset that it determines almost surely, and conclude by ordering the points uniformly at
random.
\item[Combinatorial projection:] pick an affine hemiset equiprobably, read
  off its order type and order uniformly at random.
\end{description}
In other words, the geometric projection selects an affine hemiset
with probability proportional to the area of its dual cell (rather
than with equiprobability).
\begin{question}
  Does the distribution on affine order types given by the geometric
  or combinatorial projection of the uniform measure on $\SS^2$
  exhibit concentration?
\end{question}
\noindent
For the geometric projection, concentration would follow from our
Conjecture~\ref{c:strong}. Note that order types obtained from the
geometric projection already have a constant number of extreme points
on average~\cite{barany2017random,kabluchko2019cones}, so
Conjecture~\ref{c:weak} would not suffice.

\bibliographystyle{acm} \bibliography{ExtrInOrderJACM}{}

\end{document}